\documentclass[a4paper, 11pt]{article} 
\pdfoutput=1
\usepackage[utf8]{inputenc}
\usepackage[T1]{fontenc}
\usepackage[english]{babel}
\usepackage{lmodern, geometry,
	jheppub, amsmath, graphicx, xcolor, subfigure,
	feynmp, slashed, adjustbox, rotating, pdflscape}

\graphicspath{{figs/}}

\makeatletter\g@addto@macro\bfseries{\boldmath}\makeatother

\allowdisplaybreaks

\newcommand{\epem}{\ensuremath{e^+e^-}}
\newcommand{\eett}{\ensuremath{e^+e^-\to t\:\bar t}}
\newcommand{\bwbw}{\ensuremath{bW^+\:\bar bW^-}}
\newcommand{\eebwbw}{\ensuremath{e^+e^-\to \bwbw}}

\def\TeV{\ifmmode {\mathrm{\,Te\kern -0.1em V}}\else
                   \textrm{\,Te\kern -0.1em V}\fi}%
\def\GeV{\ifmmode {\mathrm{\,Ge\kern -0.1em V}}\else
                   \textrm{\,Ge\kern -0.1em V}\fi}%
\def\MeV{\ifmmode {\mathrm{\,Me\kern -0.1em V}}\else
                   \textrm{\,Me\kern -0.1em V}\fi}%
\def\keV{\ifmmode {\mathrm{\,ke\kern -0.1em V}}\else
                   \textrm{\,ke\kern -0.1em V}\fi}%
\def\eV{\ifmmode  {\mathrm{\,e\kern -0.1em V}}\else
                   \textrm{\,e\kern -0.1em V}\fi}%
\let\tev=\TeV
\let\gev=\GeV
\let\mev=\MeV

\def\iab{\mbox{\,ab$^{-1}$}}
\def\ifb{\mbox{\,fb$^{-1}$}}

\newcommand{\mg}{\texttt{MG5\_aMC@NLO}}

\newcommand{\FDF}[1][]{\varphi^\dagger #1\!\overleftrightarrow{D}\!_\mu\varphi}
\newcommand{\FDFI}[1][]{\varphi^\dagger #1\!\overleftrightarrow{D}^I\!\!\!_\mu\:\varphi}

\DeclareMathOperator{\sign}{sign}
\let\Re\undefined
\let\Im\undefined
\DeclareMathOperator{\Re}{Re}
\DeclareMathOperator{\Im}{Im}

\newcommand{\ckm}{\ensuremath{V_\text{\tiny CKM}}}
\newcommand{\pmns}{\ensuremath{V_\text{\tiny PMNS}}}

\DeclareMathAlphabet{\mathsfit}{\encodingdefault}{\sfdefault}{m}{sl}
\newcommand{\ges}[1]{\mathsfit{#1}}

\newcommand{\n}{1.8mm}
\newcommand{\nn}{1.8mm}
\newcommand{\ph}{\phantom{33}\cdot\phantom{33}}
\newcommand*{\mysec}[6]{
	\ensuremath{%
	\begin{aligned}
	#1%
	=
	#2%
	+
	\Bigg(\frac{1\text{ TeV}}{\Lambda}\Bigg)^2
	\begin{pmatrix}
		C_{lq}^{A}	\\[\nn]
		C_{eq}^{A}	\\[\nn]
		C_{\varphi q}^{A}	\\[\nn]
		C_{lq}^{V}	\\[\nn]
		C_{eq}^{V}	\\[\nn]
		C_{\varphi q}^{V}\\[\nn]
		C_{uZ}^R	\\[\nn]
		C_{uA}^R	\\[\nn]
		C_{uZ}^I	\\[\nn]
		C_{uA}^I
	\end{pmatrix}^{\hspace{-1mm}T}
	\begin{pmatrix}
	#3
	\end{pmatrix}
	&+\Bigg(\frac{1\text{ TeV}}{\Lambda}\Bigg)^4
	\begin{pmatrix}
		C_{lq}^{A}	\\[\nn]
		C_{eq}^{A}	\\[\nn]
		C_{\varphi q}^{A}	\\[\nn]
		C_{lq}^{V}	\\[\nn]
		C_{eq}^{V}	\\[\nn]
		C_{\varphi q}^{V}\\[\nn]
		C_{uZ}^R	\\[\nn]
		C_{uA}^R	\\[\nn]
		C_{uZ}^I	\\[\nn]
		C_{uA}^I
	\end{pmatrix}^{\hspace{-1mm}T}
	\begin{pmatrix}
	#4
	\end{pmatrix}
	\begin{pmatrix}
		C_{lq}^{A}	\\[\nn]
		C_{eq}^{A}	\\[\nn]
		C_{\varphi q}^{A}	\\[\nn]
		C_{lq}^{V}	\\[\nn]
		C_{eq}^{V}	\\[\nn]
		C_{\varphi q}^{V}\\[\nn]
		C_{uZ}^R	\\[\nn]
		C_{uA}^R	\\[\nn]
		C_{uZ}^I	\\[\nn]
		C_{uA}^I
	\end{pmatrix}
	\\
	&+\Bigg(\frac{1\text{ TeV}}{\Lambda}\Bigg)^4
	\Bigg\{
		\def\tmp{#6}
		\ifx\tmp\empty
		#5	\:\Re\Big(C_{lequ}^{S*}C_{lequ}^T\Big)
		\else
		#5	\;\Big|C_{lequ}^S\Big|^2
		#6	\;\Big|C_{lequ}^T\Big|^2
		\fi
	\Bigg\}
	\end{aligned}%
	}
}

\title{Global and optimal probes for the top-quark effective field theory at future lepton colliders}
\author[a]{Gauthier Durieux,}
\author[b]{Mart{\'i}n Perell{\'o},}
\author[b]{Marcel Vos}
\author[c,d]{and Cen Zhang}

\affiliation[a]{DESY Notkestraße 85, D-22607, Hamburg, Germany}
\affiliation[b]{IFIC, Universitat de València and CSIC, Catedrático Jose Beltrán 2, E-46980 Paterna, Spain}
\affiliation[c]{Department of Physics, Brookhaven National Laboratory, Upton, NY 11973, USA}
\affiliation[d]{Institute of High Energy Physics, Chinese Academy of Sciences, Beijing, 100049, China}

\emailAdd{gauthier.durieux@desy.de}
\emailAdd{martin.perello@ific.uv.es}
\emailAdd{marcel.vos@ific.uv.es}
\emailAdd{cenzhang@ihep.ac.cn}

\abstract{
We study the sensitivity to physics beyond the standard model of precise top-quark pair production measurements at future lepton colliders. A global effective-field-theory approach is employed, including all dimension-six operators of the Warsaw basis which involve a top-quark and give rise to tree-level amplitudes that interfere with standard-model $\eett\to\bwbw$ ones in the limit of vanishing $b$-quark mass. Four-fermion and CP-violating contributions are taken into account. Circular-collider-, ILC- and CLIC-like benchmark run scenarios are examined. We compare the constraining power of various observables to a set of statistically optimal ones which maximally exploit the information contained in the fully differential \bwbw\ distribution. The enhanced sensitivity gained on the linear contributions of dimension-six operators leads to bounds that are insensitive to quadratic ones. Even with statistically optimal observables, two centre-of-mass energies are required for constraining simultaneously two- and four-fermion operators. The impact of the centre-of-mass energy lever arm is discussed, that of beam polarization as well. A realistic estimate of the precision that can be achieved in ILC- and CLIC-like operating scenarios yields individual limits on the electroweak couplings of the top quark that are one to three orders of magnitude better than constraints set with Tevatron and LHC run~I data, and three to two hundred times better than the most optimistic projections made for the high-luminosity phase of the LHC. Clean global constraints can moreover be obtained at lepton colliders, robustly covering the multidimensional effective-field-theory space with minimal model dependence.
}

\preprint{\begin{tabular}{@{}r@{}}DESY 18-096\\IFIC 18-27\end{tabular}}

\begin{document}

\def\figureautorefname~#1\null{Fig.\,#1\null}
\def\subfigureautorefname~#1\null{Fig.\,#1\null}
\def\equationautorefname~#1\null{Eq.\,(#1)\null}
\def\sectionautorefname~#1\null{Sec.\,#1\null}
\def\subsectionautorefname~#1\null{Sec.\,#1\null}
\newcommand{\fullref}[2]{\hyperref[#2]{#1\,\ref*{#2}}}
\makeatletter
\newcommand*{\lcite }[1]{\hyper@@link[cite]{}{cite.#1}{\cite{#1}}}
\newcommand*{\lcites}[2]{\hyper@@link[cite]{}{cite.#1}{\cite{#2}}}
\newcommand*{\rcite }[1]{\hyper@@link[cite]{}{cite.#1}{Ref.\,\cite{#1}}}
\newcommand*{\rcites}[2]{\hyper@@link[cite]{}{cite.#1}{Refs.\,\cite{#2}}}
\makeatother%

\begin{fmffile}{figs/fgraph}
\setlength{\unitlength}{1mm}
\DeclareGraphicsRule{*}{mps}{*}{}
\fmfcmd{%
	prologues:=3;
	arrow_ang := 12;
	arrow_len := 3.5thick;
}

\maketitle

\section{Introduction}
\label{sec:intro}

The particle content of the standard model (SM) is experimentally confirmed through the discovery of a state with properties compatible with that of its Higgs boson. Searches at the LHC exploring the $\tev$ scale for signs of new physics have so far come up empty-handed. A complementary probe for physics beyond the standard model (BSM) is found in a precise characterization of standard-model processes with a pronounced BSM sensitivity. Using a global effective field theory, we evaluate in this paper the potential of future electron-positron colliders to constrain new physics affecting top-quark interactions.

The electroweak couplings of the top quark constitute one of the least precisely constrained quantities in the standard model. At the Tevatron and LHC, the
electroweak $q\bar{q} \rightarrow Z^*/\gamma^* \to t\, \bar{t}$ process is inaccessible. The hadron-collider experiments can probe the
charged-current electroweak interactions of the top quark in its decay and single production. They have also started to probe the couplings
of the top quark to the photon and $Z$ boson in associated production.

Lepton colliders can probe top-quark couplings with neutral electroweak gauge bosons directly in the \eett\ pair production process. ILC studies~\cite{Amjad:2015mma, Amjad:2013tlv} relying on a full simulation of the detector response and on estimates for the main systematic uncertainties have shown that cross section and forward-backward asymmetry measurements would yield percent-level determinations of the anomalous couplings of the top quark to the photon and $Z$ boson, with $500\ifb$ of integrated luminosity shared between two beam polarization configurations at a centre-of-mass energy of $500\gev$. Other studies on this topic include Refs.~\cite{Kane:1991bg, Atwood:1991ka, Schmidt:1995mr, Grzadkowski:1997cj, Brzezinski:1997av, Grzadkowski:1998bh, Boos:1999ca, Jezabek:2000gr, Grzadkowski:2000nx, Devetak:2010na, AguilarSaavedra:2012vh, Rontsch:2015una, Janot:2015yza, Khiem:2015ofa, Englert:2017dev}.

Improving on the anomalous coupling description, we parametrize deviations from the standard model in top-quark pair production using a broad set of dimension-six effective operators. We include all ten dimension-six operators of the so-called Warsaw basis~\cite{Grzadkowski:2010es} involving a top quark and interfering with the resonant standard-model $\eett\to\bwbw$ amplitudes, at leading order and in the limit of vanishing bottom-quark mass.
This set contains four-fermion operators that are are absent in anomalous coupling descriptions as well as the CP-violating components of top-quark electroweak dipole operators.
Four-fermion operators with a scalar or tensor Lorentz structure are treated separately. Various observables are discussed, in particular, the statistically optimal observables defined on the fully differential \bwbw\ final state. They form a discrete set exactly sufficient to maximally exploit the information contained in distributions. For simplicity, the narrow top-quark width approximation and a vanishing bottom-quark mass are used in their definitions. The impact of non-resonant and higher-order corrections or detector effects should be determined before a comparison with real data. No major obstacle seems to prevent the achievement of those tasks. A detailed study of the detector response relying of full simulation will be presented elsewhere (see also Ref.\,\cite{clictop}). We have extended the existing implementation of top-quark electroweak operators at next-to-leading order (NLO) accuracy in QCD~\cite{Bylund:2016phk} that was already available in the \mg\ software suite~\cite{Alwall:2014hca} to include four-fermion and CP-violating electroweak dipole operators. Prediction at that order, matched to parton shower, and including off-shell top-quark effects are thus made available for these operators in the \eebwbw\ process.

The impact of runs at various centre-of-mass energies and for several beam polarizations is examined. We derive global constraints on the whole ten-dimensional effective-field-theory parameter space considered for different collider programmes. A special focus is devoted to two benchmark run scenarios covering the ranges of possibilities contemplated by future linear colliders, with runs at centre-of-mass energies of $500\gev$ and $1\tev$ and $P(e^+,e^-)=(\pm30\%,\mp80\%)$ beam polarizations for the ILC~\cite{Baer:2013cma, Djouadi:2007ik}, and $380\gev$, $1.4\tev$ and $3\tev$ runs with $P(e^+,e^-)=(0\%,\mp80\%)$ beam polarizations for CLIC~\cite{Linssen:2012hp}. Circular lepton collider could also access top-quark pair production, collecting for instance $1.5\,\iab$ at a centre-of-mass energy of $365\gev$, in addition to $200\,\ifb$ at the top-quark pair production threshold, without beam polarization~\cite{Benedikt:2018}. We will also briefly discuss this scenario.

This paper is organized as follows. Top-quark production at lepton colliders is introduced in \autoref{sec:top_production}. \hyperref[sec:eft]{Section\,\ref{sec:eft}} describes our effective-field-theory setup. Various observables are then discussed in \autoref{sec:obs}, together with their sensitivity to operator coefficients. Statistically optimal observables are treated in a separate \autoref{sec:oo}. The global reach of future lepton collider is then finally presented in \autoref{sec:fit}. Our main results appear in \hyperref[fig:fit_opt_benchmark_cc]{Figs.\,\ref{fig:fit_opt_benchmark_cc}}, \ref{fig:fit_opt_benchmark}, and \ref{fig:fit_opt_benchmark_clic} for the three benchmark run scenarios considered.
Comparisons with existing constraints and various HL-LHC prospects are provided \autoref{sec:lhc}. A few appendices include additional material, notably a conversion of our results into the effective-field-theory conventions established by the LHC TOP WG~\cite{AguilarSaavedra:2018nen}. Useful computer codes and numerical results are made available at \url{https://github.com/gdurieux/optimal_observables_ee2tt2bwbw}.

\section{Top-quark production at lepton colliders}
\label{sec:top_production}

\begin{figure}
\vskip 1cm
\centering
\subfigure[\label{fig:feynmandiag_a}]{
	\begin{fmfgraph*}(30,15) 
	\fmfleftn{i}{2}
	\fmfrightn{o}{2}
	\fmf{fermion}{i1,v1,i2}
	\fmf{photon, lab=$Z^*,,\gamma^*$, lab.side=left}{v1,v2}
	\fmf{fermion}{o1,v2,o2}
	\fmflabel{$\bar{t}$}{o1}
	\fmflabel{$t$}{o2}
	\fmflabel{$e^+$}{i2}
	\fmflabel{$e^-$}{i1}
	\end{fmfgraph*}
}
\hfil
\subfigure[\label{fig:feynmandiag_b}]{
	\begin{fmfgraph*}(30,15)
	\fmfleftn{i}{2}
	\fmfrightn{f}{3}
	\fmf{fermion}{i1,v1}
	\fmf{fermion}{i2,v2}
	\fmf{fermion, lab=$\nu$, lab.side=left}{v1,v2}
	\fmf{phantom}{v2,f3}
	\fmf{photon}{v1,f1}
	\fmffreeze
	\fmf{photon,tens=1.5, lab=$W^*$, lab.side=left}{v2,o2}
	\fmf{fermion}{o2,f2}
	\fmf{fermion}{o2,f3}
	\fmflabel{$t$}{f3}
	\fmflabel{$\bar{b}$}{f2}
	\fmflabel{$W^-$}{f1}
	\fmflabel{$e^+$}{i2}
	\fmflabel{$e^-$}{i1}
	\end{fmfgraph*}
}
\caption{Tree-level diagrams contributing to the $\epem\to\bwbw$
         production at lepton colliders: (a) the top-quark pair production 
         through a $s$-channel $Z$ boson or photon exchange, (b) representative diagram for single top-quark production $\epem\to t\bar{b} W^-$.}
\label{fig:feynmandiag}
\end{figure}
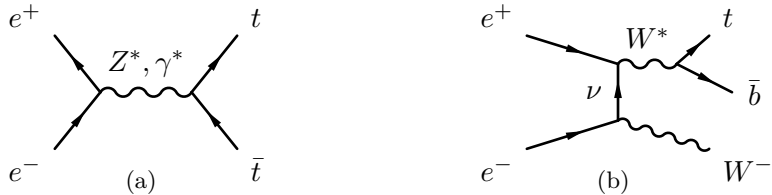

In the standard model, lepton colliders primarily produce top quarks in pairs through a $s$-channel $Z$ boson or photon, as pictured in \autoref{fig:feynmandiag_a}. A number of other processes, including single top-quark production illustrated in \autoref{fig:feynmandiag_b}, also contribute to the \bwbw\ final state. Although certain regions of the $bW$ energies and invariant
masses are enriched in double- and single-resonant
processes~\cite{Fuster:2015jva, Liebler:2015ipp}, a clean separation is
generally not achievable. It is therefore in principle preferable to consider the inclusive
\eebwbw\ process.

\begin{figure}
\centering
\includegraphics{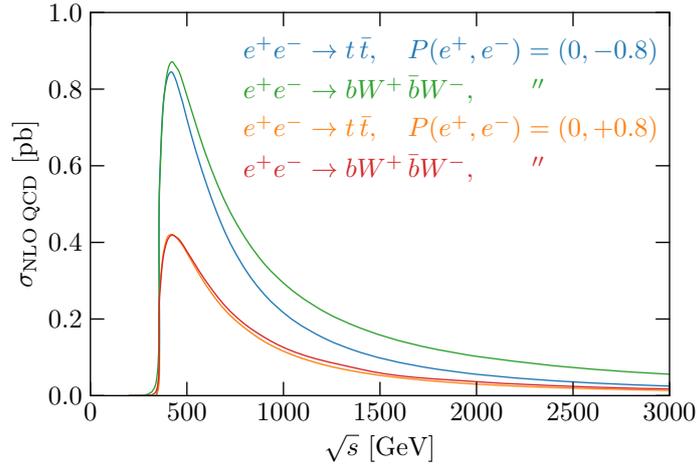}
\caption{The production cross sections for the $e^+e^- \rightarrow t\,\bar{t}$ and
         \eebwbw\ processes, at NLO in QCD, as functions of the centre-of-mass
         energy for two choices of electron beam polarization. Bound-state
         effects that significantly enhance the cross section in the threshold
         region are not included.}
\label{fig:xsecsqrts}
\end{figure}

\fullref{Figure}{fig:xsecsqrts} displays the \eett\ and \eebwbw\ production
cross sections as functions of the centre-of-mass energy. Those results,
accurate to next-to-leading order in QCD, are obtained with
\texttt{MG5\_aMC@NLO}~\cite{Alwall:2014hca}. The significant enhancement
occurring at threshold is not accounted for. Two beam
polarizations are chosen for illustration, with either $80\%$ left-handed
or $80\%$ right-handed electrons.\footnote{Polarized positron beams are
envisaged in the baseline ILC project, with relatively modest polarization
($30\%$ at $500\gev$, $20\%$ at $1\tev$). Using the $P(e^+,e^-) = (+,-)$ and
$(-,+)$ configurations, the effective electron polarization reaches $90\%$,
enhancing the effect observed in \autoref{fig:xsecsqrts}. As the degree to which
the positron beam can be polarized depends on $\sqrt{s}$, positron polarization
is not considered in this figure.} The left-handed polarized electron beam leads
to a significantly larger $t\,\bar{t}$ production cross section. The enhancement is
even more pronounced for single top-quark production, as the
neutrino-exchange diagram in \autoref{fig:feynmandiag_b} is absent for a
right-handed electron.
The pair production process is seen to provide the dominant contribution to
\eebwbw\ production for centre-of-mass energies below about $1\tev$. At higher
energies, single top-quark production overtakes the $s$-channel pair production whose rate approximately falls off as $1/s$. Single and pair production contributions have comparable magnitudes at about $\sqrt{s} \simeq 3\tev$. On the other hand, the \bwbw\ production remains measurable below the pair production threshold down to
$\sqrt{s} \simeq 300\gev$.

\section{Top-quark effective field theory}
\label{sec:eft}

We adopt an effective-field-theory approach to deviations from the standard model in top-quark interactions. Remarkably, it is able to parametrize systematically the theory space
in direct vicinity of the standard model, provided new states are not directly
producible. Unlike anomalous couplings, it also preserves ---by construction---
the full standard-model gauge symmetry and constitutes a proper quantum field
theory whose predictions are improvable perturbatively, order by order.
Importantly, its systematic character is retained only when all operators
contributing to the studied process are simultaneously considered, up to a given
order in the effective-field-theory expansion counted in powers of $1/\Lambda$. Constraining
effectively all directions in this sizeable parameter space requires the
combination of various measurements. Performing global analyses is not only
required but also permitted when using an effective field theory which is able
to consistently make prediction in various sectors. Although more modest
objectives are to be addressed first, a combination of measurements involving the Higgs boson, top and bottom quarks could ultimately be performed in the very same framework.

\subsection{Operators}
In this work, we limit ourselves to effective operators involving at least a top
quark. Although other operators could also contribute to the $\eett\to\bwbw$ process we study, they can presumably be constrained sufficiently well in other processes to have only a small impact on the marginalized bounds in the space of top-quark operator coefficients.

In terms of physical eigenstates, we define the fermionic gauge eigenstate out
of which the gauge-invariant operators are constructed as
\begin{equation*}
	\ges q \equiv (u_L,\ckm d_L)^T,		\qquad
	\ges u \equiv u_R,			\qquad
	\ges d \equiv d_R,			\qquad\quad
	\ges l \equiv (\pmns \nu_L, e_L)^T,	\qquad
	\ges e \equiv e_R.
\end{equation*}
where \ckm\ and \pmns\ are respectively the Cabibbo, Kobayashi,
Maskawa~\cite{Cabibbo:1963yz, Kobayashi:1973fv} and Pontecorvo, Maki, Nakagawa,
Sakata~\cite{Pontecorvo:1957cp, Maki:1962mu, Pontecorvo:1967fh} mixing matrices.
In this work, we approximate both of them by unit matrices. Fermionic mass and
gauge eigenstates are therefore not distinguished. Where they are unspecified,
the generation indices of fields and operator coefficients will be set to `3'
for quarks and `1' for leptons.

We list, in their flavour-generic form, the operators of the chosen basis that
affect the top-quark interactions at leading order. Given that the parity of an
operator dimension is that of $(\Delta B-\Delta L)/2$~\cite{Kobach:2016ami}, all
operator conserving baryon and lepton numbers are of even dimension. We restrict
ourselves to dimension-six operators and rely on the so-called Warsaw basis~\cite{Grzadkowski:2010es} (see also Refs.~\cite{AguilarSaavedra:2008zc,
Zhang:2010dr}), with normalizations convenient for our specific application. The correspondence between our conventions and the standards established within the LHC TOP Working Group~\cite{AguilarSaavedra:2018nen} is detailed in \autoref{sec:lhc_top_wg}.
Two-fermion operators affecting top-quark interactions have vector, tensor, or
scalar Lorentz structures:
\begin{equation}
\begin{array}{@{}rlcc@{}}
	O_{\varphi q}^1
		&\equiv \frac{y_t^2}{2}
		&\bar{\ges q}\gamma^\mu \ges q
		&\FDF[i]
	,\\
	O_{\varphi q}^3
		&\equiv \frac{y_t^2}{2}
		&\bar{\ges q}\tau^I\gamma^\mu \ges q
		&\FDFI[i]
	,\\
	O_{\varphi u}
		&\equiv \frac{y_t^2}{2}
		&\bar{\ges u}\gamma^\mu \ges u
		&\FDF[i]
	,\\
	O_{\varphi ud}
		&\equiv \frac{y_t^2}{2}
		&\bar{\ges u}\gamma^\mu \ges d
		&\varphi^T\!\epsilon\: iD_\mu\varphi
	,
\end{array}
\quad
\begin{array}{@{}rlcc@{}}
	O_{uG}
		&\equiv y_t g_s
		&\bar{\ges q}T^A\sigma^{\mu\nu} \ges u
		&\epsilon\varphi^* G_{\mu\nu}^A
	,\\
	O_{uW}
		&\equiv y_t g_W
		&\bar{\ges q}\tau^I\sigma^{\mu\nu} \ges u
		&\epsilon\varphi^* W_{\mu\nu}^I
	,\\
	O_{dW}
		&\equiv y_t g_W
		&\bar{\ges q}\tau^I\sigma^{\mu\nu} \ges d
		&\epsilon\varphi^* W_{\mu\nu}^I
	,\\
	O_{uB}
		&\equiv y_t g_Y
		&\bar{\ges q}\sigma^{\mu\nu} \ges u
		&\epsilon\varphi^* B_{\mu\nu}
	,
\end{array}
\quad
\begin{array}{@{}rlcc@{}}
	O_{u\varphi}
		&\equiv y_t^3 
		&\bar{\ges q} \ges u
		&\epsilon\varphi^* \; \varphi^\dagger\varphi
	,
\end{array}
\label{eq:op_2q}
\end{equation}
where $\epsilon\equiv(^{\;\;\:0}_{-1}{}^1_0)$ acts on $SU(2)_L$ indices. The
scalar Yukawa operator will however not be relevant for our purpose. The
$O^\pm_{\varphi q} \equiv (O_{\varphi q}^1 \pm O_{\varphi q}^3)/2$ combinations
respectively modify the coupling of the $Z$ to down- and up-type quarks, at
leading order and in the unitary gauge. The corresponding operator coefficients
are $C_{\varphi q}^\pm \equiv C_{\varphi q}^1\pm C_{\varphi q}^3$, so that
schematically $C^1 O^1 + C^3 O^3 = C^+ O^+ + C^- O^-$. Similarly, the
$O_{uA}\equiv s_W^2 O_{uW} + c_W^2 O_{uB}$ and $O_{uZ} \equiv s_Wc_W (O_{uW}
-O_{uB})$ combinations respectively give rise to a tensor coupling of the photon
and $Z$ boson to up-type quarks. In order to satisfy $C_{uW} O_{uW}+ C_{uB}
O_{uB} = C_{uA} O_{uA}+ C_{uZ} O_{uZ}$, the corresponding operator coefficients
are $C_{uA} \equiv C_{uW} + C_{uB}$ and $C_{uZ} = (c_W^2 C_{uW} -s_W^2
C_{uB})/s_Wc_W$, where we have used the $s_W$, $c_W$ shorthands for the sine and
cosine of the weak mixing angle. Additionally, the $O^3_{\varphi q}$ and
$O_{uW}$ operators also modify the interactions of the top quark, left-handed
bottom quark and $W$ boson. The $O_{\varphi ud}$ and $O_{dW}$ operators give rise to interactions between the top quark, right-handed bottom quark, and $W$ boson. In the massless bottom-quark limit we use, these last two operators do not appear in interferences with standard-model amplitudes.

The two-quark--two-lepton operators of the Warsaw basis affecting top-quark interactions can also be grouped according to their Lorentz structures:
\begin{equation}
\begin{array}{rl@{\,}cc}
	O_{lq}^1
		&\equiv \frac12
		&\bar{\ges q}\gamma_\mu \ges q
		&\bar{\ges l}\gamma^\mu \ges l
	,\\
	O_{lq}^3
		&\equiv \frac12
		&\bar{\ges q}\tau^I\gamma_\mu \ges q
		&\bar{\ges l}\tau^I\gamma^\mu \ges l
	,\\
	O_{lu}
		&\equiv \frac12
		&\bar{\ges u}\gamma_\mu \ges u
		&\bar{\ges l}\gamma^\mu \ges l
	,\\
	O_{eq}
		&\equiv \frac12
		&\bar{\ges q}\gamma_\mu \ges q
		&\bar{\ges e}\gamma^\mu \ges e
	,\\
	O_{eu}
		&\equiv \frac12
		&\bar{\ges u}\gamma_\mu \ges u
		&\bar{\ges e}\gamma^\mu \ges e
	,
\end{array}
\qquad
\begin{array}{rlccc}
	O_{lequ}^T
		&\equiv
		&\bar{\ges q}\sigma^{\mu\nu} \ges u
		&\epsilon
		&\bar{\ges l}\sigma_{\mu\nu} \ges e
	,
\end{array}
\qquad
\begin{array}{rlcc}
	O_{lequ}^S
		&\equiv
		&\bar{\ges q}\ges u
		&\epsilon\;
		\bar{\ges l}\ges e
	,\\
	O_{ledq}
		&\equiv
		&\bar{\ges d}\ges q
		&\bar{\ges l}\ges e
	.
\end{array}
\label{eq:op_2q2l}
\end{equation}
As for the two-quark operators, we define the combinations $O^\pm_{lq} \equiv
(O^1_{lq} \pm O^3_{lq})/2$ and $C^\pm_{lq} = C^1_{lq}\pm C^3_{lq}$. The
$O^\pm_{lq}$ operators respectively give rise to interactions between down-type
quarks and charged leptons, and up-type quarks and charged leptons. Because of
their chirality structures, the tensor and scalar operators above do not
lead to amplitudes interfering with standard-model ones in the limit of vanishing lepton masses. All the four-fermion operators above but $O_{ledq}$ contribute at the Born level to the \eett\
process (see \autoref{fig:diag_4f_ee2tt}). The $O^3_{lq}$, tensor and scalar
operators also give rise to single top-quark production (see
\autoref{fig:diag_4f_ee2tbw}) and $t\to b\,\ell^+\nu_\ell$ decay
(see \autoref{fig:diag_4f_t2bev}).

\begin{figure}
\centering
\subfigure[\label{fig:diag_4f_ee2tt}]{
\fmfframe(5,5)(5,5){
\begin{fmfgraph*}(20,10)
	\fmfstraight
	\fmfleftn{l}{2}
	\fmfrightn{r}{2}
	\fmf{fermion}{l1,v1}	\fmflabel{$e^-$}{l1}
	\fmf{fermion}{v1,l2}	\fmflabel{$e^+$}{l2}
	\fmf{fermion}{v1,r2}	\fmflabel{$t$}{r2}
	\fmf{fermion}{r1,v1}	\fmflabel{$t$}{r1}
	\fmfdot{v1}
\end{fmfgraph*}}
}
\hfil
\subfigure[\label{fig:diag_4f_ee2tbw}]{
\fmfframe(5,5)(5,5){
\begin{fmfgraph*}(20,10)
	\fmfstraight
	\fmfleftn{l}{2}
	\fmfrightn{r}{3}
	\fmf{fermion}{l1,v1}		\fmflabel{$e^-$}{l1}
	\fmf{fermion, tens=.5, lab=$\nu$, lab.side=left}{v1,v2}
	\fmf{fermion, tens=2}{v2,l2}	\fmflabel{$e^+$}{l2}
	\fmf{fermion}{v2,r3}		\fmflabel{$t$}{r3}
	\fmf{fermion}{r2,v2}		\fmflabel{$\bar{b}$}{r2}
	\fmf{photon}{v1,r1}		\fmflabel{$W^-$}{r1}
	\fmfdot{v2}
\end{fmfgraph*}}
}
\hfil
\subfigure[\label{fig:diag_4f_t2bev}]{
\fmfframe(5,5)(5,5){
\begin{fmfgraph*}(20,10)
	\fmfstraight
	\fmfleftn{l}{1}
	\fmfrightn{r}{3}
	\fmf{fermion, tens=3}{l1,v1}	\fmflabel{$t$}{l1}
	\fmf{fermion}{v1,r3}		\fmflabel{$b$}{r3}
	\fmf{fermion}{r2,v1}		\fmflabel{$\ell^+$}{r2}
	\fmf{photon}{v1,r1}		\fmflabel{$\nu_\ell$}{r1}
	\fmfdot{v1}
\end{fmfgraph*}}
}
\caption{Some of the various processes to which the two-quark--two-lepton
operators of \autoref{eq:op_2q2l} contribute.}
\end{figure}
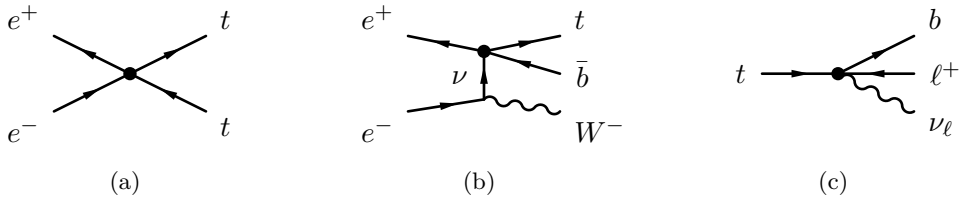

By convention, our effective Lagrangian includes the Hermitian conjugate of all
operators, even though the flavour-diagonal component of some of them
are already Hermitian: $\mathcal{L}_\text{EFT} = \sum_i
\left(\frac{C_i}{\Lambda^2} O_i + \text{h.c.}\right)$. This for instance compensates for the factors of $\frac12$ appearing in front of vector four-fermion operators. 
The energy scale
$\Lambda$ is conventionally set to $1\tev$. We
treat the coefficients of the $O_{\varphi ud}$, tensor ($O_{uG}$ excepted), and
scalar operators as complex. Their imaginary parts produce CP-violating effects
and have no interference with the standard model, in motion-reversal-even
observables like total rates, at leading order, and in the $\ckm=\mathbb{I}$
approximation we use.

For convenience, we distinguish the real and
imaginary parts of the weak dipole operator coefficients as two different
real degrees of freedom:
\begin{align*}
	C^{R,I}_{uA}
		&= \Re,\Im\{C_{uA}\}
		= \Re,\Im\{ C_{uW} + C_{uB} \},
	\\
	C^{R,I}_{uZ}
		&= \Re,\Im\{C_{uZ}\}
		= \Re,\Im\{ c_W^2 C_{uW} - s_W^2 C_{uB} \}/s_Wc_W.
\end{align*}
The different parity transformation properties and sensitivities of operator combinations featuring vector and axial-vector quark currents also motivate us to define:
\begin{equation*}
	\begin{aligned}
		C_{lq}^V	&\equiv C_{lu} + C_{lq}^-,	\\
		C_{lq}^A	&\equiv C_{lu} - C_{lq}^-,
	\end{aligned}
	\qquad\quad
	\begin{aligned}
		C_{eq}^V	&\equiv C_{eu} + C_{eq}	,	\\
		C_{eq}^A	&\equiv C_{eu} - C_{eq}	,
	\end{aligned}
	\qquad\quad
	\begin{aligned}
		C_{\varphi q}^V	&\equiv C_{\varphi u} + C_{\varphi q}^-,	\\
		C_{\varphi q}^A	&\equiv C_{\varphi u} - C_{\varphi q}^-.
	\end{aligned}
\end{equation*}

When focusing on operators that appear in top-quark pair production at the linear $1/\Lambda^2$ level and at leading order, one counts $8$ CP-conserving and $2$ CP-violating parameters:
\begin{equation*}
	\begin{array}{c}
	C_{lq}^A
	,\quad
	C_{eq}^A
	,\quad
	C_{\varphi q}^A
	,\\
	C_{lq}^V
	,\quad
	C_{eq}^V
	,\quad
	C_{\varphi q}^V
	,
	\end{array}
	\qquad\;
	\begin{array}{c}
	C_{uZ}^R
	,\quad
	C_{uA}^R
	,\\
	C_{uZ}^I
	,\quad
	C_{uA}^I.
	\end{array}
\end{equation*}
We will mostly focus on these, discussing also the sensitivity to four-fermion scalar and tensor operators $O_{lequ}^{S,T}$. Note that the linear dependence on $C_{\varphi q}$ operators drops out in the differential $t\to bW$ branching fraction in the narrow top-quark width approximation. The expressions of anomalous vertices in terms of effective-field-theory coefficients are provided in \autoref{sec:anomalous_vertices}.

\subsection{Specific models}
\label{sec:concrete}

The effective field theory described above can be matched to specific
new-physics models. Effective-operator coefficients then become functions of the
underlying model parameters which therefore inherit their constraints. Different
patterns of correlations between operator coefficients are produced depending on
the model. Any evidence for non-vanishing operator coefficients can thus also
point at particular extensions of the standard model. For illustration, we
consider in this section two kinds of models: the first one produces two-quark operators only and the second one also yields two-quark--two-lepton operators with vector Lorentz structures. Scalar and tensor four-fermion operators could be obtained from the exchange of heavy scalars or vector leptoquarks (after Fierz rearrangement).

Two-quark operators like $O_{\varphi q}^{1,3}$ and $O_{\varphi u}$ can be
generated at tree level by mixing of standard-model particles with new fields
of identical quantum numbers: $W'$, $Z'$, or heavy quarks. The $O_{uW}$ and
$O_{uB}$ operators are in general generated at the loop level in scenarios like
two-Higgs-doublet or supersymmetric models.

Let us consider, for illustration,
the mixing of new vector-like quarks with the third generation. The complete
matching for such scenarios has been carried out in Ref.~\cite{delAguila:2000rc}. A singlet $U$ with quantum numbers identical to that of $\ges u$ could mix with
the third-generation quark doublet via a $\lambda\;\bar U_R\ges q\; \epsilon
\varphi$ interaction. Integrating out $U$ then gives
\begin{align}
	\frac{C_{\varphi q}^1}{\Lambda^2}=\frac{\lambda^2y_t^{-2}}{4M_U^2},
	&&
	\frac{C_{\varphi q}^3}{\Lambda^2}=-\frac{\lambda^2y_t^{-2}}{4M_U^2}.
\end{align}
So limits on $C_{\varphi q}^-/\Lambda^2$ can be turned into limits on
$\lambda^2/M_U^2$. Direct LHC constraints on $M_U$ are around $1.2\tev$
regardless of $\lambda$~\cite{Aaboud:2017zfn}. Together with a limit of the order of $C_{\varphi q}/\Lambda^2 \lesssim 0.2\tev^{-2}$ which can realistically be obtained at future lepton colliders as we will see below, this would imply $\lambda\lesssim 1.1$.

Alternatively, a $SU(2)_L$ doublet $T=(X,U)$ with hypercharge $7/6$ could mix
with the third-generation quark singlet via $\lambda\; \bar T\ges u\; \varphi$.
In this case, a different operator is generated:
\begin{flalign}
	\frac{C_{\varphi u}}{\Lambda^2}=\frac{\lambda^2y_t^{-2}}{2M_U^2}.
\end{flalign}
Again, limits on $C_{\varphi u}/\Lambda^2$ can expressed in terms of
$\lambda^2/M_U^2$. Furthermore, the pattern of deviation yields information about the underlying new physics. Direct limits on $M_U$ for a $SU(2)_L$ doublet are of the order of $850\gev$~\cite{Aaboud:2017zfn}. Again, combined with $C_{\varphi u}/\Lambda^2 \lesssim 0.2\tev^{-2}$, this would constrain $\lambda\lesssim 0.54$.

Two-quark--two-lepton operators of vector Lorentz structure (like
$O_{lq}^{1,3}$, $O_{lu}$, $O_{eq}$, or $O_{eu}$) could for instance be generated
in models featuring new heavy gauge bosons, such as a $Z'$. In
non-flavour-universal scenarios, couplings to the third generation could be
enhanced. Randall-Sundrum (RS1) models of a warped extra dimension in which
standard-model fermions and gauge fields
propagate~\cite{Agashe:2003zs,Agashe:2007ki} belong to this category. Because
it is localized near the TeV brane, the top-quark has its couplings to the KK
modes of the electroweak gauge bosons enhanced by a factor of
$\mathcal{O}(\xi)$, where $\xi\equiv\sqrt{k\pi r_c}\simeq 6$. On the contrary,
the light fermions are localized near the Planck brane so that their couplings
are suppressed by $\mathcal{O}(\xi^{-1})$.

The electroweak gauge group in the bulk is $SU(2)_L\times SU(2)_R \times U(1)_X$. So we will consider the first KK mode of photon, $Z$, and $Z_X$ ---which is the combination of $U(1)_{R,X}$ orthogonal to $U(1)_Y$--- denoted respectively as $A_1$, $Z_1$ and $Z_{X1}$. The couplings of these three modes to the third-generation $t_{L,R}$ and $b_L$ are enhanced. Considering the benchmark model depicted in appendix A.3 of Ref.~\cite{Agashe:2007ki}, i.e~setting $g_L=g_R$, $c_{t_R}=0$, $c_{t,b_L}=0.4$, the couplings of the three KK modes to the SM fermions can be written as
\begin{flalign}	
	\mathcal{L}\subset eA_{1\mu}J_{A_1}^\mu+g_ZZ_{1\mu}J_{Z_1}^{\mu}
	+g_{Z'}Z_{X1\mu}J_{Z_{X1}}^\mu
\end{flalign}
where $g_Z=g/c_W$, $g_{Z'}=\sqrt{g_R^2+g_X^2}$, and
\begin{flalign}
	J_X^\mu=\sum_f \kappa^X_f\left( \bar f\gamma^\mu f \right)
\end{flalign}
with
\newcommand{\xx}{\frac{-1.13}{\xi}}
\begin{flalign}
	\begin{array}{@{}llll@{}}
		\kappa^{A_1}_{t_L}=\left( \xx+0.2\xi \right)\frac{2}{3},
		&\kappa^{Z_1}_{t_L}=\left( \xx+0.2\xi \right)\left(
		\frac{1}{2}-\frac{2}{3}s_W^2 \right),
		&\kappa^{Z_{X1}}_{t_L}=\left( 0.2\xi \right)\left(
		-\frac{1}{6}{s'_W}^2 \right),
		\\
		\kappa^{A_1}_{t_R}=\left( \xx+0.7\xi \right)\frac{2}{3},
		&\kappa^{Z_1}_{t_R}=\left( \xx+0.7\xi \right)\left(
		-\frac{2}{3}s_W^2 \right),
		&\kappa^{Z_{X1}}_{t_R}=\left( 0.7\xi \right)\left(
		\frac{1}{2}-\frac{2}{3}{s'_W}^2 \right),
		\\
		\kappa^{A_1}_{b_L}=\left( \xx+0.2\xi \right)\left(-\frac{1}{3}\right),
		&\kappa^{Z_1}_{b_L}=\left( \xx+0.2\xi \right)\left(
		-\frac{1}{2}+\frac{1}{3}s_W^2 \right),
		&\kappa^{Z_{X1}}_{b_L}=\left( 0.2\xi \right)\left(
		-\frac{1}{6}{s'_W}^2 \right),
		\\
		\kappa^{A_1}_f=\left( \xx \right)Q,
		&\kappa^{Z_1}_f=\left( \xx \right)\left( T_3-s_W^2Q \right),
		&\kappa^{Z_{X1}}_f=0,
		\; \text{for} f\neq t_L,t_R,b_L
		,
	\end{array}
\end{flalign}
where $s_W'=g_X/g_{Z'}$. Once the KK modes are integrated out, at leading order,
the effective Lagrangian is
\begin{flalign}
	\mathcal{L}_\text{eff}=-\frac{1}{2M_{A_1}^2}e^2J_{A_1\mu}J_{A_1}^\mu
	-\frac{1}{2M_{Z_1}^2}g_Z^2J_{Z_1\mu}J_{Z_1}^\mu
	-\frac{1}{2M_{Z_{X1}}^2}g_{Z'}^2J_{Z_{X1}\mu}J_{Z_{X1}}^\mu
	.
\end{flalign}

The four-fermion operators involving only light fermions are suppressed
by $\mathcal{O}(\xi^{-2})$ and negligible. We focus on the operators
that involve two light fermions and two top quarks, for which the suppression
factor $\xi^{-1}$ on the light fermion side cancels the enhancement factor
$\xi$ on the top-quark side.
Using $M_{A_1,Z_1}=m_{KK}$ and $M_{Z_{X1}}\simeq 0.981 m_{KK}$, and inserting
the values of $\kappa^X_f$ for electron and top-quark fields, we find
\begin{equation*}
	\frac{C_{lq}^{-}}{\Lambda^2}=\frac{-0.022}{m_{KK}^2}
	,\qquad
	\frac{C_{lu}}{\Lambda^2}=\frac{-0.032}{m_{KK}^2}
	,\qquad
	\frac{C_{eq}}{\Lambda^2}=\frac{-0.004}{m_{KK}^2}
	,\qquad
	\frac{C_{eu}}{\Lambda^2}=\frac{-0.064}{m_{KK}^2}
	.
\end{equation*}
Indirect constraints on the $KK$ mass can thereby be derived. Limits of the order of $10^{-1}$, $10^{-3}$ and $10^{-4}$ ---which can be obtained at circular, ILC- and CLIC-like colliders, as we will see--- would respectively translate into $m_{KK}\gtrsim 0.5$, $5$ and $15\tev$.

Note however that the generated two-fermion operator coefficients are not
negligible. The modification if light-fermion couplings are suppressed, and can
be captured by a redefinition of the oblique parameters \cite{Agashe:2003zs}, so
we will not consider them here. We need to take into account the contribution to
$Ztt$ couplings, which correspond to the operators $O_{\varphi q}^{-}$ and
$O_{\varphi u}$. It is interesting to ask which kind of operator better
probes the proposed scenario, at a given collider. The four-fermion operators describe the underlying process $e^+e^-\to Z'\to t\, \bar t$. As mentioned above, this is neither suppressed nor enhanced by $\xi$, and its amplitude is of order $s/m_{KK}^2$. The vertex operators, on the other hand, describe the process $e^+e^-\to Z\to Z'\to t\,\bar t$ through $ZZ'$ mixing, which is of order $\xi m_Z^2/m_{KK}^2$. Furthermore the $Z' t\bar t$ vertex is enhanced by a factor of $\xi$. Altogether, this amplitude is of order $\xi^2 m_Z^2/m_{KK}^2$. Therefore the relative contributions of two- and four-fermion operators depend approximately on the relative size of collider energy $\sqrt{s}$ and $\xi m_Z$. As a comparison, in the same scenario described above, we find
\begin{flalign}
	\frac{C_{\varphi q}^{-}}{\Lambda^2}=\frac{1.2}{m_{KK}^2}
	,\qquad
	\frac{C_{\varphi u}}{\Lambda^2}=\frac{-5.1}{m_{KK}^2}
\end{flalign}
Which would translate into $m_{KK}\gtrsim 4\tev$ for $C_{\varphi q,u}/\Lambda^2\lesssim 0.2\tev^{-2}$. Of course, in practice both effects should be accounted for simultaneously and a global analysis performed.

The top-quark also plays a distinct role in composite Higgs models. Pair production measurements at lepton colliders are therefore especially sensitive to such scenarios. A dedicated discussion will be presented elsewhere~\cite{DurieuxMatsedonskyi2018}.

\subsection{Validity}
\label{sec:validity}
The use of effective field theories relies on a low-energy decoupling
theorem~\cite{Appelquist:1974tg}. It states that, in unbroken gauge theories,
the effects of heavy fields on phenomena observed at energies scales $E$ much
smaller than their masses $M$ are suppressed by powers of $E/M$. In this
low-energy limit, these effects can be parametrized by a tower of
higher-and-higher dimensional operators suppressed by higher-and-higher powers
of $1/M$. Retaining only the operators of lowest dimensions is then justified
when $E/M\ll 1$.

In a bottom-up approach that does not assume some specific heavy new physics, the operator coefficients $C$ and mass scale $\Lambda$ stand for unknown combinations of couplings and masses. They always appear as $C/\Lambda^2$ ratios (for dimension-six operators), on which the only model-independent information arises from the experimentally imposed constraints. The validity of the low-energy limit is then rather intangible: it requires the physical scale $M$ of the underlying theory to be significantly higher than the energies $E$ probed in the process described through our effective field theory. Without explicit model or power counting, this condition cannot be translated in terms of $C$ and $\Lambda$, and no statement about the relative magnitude of different operator coefficients ---of identical or different dimension--- can be made a priori either. As we will not consider the dependence on operators of higher dimensions, our results will only be interpretable in models where dimension-six operators provide the leading contributions to the observables considered.

Under the assumption that higher-dimensional operators are subleading, one may  examine what truncation of the series in powers of dimension-six operator coefficients is sensible. The expansion of an observable in dimension-six operator coefficients can be expected to contain higher-and-higher powers of $CE^2/\Lambda^2$ where, again, $E$ is the characteristic energy scale of the process considered. For tree-level operator insertions and a coefficient normalization that is natural to the observable considered, numerical prefactors in this series can naively be expected to be of order one, schematically:
\begin{equation*}
	\frac{O}{O^\text{SM}} =
		1
		+\mathcal{O}(1) \frac{CE^2}{\Lambda^2}
		+\mathcal{O}(1) \left(\frac{CE^2}{\Lambda^2}\right)^2
		+\ldots
\end{equation*}
More moderate growths with energy are also possible, especially for the linear terms which arises from interferences between effective-field-theory and standard-model amplitudes. This first term roughly dominates the expansion when $CE^2/\Lambda^2\lesssim 1$. Given the experimental bound on $C/\Lambda^2$ and the energy $E$ of the process considered, one can explicitly check whether this condition is satisfied. If it is not, higher powers of $C/\Lambda^2$ should also be included. Note that dimension-eight operators in principle start contributing at order $1/\Lambda^4$. Note also that the renormalizability of quadratic dimension-six contributions in general requires counterterms from dimension-eight operators. Quantum perturbativity restrictions on the dimension-six effective field theory can also be derived by considering the insertion of more-and-more operators in higher-and-higher numbers of loops. For one additional insertion in each loop and a suitable coefficient normalization, such a series contains higher-and-higher powers of $CE^2/(4\pi\Lambda)^2$. Perturbativity is therefore lost when $CE^2/\Lambda^2$ approximately exceeds $(4\pi)^2$.

At lepton colliders, one can expect precision measurements at each
centre-of-mass energy (which constitutes the characteristic scale of a production
process) to reach the percent level. The combination of such
percent-level measurements, provided they are sufficiently complementary, should be able to constrain all directions
of the effective-field-theory parameter space at the $CE^2/\Lambda^2<10^{-2}$
level. Terms beyond the leading one are then naively expected to induce
only percent-level corrections to the constraints set with a linearised effective field theory.
Complications invalidating this general reasoning are however bound to arise in
some specific cases. In particular, some dimension-six operators are known to
have (highly) suppressed interferences with the standard model. Helicity
selection rules can for instance cause such suppressions~\cite{Azatov:2016sqh}.
As mentioned earlier, they also occur for operator containing a tensor or scalar
lepton current, involving a right-handed $b$-quark, or violating CP. The
dominant contributions of such operators to observables can then arise from the
square of amplitudes in which they are inserted once. 
In practice, we will focus mostly on linear effective-field-theory dependences and examine the quadratic dimension-six contributions to ensure they are subleading. The quadratic dependence of scalar and tensor four-fermion operators will be discussed briefly.

\section{Observables and sensitivities}
\label{sec:obs}
\label{sec:sensitivity}

\newcommand{\twofig}[2]{\parbox{.5\textwidth}{#1\hfill}\parbox{.5\textwidth}{#2\hfill}}

\newcommand{\nol}[1]{\text{---}}

\newcommand{\vall}[5]{%
	\def\central{#1}
	\def\myzero{0}
	\ifx\central\myzero%
		\text{---}%
	\else%
		\underset{#2}{#1}%
		\begin{array}{@{\:}l@{}}
			\scriptstyle #4\%\\[-1.3mm]
			\scriptstyle \pm #5\%\\[-1.3mm]
			\scriptstyle #3\%
		\end{array}%
	\fi%
}%

We examine the sensitivity of several observables to the operators affecting top-quark pair production at lepton colliders. To efficiently constrain all directions in the multidimensional space of operator coefficients, various observables are to be combined. Runs with different beam polarizations and centre-of-mass energies provide complementary handles.

\subsection{Cross section and forward-backward asymmetry}
\paragraph{Definitions}
Let us consider first the simplest observables one can define in the \eett\ process, namely the total
cross section $\sigma$ and forward-backward asymmetry $A^\text{FB}$ which is defined as follows:
\begin{equation*}
	A^\text{FB}\equiv \frac{\sigma^\text{FB}}{\sigma}
	\qquad\text{with }
	\sigma^\text{FB}	\equiv
		\int_{-1}^{+1} \text{d}\cos\theta_t	\quad
		\sign\{\cos\theta_t\}			\quad
		\frac{ \text{d}\sigma }{ \text{d}\cos\theta_t },
\end{equation*}
where $\theta_t$ is the angle between the positron and top-quark momenta in the
centre-of-mass frame.

\paragraph{Computation}
We computed the standard-model, linear, and quadratic dependences of the total and forward-backward cross sections both analytically at leading order, and using \mg~\cite{Alwall:2014hca} at next-to-leading order in QCD, for the various initial-state helicities. For illustration, the leading-order total and forward-backward \eett\ cross sections, for unpolarized beams and $\sqrt{s}=500\gev$, are displayed in \autoref{tab:lo_500_00}. The linear effective-field-theory dependences computed at NLO in QCD are provided in \autoref{app:numerics_nlo}.

As can be seen in \autoref{fig:tt_kfactors} for total (left) and forward-backward integrated (right) cross sections, the NLO QCD $k$-factors in the standard model and in the linearised effective field theory are fairly similar, diverging from each other mostly at higher centre-of-mass energies. The ratios of those two types of contributions, which assess the sensitivity to operator coefficients, are therefore only marginally affected by QCD corrections. In the following, central values will always be assumed to confirm the standard-model hypothesis, and deviations will be considered at leading order everywhere but in \autoref{sec:oo_theoretical_robustness}.

\begin{table}\centering
\adjustbox{max width = \textwidth}{%
\mysec{\sigma_{500\gev} \text{ [fb]}}{
	+568   
}{
	+221   \\[\n]
	-194   \\[\n]
	+7.01  \\[\n]
	-1110  \\[\n]
	-737   \\[\n]
	-8.24  \\[\n]
	+33.8  \\[\n]
	+209   \\[\n]
	\ph    \\[\n]
	\ph    
}{
	+367   & \ph    & +13.2  & \ph    & \ph    & \ph    & \ph    & \ph    & \ph    & \ph    \\[\n]
	\ph    & +367   & -11.5  & \ph    & \ph    & \ph    & \ph    & \ph    & \ph    & \ph    \\[\n]
	\ph    & \ph    & +0.209 & \ph    & \ph    & \ph    & \ph    & \ph    & \ph    & \ph    \\[\n]
	\ph    & \ph    & \ph    & +868   & \ph    & +31.1  & -128   & -197   & \ph    & \ph    \\[\n]
	\ph    & \ph    & \ph    & \ph    & +868   & -27.3  & +112   & -197   & \ph    & \ph    \\[\n]
	\ph    & \ph    & \ph    & \ph    & \ph    & +0.493 & -4.05  & -0.432 & \ph    & \ph    \\[\n]
	\ph    & \ph    & \ph    & \ph    & \ph    & \ph    & +9.36  & +2     & \ph    & \ph    \\[\n]
	\ph    & \ph    & \ph    & \ph    & \ph    & \ph    & \ph    & +25.2  & \ph    & \ph    \\[\n]
	\ph    & \ph    & \ph    & \ph    & \ph    & \ph    & \ph    & \ph    & +2.51  & +0.536 \\[\n]
	\ph    & \ph    & \ph    & \ph    & \ph    & \ph    & \ph    & \ph    & \ph    & +6.75  
}{
	+1600  
}{	+13900 
}%
,}
\\[1mm]
\adjustbox{max width = \textwidth}{%
\mysec{\sigma^\text{FB}_{500\gev} \text{ [fb]}}{
	-231   
}{
	-485   \\[\n]
	+323   \\[\n]
	-13.8  \\[\n]
	+229   \\[\n]
	+201   \\[\n]
	+0.944 \\[\n]
	-3.2   \\[\n]
	-40.2  \\[\n]
	\ph    \\[\n]
	\ph    
}{
	\ph    & \ph    & \ph    & +761   & \ph    & +13.6  & -46.2  & -71.2  & \ph    & \ph    \\[\n]
	\ph    & \ph    & \ph    & \ph    & -761   & +12    & -40.5  & +71.2  & \ph    & \ph    \\[\n]
	\ph    & \ph    & \ph    & +13.6  & +12    & +0.0562& -0.19  & -2.4   & \ph    & \ph    \\[\n]
	\ph    & \ph    & \ph    & \ph    & \ph    & \ph    & \ph    & \ph    & \ph    & \ph    \\[\n]
	\ph    & \ph    & \ph    & \ph    & \ph    & \ph    & \ph    & \ph    & \ph    & \ph    \\[\n]
	\ph    & \ph    & \ph    & \ph    & \ph    & \ph    & \ph    & \ph    & \ph    & \ph    \\[\n]
	\ph    & \ph    & \ph    & \ph    & \ph    & \ph    & \ph    & \ph    & \ph    & \ph    \\[\n]
	\ph    & \ph    & \ph    & \ph    & \ph    & \ph    & \ph    & \ph    & \ph    & \ph    \\[\n]
	\ph    & \ph    & \ph    & \ph    & \ph    & \ph    & \ph    & \ph    & \ph    & \ph    \\[\n]
	\ph    & \ph    & \ph    & \ph    & \ph    & \ph    & \ph    & \ph    & \ph    & \ph    
}{
	+6090  
}{}.%
}
\caption{Leading-order dependence on the effective operator coefficients of the
unpolarized total and forward-backward \eett\ cross sections at
$\sqrt{s}=500\gev$.}
\label{tab:lo_500_00}
\end{table}

\begin{figure}
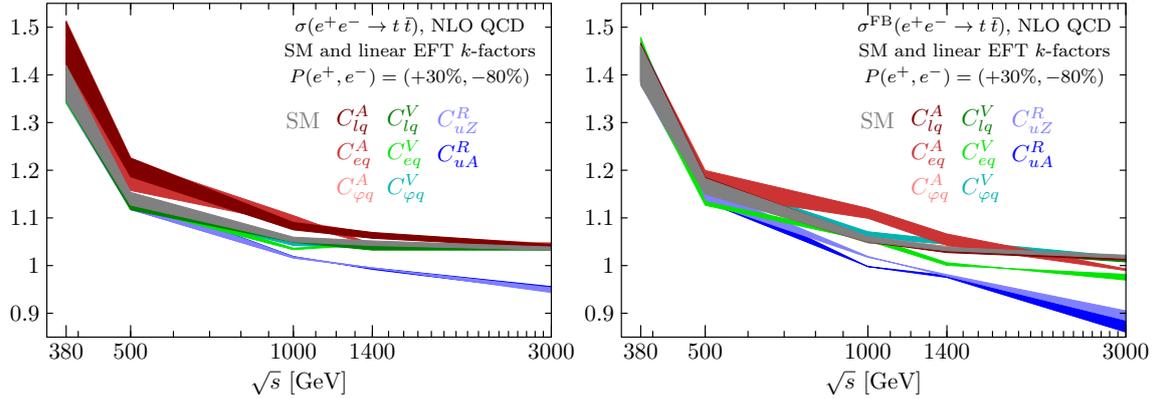
\centering
{\includegraphics[width=.5\textwidth]{kfactor_tt_s_lin.mps}}%
{\includegraphics[width=.5\textwidth]{kfactor_tt_a_lin.mps}}
\caption{The NLO QCD $k$-factors for the linear effective-field-theory dependences of the total and forward-backward cross sections compared to the standard-model one, as functions of the centre-of-mass energy. A mostly left-handed $P(e^+,e^-)=(+30\%,-80\%)$ beam polarization is assumed. The bands' thickness covers QCD renormalization scale variation between $m_t/2$ and $2m_t$.}
\label{fig:tt_kfactors}
\end{figure}

\paragraph{Sensitivity}
\label{sec:sqrts}

We define the sensitivity of an observable $o$ to an operator coefficient $C_i$ as its normalized variation in that direction, around the standard-model point:
\begin{equation}
	S_i^o = \left.\frac{1}{o}
	\frac{\partial o}{\partial C_i}
	\right|_{C_i=0,\;\forall i}
	= \frac{o_i}{o_\text{SM}}
	\qquad\text{with}\quad
	o=o_\text{SM} + C_i o_i + C_i C_j o_{ij} + ...
\label{eq:sensitivity}
\end{equation}
The scale $\Lambda$, conventionally set to $1\tev$ as noted earlier, is absorbed into the definition of $o_i$, $o_{ij}$, etc. The left panel of \autoref{fig:sensitivities} shows the sensitivity of the cross section to operator coefficients, as a function of the centre-of-mass energy, for a mostly left-handed electron beam polarization $P(e^+, e^-) = (+30\%,-80\%)$. It tends to a constant value at high energies for the two-quark operators: $C_{\varphi q}^A$, $C_{\varphi
q}^V$, $C_{uZ}^R$ and $C_{uA}^R$. This behaviour can be understood given that
the $\varphi q$ operators induce $t\bar tZ$ couplings which scale as
$v^2/\Lambda^2$ once the two Higgs fields they contain condense to their vacuum
expectation value. The sensitivity of the cross
section to the $O^V_{\varphi q}$ operator actually slightly decreases with
energy, as $1+2m_t^2/s$. On the other hand, the two $uA$ and
$uZ$ electroweak dipole operators generate three-point interactions scaling as
$Ev/\Lambda^2$, where $E$ is an energy scale characteristic of the momentum transfer in the associated vertex. Their interference with standard-model amplitudes of identical
top-quark helicities however requires a flip of chirality along the quark line, and
thus a top-quark mass insertion.\footnote{It is formally seen that none of the individual helicity amplitude
squared provided in Eq.\,(4) of Ref.~\cite{Schmidt:1995mr} leads to a $\Re\{\mathcal{F}_{1V,1A}^* \mathcal{F}_{2V}\}$ term
proportional to $\gamma\equiv \sqrt{s}/2m_t$.} The resulting linear effective-field-theory contributions
therefore scale with energy exactly as the standard-model cross section and
the sensitivity tends to a constant. As will be discussed below, a sensitivity to
the dipole operators that grows with energy can be recovered through
the interference of different helicity amplitudes once the angular distributions
of the top-quark decay products are considered. The sensitivity of the cross section
to four-fermion operator coefficients $C_{lq}^V$, $C_{lq}^A$, $C_{eq}^V$, $C_{eq}^A$ 
shows the naive $s/\Lambda^2$ increase with energy
expected from dimensional analysis (see dashed black line). The constraints on those operators therefore highly benefits from increased centre-of-mass energies.

\begin{figure}
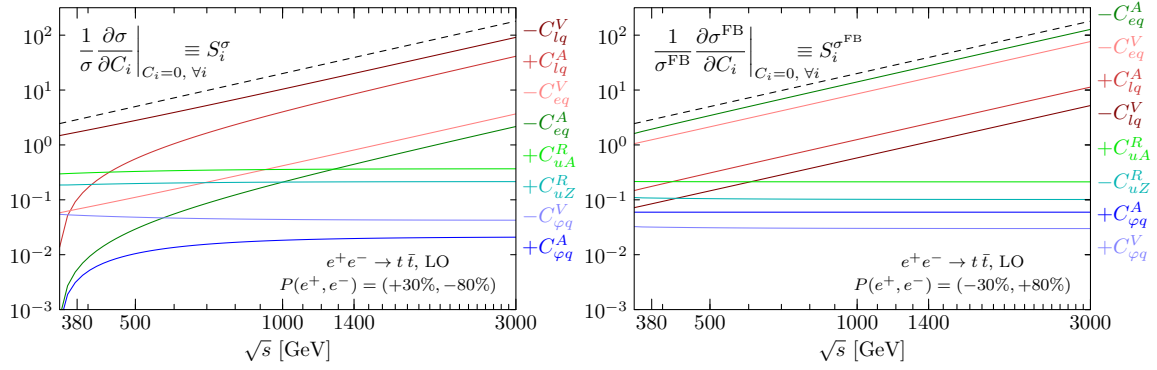
\centering
{\includegraphics[width=.5\textwidth]{sensitivities_left_s.mps}}%
{\includegraphics[width=.5\textwidth]{sensitivities_right_a.mps}}%
\vspace{-3mm}
\caption{Sensitivity of the total (left) and forward-backward (right) \eett\
         cross sections to various operator coefficients, as a function of the
         centre-of-mass energy, for a mostly left-handed (left) and
         right-handed (right) electron beam polarization. The dashed black line
         indicates the slope of a sensitivity scaling as the centre-of-mass
         energy squared.}
\label{fig:sensitivities}
\end{figure}

The right panel of
\autoref{fig:sensitivities} shows the sensitivity of the forward-backward
cross section, for a mostly right-handed electron beam polarization $P(e^+, e^-)= (-30\%,+80\%)$. 
The sensitivity of the forward-backward asymmetry is simply given by
$S^{A^\text{FB}}_i = S^{\sigma^\text{FB}}_i - S^{\sigma}_i$ and is qualitatively
similar to that of $\sigma^\text{FB}$. The mostly right-handed electron polarization 
enhances the sensitivity to $O_{eq}$ operators compared to the $O_{lq}$ one, 
whereas the opposite is true for the mostly left-handed electron polarization.
Interestingly also, the change of polarization reverts the
sign of the $O_{uZ}$ and $O_{\varphi q}^V$ interferences with standard-model
amplitudes. The standard-model couplings of the $Z$ to left- and right-handed electrons, $\frac{e}{2s_Wc_W}(-1+2s_W^2)$ and
$\frac{e}{2s_Wc_W}(2s_W^2)$ respectively, indeed
have different signs. A combination of the two polarizations
therefore provides complementary information on different combinations of
operators. The forward-backward cross section also has an enhanced sensitivity
to the axial-vector combinations of operators,\footnote{Again, from $t\,\bar{t}$ helicity amplitudes in Eq.\,(4) of Ref.~\cite{Schmidt:1995mr}, the forward-backward cross section is proportional to the $|+-|^2-|-+|^2$ difference which is in turn proportional to $\beta\Re\{\mathcal{F}_{1A}(\mathcal{F}_{1V}+\mathcal{F}_{2V})^*\}$ combination of couplings.} while the total cross section is more
sensitive to vector operators. This is especially true at lower energies where
the sensitivity of the total cross section to the $O_{\varphi q}^A$, $O_{eq}^A$,
and $C_{lq}^A$ operators suffers from a so-called \emph{$p$-wave} suppression and falls off as $\beta\equiv(1- 4m_t^2/s)^{1/2}$. In the forward-backward asymmetry, both the standard-model and linear effective-field-theory dependences are proportional to $\beta$ so that this suppression drops out in their ratio.

\begin{figure}
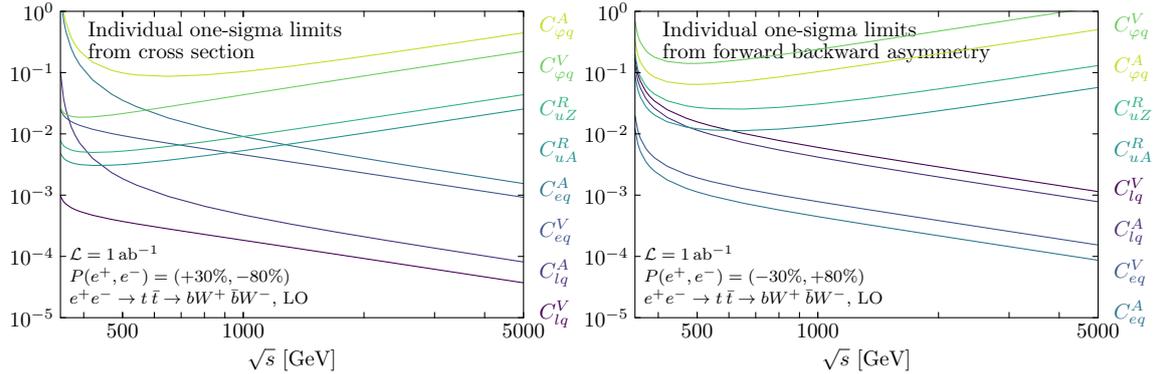
\centering
\includegraphics[width=.5\textwidth]{individual_left_s.mps}%
\includegraphics[width=.5\textwidth]{individual_right_a.mps}%
\vspace{-3mm}
\caption{Individual one-sigma limits on operator
         coefficients as functions of the centre-of-mass energy, with either
         mostly left-handed (left) and mostly right-handed (right) electron beam
         polarizations, from either cross section (left) or forward-backward
         asymmetry (right) measurements, for a fixed integrated luminosity times
         efficiency of $1\,\iab$. Different integrated luminosities are
         trivially obtained through a $1/\sqrt{\mathcal{L}\: [\iab]}$
         rescaling.}
\label{fig:individual_def}
\end{figure}

Individual statistical constraints deriving from the measurements of cross
sections and forward-backward asymmetries are displayed in
\autoref{fig:individual_def} as functions of the centre-of-mass energy. They are
arbitrarily normalized to an integrated luminosity times efficiency of
$1\,\iab$. Note however that, at linear colliders, the instantaneous luminosity which can be achieved scales approximately linearly with the centre-of-mass energy,
while it falls off as the fourth power with the centre-of-mass energy at circular lepton colliders for a constant power of synchrotron radiation emission. Unlike the sensitivity, these idealised individual limits also account for the statistical precision to which cross sections and forward-backward asymmetries can be measured. Quite naturally, the operators whose sensitivity does not grow with energy are more efficiently constrained at lower centre-of-mass energies, where the top-quark pair production cross section is larger. For those, the optimal centre-of-mass energy lies roughly between $400$ and $600\gev$.

\paragraph{Complementarity}

\begin{figure}
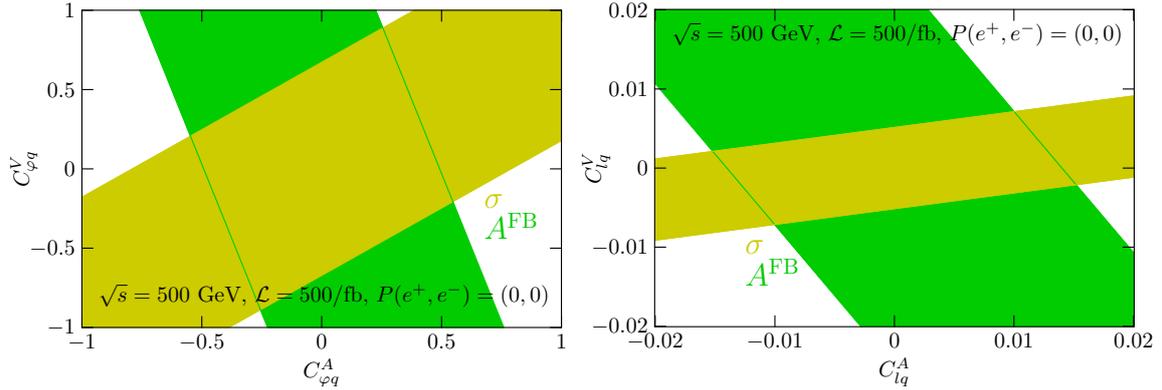
 
\centering
\adjustbox{max width=\textwidth}{%
\includegraphics{s-afb.mps}
\includegraphics{s-afb-1.mps}}
\caption{The $68\%$~C.L. regions allowed by measurements of the cross section and forward-backward asymmetry in \eett\ production. An integrated luminosity of $500\,\ifb$ at a centre-of-mass energy of $500\gev$ is considered, with unpolarized beams. Central values are assumed to confirm the standard model.}
\label{fig:complementarity_xsec_afb}
\end{figure}

With unpolarized beams, the combination of cross section and for\-ward-backward asymmetry measurements allows to simultaneously constrain pairs of operator coefficients, as illustrated in \autoref{fig:complementarity_xsec_afb}.
Runs with two different beam polarizations effectively duplicate the number of observables.
Polarization was shown to effectively provide separate sensitivity to the photon and $Z$-boson form factors in Refs.~\cite{Amjad:2015mma, Amjad:2013tlv}. Similarly, in an effective field theory, dipole operator coefficients $C_{uZ}^R$ and $C_{uA}^R$ can be disentangled very effectively by taking
data in two different beam polarization configurations.
The combination of the cross section and $A^\text{FB}$ measurements with 
unpolarized beams are largely degenerate in this two-dimensional parameter subspace, see \autoref{fig:complementarity_polarization} (left). 
The combination of measurements with different beam polarizations,
on the other hand, yields the tight constraint shown in 
\autoref{fig:complementarity_polarization} (right).

\begin{figure}
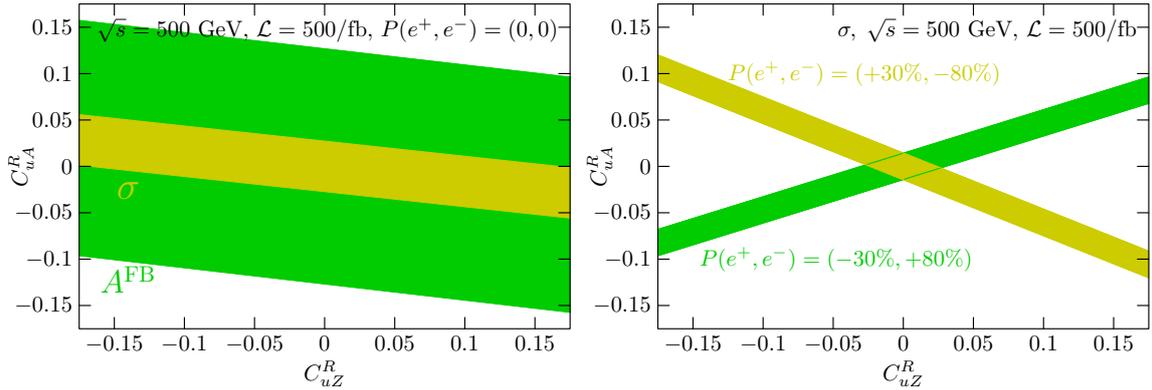
 
\centering
\adjustbox{max width = \textwidth}{%
\includegraphics{pol-0.mps}
\includegraphics{pol-1.mps}%
}
\caption{The 68\% C.L. regions allowed by measurements of the cross section and forward-backward asymmetry in \eett\ production with unpolarized beams (left) and that of the cross sections with two different configurations of the beam polarization (right). A total luminosity of $500\,\ifb$ collected at $500\gev$ is split evenly among two beam polarization configurations. The central values of measurements are assumed to match standard model predictions.}
\label{fig:complementarity_polarization}
\end{figure}

\subsection{Top-quark polarization}
\label{subsec:dipolereal}

We have seen in the previous section that the sensitivity to the electroweak dipole operator coefficients $C_{uA}$, $C_{uZ}$ of the cross section and forward-backward asymmetry is approximately constant as a function of centre-of-mass energy. To achieve a sensitivity that grows with energy we must consider the interference between amplitudes with top quarks of different helicities. This can therefore only be observed through observables incorporating top decay product distributions. Statistically optimal observables based on the $\eett\to\bwbw$ kinematics as well as on the $W$ polarizations were already shown to exhibit such a sensitivity growing with the centre-of-mass energy in Ref.~\cite{Atwood:1991ka}. This technique will be revisited in \autoref{sec:oo}. We examine in this section the sensitivity of a number of alternative observables.

One could study the polarization of the top quark along its direction of motion in the centre-of-mass frame, by examining the angular distribution between one of the top-quark decay products in its rest frame. This so-called \emph{helicity angle} distribution takes the form
\begin{equation}
	\frac{1}{\sigma}
	\frac{\text{d}\sigma}{\text{d}\cos\theta_i}
	= \frac{1}{2} (1+\alpha_i P\cos\theta_i)
\label{eq:top-ang-dis}
\end{equation}
where $P$ is the polarization and $\alpha_i$ is the so-called \emph{analysing power} of the decay product $i$. The charged lepton and down-type quark arising from the $W$ decay are known to have the largest $\alpha_i$.
At leading order and in the absence of $C_{uZ,uA}^I$ and $C_{lequ}^{S,T}$ operator coefficients, $P$ is however not independent of the forward-backward production asymmetry. In terms of $t\,\bar{t}$ helicity amplitudes (see e.g.\ Refs.~\cite{Parke:1996pr, Schmidt:1995mr}), the forward-backward asymmetry is sensitive to the $|{\rm+-}|^2-|{\rm-+}|^2$ combination, while $P$ involves $|{\rm+-}|^2-|{\rm-+}|^2+|{\rm++}|^2-|{\rm--}|^2$. In the standard model, however, $|{\rm++}|^2=|{\rm--}|^2$. This remains true when introducing CP-conserving dipole operators, or two- and four-fermion operators having (axial) vector Lorentz structures.

Generalisations of the $W$ helicity fractions (see \autoref{sec:helicity_frac}) have been proposed in Ref.~\cite{AguilarSaavedra:2010nx}. They are based on the definition of two additional axes in the top-quark rest frame \mbox{---besides} the direction of motion of the $W$ boson--- with respect to which the angle of the charged lepton momentum in the $W$ rest frame could be measured. For this purpose, it is prescribed to use a reference direction along which most of the top-quark polarization lies. Reference~\cite{Parke:1996pr} demonstrated that a convenient such direction is that the incoming positron for the top and that of the incoming electron for the anti-top (in the respective top and anti-top rest frames). One can then define two new axes,
$
\hat{e}\times \hat{W}
\text{ and }
\hat{W}\times (\hat{e}\times \hat{W}),
$
from the directions of the $W$ boson and electron (positron) beam in the top (anti-top) rest frame. One can for instance construct asymmetries based on the sign of the cosine of the angle between either of these directions and the direction of the charged lepton arising from the leptonic $W$ decay measured in that $W$ rest frame.

Inspired by this proposal but aiming to obtain sensitivity to the top-quark polarization instead of that of the $W$, we define \emph{normal} and \emph{transverse} axes in the centre-of-mass frame:
\begin{equation*}
\hat{N}\equiv \hat{t}\times \hat{e}
\quad\text{ and }\quad
\hat{T}\equiv \hat{t}\times \hat{N}
\end{equation*}
from the direction of motion of the (anti-)top $\hat{t}$ and that of the electron (positron) beam $\hat{e}$. Being orthogonal to $\hat{t}$, those vectors are not affected by a subsequent boost in the (anti-)top rest frame. Observables generalizing $P$ can then be constructed from the distribution of the angles between $\hat{N}$ or $\hat{T}$ and one of the top-quark decay product, in the top-quark rest frame. The direction of the charged lepton arising from a semi-leptonic top-quark decay will be employed in the following. Asymmetries based on the sign of the cosine of either of these angles will be named \emph{normal} and \emph{transverse polarization asymmetries}, $A^N$ and $A^T$.

\begin{figure}
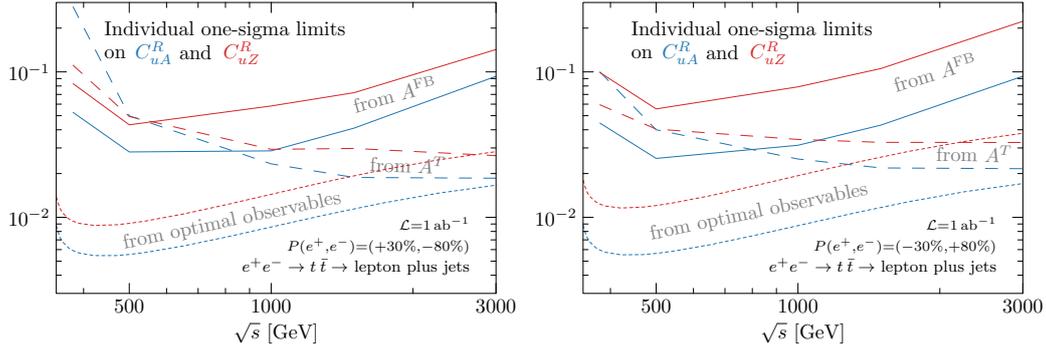

\centering
\includegraphics[width=0.45\linewidth]{real_VA_LR_OO_bis.mps}
\includegraphics[width=0.45\linewidth]{real_VA_RL_OO_bis.mps}%
\caption{Individual one-sigma limits on the coefficients of the dipole operators $O_{uA}$ and $O_{uZ}$ obtained from measurements of the forward-backward production asymmetry $A^\text{FB}$ (solid lines), transverse polarization asymmetry (dashed lines), and statistically optimal observables of \autoref{sec:oo} (dotted lines), with beam polarization $P(e^+, e^-) = (+30\%,-80\%)$ in the left panel and $(-30\%,+80\%)$ in the right panel. A data sample of $1\iab$ is assumed at centre-of-mass energies between the top-quark pair production threshold and $3\tev$.}
\label{fig:dipole_real}
\end{figure}

In the same format as \autoref{fig:individual_def}, we compare in \autoref{fig:dipole_real} the one-sigma limits obtained from measurements of the forward-backward production asymmetry (solid line), the transverse polarization asymmetry (dashed line), and the statistically optimal observables (dotted line) on the coefficients of the dipole operators $C_{uA}^R$ and $C_{uZ}^R$.
As mentioned earlier, the standard helicity angle asymmetry is fully degenerate with the forward-backward production asymmetry $A^\text{FB}$.
Here again, a constant integrated luminosity times efficiency of $1\,\iab$ is assumed at any given centre-of-mass energy.
The left and right panels respectively assume mostly left-handed and mostly right-handed electron beam polarizations.
The effective-field-theory dependences of the top-quark width and of its decay amplitudes have been included.
At low energy, the transverse polarization asymmetry yields similar limits as $A^\text{FB}$, but its added value becomes clear at high energy.
The sensitivity of $A^\text{FB}$ is approximately constant over the $\sqrt{s}$ range considered here (see right panel of \autoref{fig:sensitivities}). Therefore, with the cross section falling as $1/s$ and the luminosity assumed constant, the limits deteriorate strongly with increasing centre-of-mass energy.
The sensitivity of the transverse asymmetry to dipole operators, on the other hand, increases with centre-of-mass energy. At high energies, both the sensitivity and the statistical uncertainty increase as $\sqrt{s}$ and balance each other. The individual limits thus become independent of $\sqrt{s}$. For centre-of-mass energies above $1\tev$, the constraints derived from the transverse asymmetry are significantly stronger than those implied by $A^\text{FB}$. At $3\tev$, they are an order of magnitude better.
The optimal observables that will be introduced in \autoref{sec:oo} present a similar high-energy behaviour but perform better than the transverse polarization asymmetry at low energies.

Measurements of the transverse polarization asymmetry may add valuable information 
in a global fit of top-quark electroweak couplings. The information contained
in the usual helicity angle asymmetry overlaps with that of the forward-backward asymmetry. The sensitivity of the transverse polarization asymmetry
grows with the centre-of-mass energy and yields significantly tighter constraints for $\sqrt{s}$ above about $1\tev$.

\begin{figure}
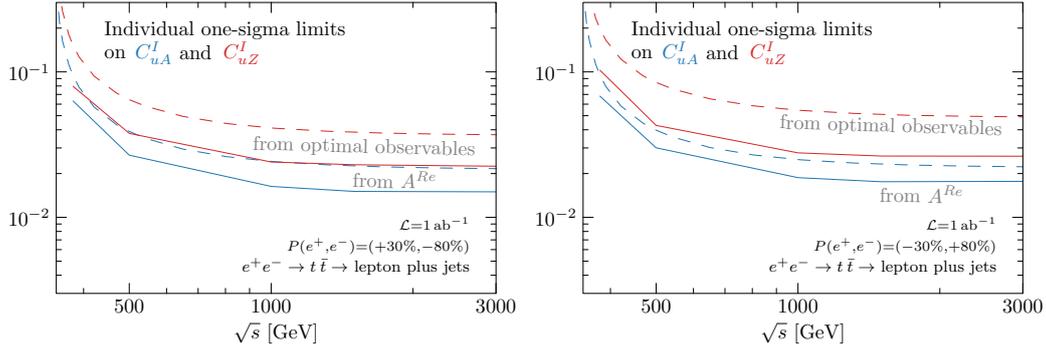

\centering
\includegraphics[width=0.45\linewidth]{imaginary_VA_LR_OO_bis.mps}
\includegraphics[width=0.45\linewidth]{imaginary_VA_RL_OO_bis.mps}%
\caption{Individual limits on the imaginary parts of the dipole operator coefficients $C_{uA}^I$ and $C_{uZ}^I$ from measurements with $P(e^+, e^-) = (-30\%,+80\%)$ and $(-30\%,+80\%)$ beam polarizations (left and right panels, respectively). The limits deriving from $A^{Re}$ measurements (solid lines) are compared to that of optimal observables defined in \autoref{sec:oo}. A data sample of $1\iab$ is assumed at centre-of-mass energies ranging between the top-quark pair production threshold and $3\tev$.}%
\label{fig:dipole_imaginary}%
\end{figure}

\subsection{CP-odd observables}

Imaginary coefficients for the electroweak dipole operators $O_{uZ}$ and $O_{uA}$ (or $O_{uW}$ and $O_{uB}$) violate the combination of charge conjugation and parity symmetries (CP). Close-to-optimal observables specifically designed to test CP conservation~\cite{Bernreuther:1995nw} provide very precise constraints on these parameters~\cite{Bernreuther:2017cyi}.

Following Ref.~\cite{Bernreuther:1995nw}, one defines first
\begin{equation*}
	\mathcal{O}_{+}^{Re} \equiv
		(\hat{\bar{t}} \times \hat{l}_+)\cdot \hat{e}_+
	\quad\text{ and }\quad
	\mathcal{O}_{-}^{Re} \equiv
		(\hat{t} \times \hat{l}_-)\cdot \hat{e}_+ \, , 
\end{equation*}
where $\hat{t}$ ($\hat{\bar t}$) and $\hat{e}_+$ are unit vectors pointing in the direction of the top (anti-top) and incoming positron beam momenta in the centre-of-mass frame. The unit vectors $\hat{l}_\pm$ point in the direction of the charge lepton momenta arising from the $W$ decay in the top and anti-top rest frames. The $\mathcal{O}_{\pm}^{Re}$ observables are CP conjugate of each other. While non-vanishing expectation values for $\mathcal{O}_{+}^{Re}$ and $\mathcal{O}_{-}^{Re}$ could be generated by absorptive parts in amplitudes, their difference $A^{Re}\equiv \mathcal{O}_{+}^{Re} - \mathcal{O}_{-}^{Re}$ is only sensitive to genuine CP-violation~\cite{Bernreuther:2017cyi}.

The limits extracted from the $A^{Re}$ asymmetry are very similar to the limits obtained from the top and anti-top normal polarization
asymmetries defined in \autoref{subsec:dipolereal}, with a slight advantage for the $A^{Re}$ asymmetry. Both observables are indeed based
on the same $\mathcal{O}_{\pm}^{Re}$ kinematic functions. We therefore only discuss the $A^{Re}$ asymmetry.

The one-sigma individual limits on the imaginary parts of the dipole operator coefficients $C_{uA}^I$ and $C_{uZ}^I$ are presented as functions of
the centre-of-mass energy in \autoref{fig:dipole_imaginary}. As before, the integrated luminosity is fixed to $1\iab$ to facilitate
comparisons. The limits extracted from the $A^{Re}$ asymmetry are displayed together with those obtained with the optimal observables that will be defined
in \autoref{sec:oo}. Both are comparable at low centre-of-mass energy, but differ by up to a factor two between $\sqrt{s}=1$ and $3\tev$. This gain may arise from the higher top-quark spin analysing power of the charged lepton compared to that of the $W$ boson (or $b$ quark) that is accessible to statistically optimal observables defined on the $\bwbw$ kinematics.
In both cases, the sensitivity grows with the centre-of-mass energy. At large energies and for a fixed integrated luminosity, the
individual limits saturate to constants. These observables are quite specific to the imaginary part of the $O_{uA}$ and $O_{uZ}$
operators and the constraints they imply have little correlations with that of the CP-conserving operator coefficients.


Other studies of CP-violation in top-physics include Refs.~\cite{Donoghue:1987ax, Gavela:1988jx, Nelson:1989uq, Ma:1991ry, Pilaftsis:1989zt, Nowakowski:1991si, Eilam:1991yv, Soares:1992kr, Korner:1990zk, Grzadkowski:1992yz, Bernreuther:1992be}.

\subsection{Top-quark decay and single production}
\label{sec:decay_single_prod}
Some of the dimension-six operators that affect the top-quark couplings to the photon and $Z$ boson also modify the $tbW$ vertex. Among those, $O_{uW}$ and $O_{\varphi q}^3$ interfere with standard-model amplitudes in the vanishing $b$-mass limit. These two operator coefficients are constrained at the LHC in measurements of top-quark decay, single-top and associated $t\bar{t}V$ production (see for instance Ref.\,\cite{Buckley:2015lku}).

In \epem\ collisions,
the $O_{uW}$ and $O_{\varphi q}^3$ operators affect
top-quark pair production, single production, and decay.
We compare in this section the sensitivity of the production and decay processes to the $C_{uW}$ operator coefficient. In the
narrow top-quark width approximation, the dependence on $O_{\varphi q}^3$
drops out from the differential $t\to bW$ branching fraction.
The $\eett$ process is
only sensitive to the difference of $O_{\varphi q}^1$ and $O_{\varphi q}^3$
(which we denote $O_{\varphi q}^-$). We explore several ways to
disentangle the contributions of $O_{\varphi q}^1$ and $O_{\varphi q}^3$.

\paragraph{Transverse polarization asymmetry}
The transverse polarization asymmetry of the top quark $A^T$, introduced in \autoref{subsec:dipolereal}, is sensitive to new physics in top-quark production. It may also be affected by new physics in top-quark decay since it relies on the distribution of the charged lepton produced in the $t\rightarrow Wb, W \rightarrow l \nu_l$ decay chain.

\paragraph{$W$ helicity fractions}\label{sec:helicity_frac}
The helicity fractions of the $W$ boson are two classical observables measured in top-quark decay. We denote $\theta_l^*$ the angle between the charged lepton momentum in the $W$-boson rest frame and the $W$-boson momentum in the top-quark rest frame. Its distribution,
\begin{equation}
\frac{1}{\Gamma}\frac{\text{d}\Gamma}{\text{d}\cos\theta_l^*} = \frac{3}{8} F_+ \left(1+\cos\theta_l^*\right)^2+\frac{3}{4}F_0 \sin^2\theta_l^*+\frac{3}{8}F_-\left(1-\cos\theta_l^*\right)^2,
\end{equation}
serves to define the positive, negative and longitudinal $W$-boson helicity fractions of unit sum: $F_+ + F_- + F_0 =1$. In the following, we consider the asymmetry formed by positive and negative helicity fractions $A_{FB}^W \equiv \frac{3}{4} (F_+ - F_- )$ extracted from a fit to the $\cos\theta_l^*$ distribution.

\begin{table}[tb]
\centering
\begin{tabular*}{\textwidth}{@{\extracolsep{\fill}}l@{\hspace{0.7cm}}cc@{\hspace{0.7cm}}cc}
\hline
\hfill$P(e^+, e^-)$  & \multicolumn{2}{c@{\hspace{0.7cm}}}{$( +30\%,-80\% )$} & \multicolumn{2}{c}{$(  -30\%, +80\%)$} \\  
\hfill observables  & $A^T $  & $A_{FB}^W $ & $A^T $ & $A_{FB}^W $\\ 
\hfill SM predictions     & $-0.6$ & $-0.17$  & 0.57 & $-0.29$ \\\hline\noalign{\vskip 1mm}
\hfill sensitivity to $C_{uW}^{R}$ [\%] \\
\qquad in production       & 38 $\pm$ 1  & 9 $\pm$ 2   & $-$25 $ \pm$ 1 & --- \\
\qquad in decay            &  ---          &  16 $\pm$ 2  &   --- & 11 $\pm$ 3  \\ 
\qquad in prod.\ \& decay  & 37 $\pm$ 1 & 26  $\pm$ 2 & $-$24 $ \pm$ 1 & 10 $\pm$ 3 \\[1mm]\hline
\end{tabular*}
\caption{Sensitivities to the $C_{uW}^{R}$ operator coefficient, artificially decomposed into \emph{production} and \emph{decay} components, and quoted in percent. A centre-of-mass energy of $500\gev$ and the two $P(e^+, e^-) = (\pm 30\%,\mp 80\%)$ beam polarizations are considered. The uncertainties displayed are due to limited Monte Carlo statistics. A sensitivity compatible with zero, within uncertainties, is replaced by a dash.}
\label{tab:helicity}
\end{table}

Leading-order standard-model estimates for  $A^T$ and $A_{FB}^W$ are presented in the first row
of \autoref{tab:helicity} for a centre-of-mass energy of $500\gev$ and the two $P(e^+, e^-) = (\pm30\%,\mp80\%)$ beam polarizations.
The remaining rows present the sensitivity of these observables to $C_{uW}^{R}$, artificially decomposed into components arising from
top-quark production and decay, in the narrow width approximation. For the purpose of this table, the sensitivity of an observable $o$ is simply defined as $o(C=1)/o(C=0) -1$
and quoted in percent.
The transverse polarization asymmetry has a marked sensitivity to $C_{uW}^{R}$. It arises dominantly from production. 
On the other hand, the sensitivity of the $W$ helicity fractions to $C_{uW}^R$ mostly arises from top-quark decay. For the centre-of-mass energy considered, it is smaller than the one achieved with the transverse polarization asymmetry.

\paragraph{Total width}
Measurements of the top-quark decay width are sensitive to dimension-six
operators. Computing its linear dependence at NLO in QCD using with \mg~\cite{Alwall:2014hca} and the input parameters specified in \autoref{app:numerics_nlo}, we obtain:
\begin{align*}
\Gamma_t &= 
\vall{1.36}{0.914}{-1.4}{+1.2}{0.043}
+ \left(\vall{0.161}{0.914}{-1.4}{+1.2}{0.027} C_{\varphi q}^3
+ \vall{0.147}{0.923}{-1.3}{+1.1}{0.03} C_{uW}^R
+ \vall{0.000225}{\nol{-74.4}}{-8.8}{+11}{1} C_{uG}^R
\right) \left(\frac{1\,\text{TeV}}{\Lambda}\right)^2
\\&+ O(\Lambda^{-4}).
\end{align*}
Central values, $k$-factors and uncertainties are displayed in the following format: 
\begin{equation*}
\vall	{\text{central value}}
	{\text{k-factor}}
	{-\text{scale down}}
	{+\text{scale up}}
	{\text{Monte Carlo}}
	.
\end{equation*}
The scale uncertainty is computed from the running of $\alpha_S(\mu)$ between $\mu=m_t/2$ and $2m_t$.

The top-quark width has only a leading-order linear dependence on $C_{\varphi q}^3$ and $C_{uW}^R$ operator coefficients. A small dependence on $C_{uG}^R$ also arises at next-to-leading order in QCD. Tight constraints on this operator coefficient can however be obtained at the LHC, or from the associated production of a top-quark pair with a hard jet at a linear collider~\cite{Rizzo:1996sy}. We therefore ignore this dependence. For constraints achievable at lepton colliders, the quadratic dependences on the coefficients which appear already at the linear order are subleading. Operators inducing a right-handed current or CP-violation only start contributing a the quadratic level.

A precise measurement of the top-quark width is possible at an \epem\ collider by scanning the centre-of-mass energy through the top-quark pair production threshold.\footnote{A determination of the total width is also possible immediately below the $t\,\bar{t}$ threshold~\cite{Batra:2006iq} or in the continuum well above the threshold~\cite{Liebler:2015ipp}.} Reference~\cite{Martinez:2002st} demonstrated that a precise determination, simultaneous to that of the top-quark mass and strong coupling constant, can be obtained from a fit of the threshold line shape. A recent analysis estimates the statistical uncertainty obtainable with an integrated luminosity of $100\ifb$ to $21\mev$~\cite{Horiguchi:2013wra}. Theory uncertainties are however likely to dominate. For uncertainties of $20$, $40$, and $80\mev$ and a standard-model central value, the regions of the parameter space spanned by the operator coefficients $C_{uW}^{R}$ and $C_{\varphi q}^{3}$ allowed at the $68\%\,$C.L.\ are presented in \autoref{fig:width_constraint}. The constraint imposed by the width measurement effectively disentangles the coefficients of the $O_{\varphi q}^{1}$ and $O_{\varphi q}^{3}$ operators.

\begin{figure}[tb]
\centering
\includegraphics[width=0.5\linewidth, trim=0 25 0 35, clip]{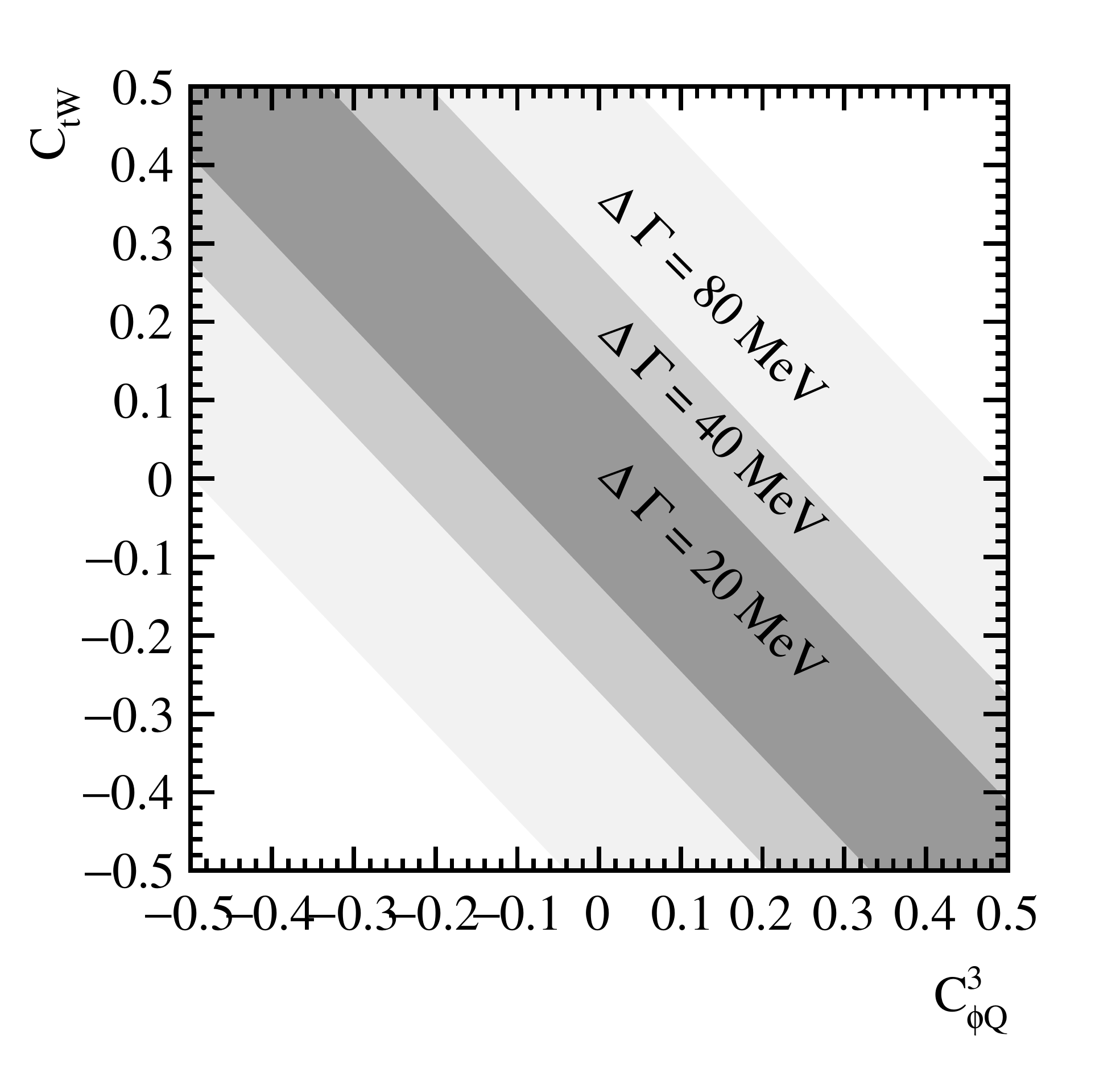}
\caption{One-sigma constraints on the $C_{uW}^{R}$ and $C_{\varphi q}^{3}$ operator coefficients that would derive from a measurement of the top-quark width with $20$, $40$, or $80\mev$ precision. The central value of the measurement is assumed to coincides with the SM prediction.}
\label{fig:width_constraint}
\end{figure}

\paragraph{Single production}
\label{sec:single_production}
Single top-quark production at \epem\ colliders is also sensitive to modification of the $tbW$ vertex.\footnote{Single production through a $e\gamma$ initial state could also be considered~\cite{Boos:1996ud, Cao:1998at}.} At high energy, its cross section becomes comparable that of pair production. A dedicated analysis was presented in Ref.~\cite{Escamilla:2017pvd}. The limits on dimension-six operator coefficients are weak compared to the ones obtained from a study of top-quark pair production. Ideally, an analysis should treat the full $\epem\to\bwbw$ process, including dimension-six operator effects on both resonant and non-resonant components. Four-fermion operators should also be included. The statistically optimal observables we employ in this work could be extended to cover non-resonant contributions and enhance the sensitivity to charged current modifications. This is left for future work.

\begin{figure}[tb]
\centering
\includegraphics[width=0.5\linewidth, trim=0 25 30 25, clip]{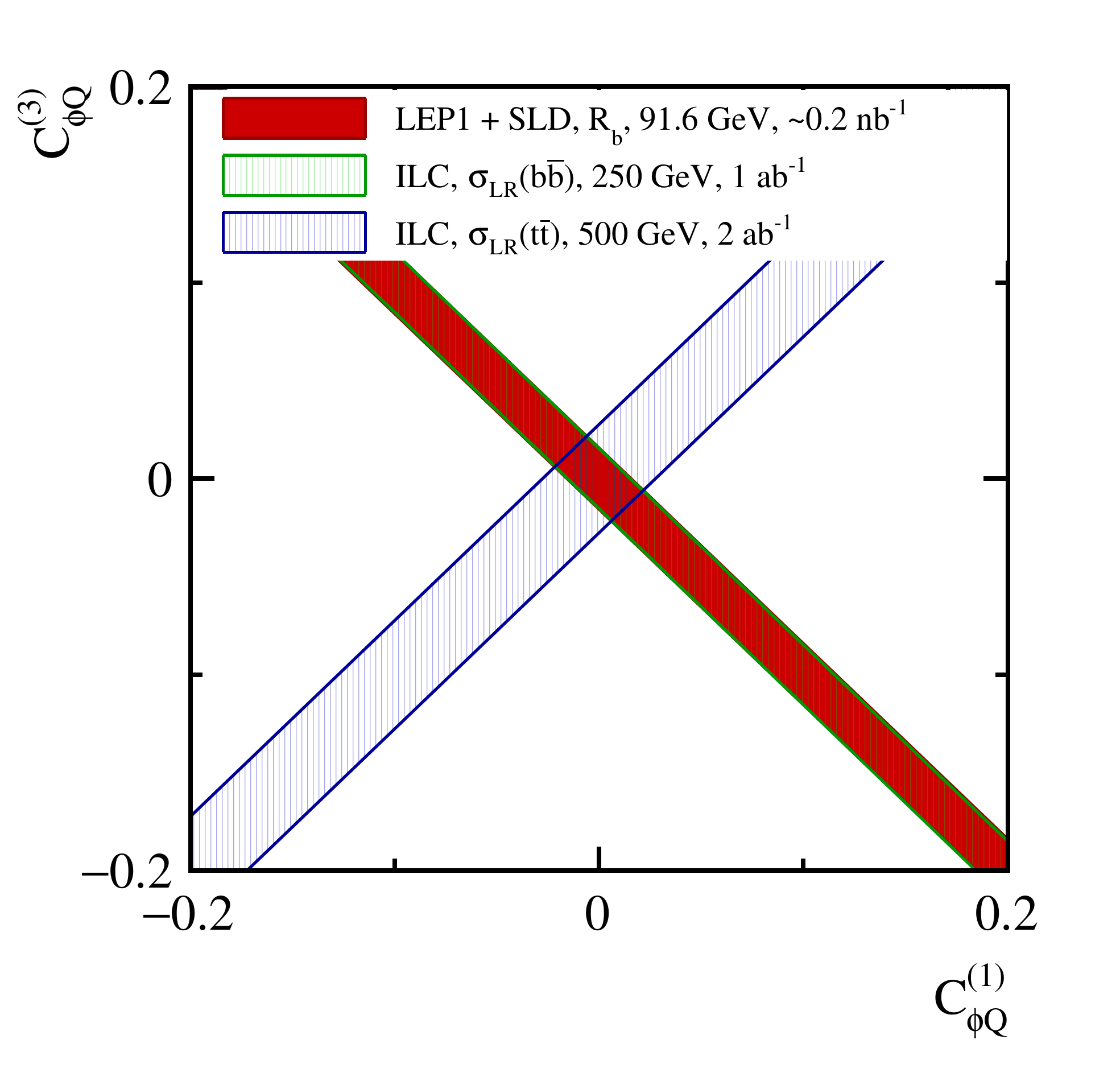}%
\includegraphics[width=0.5\linewidth, trim=0 25 30 25, clip]{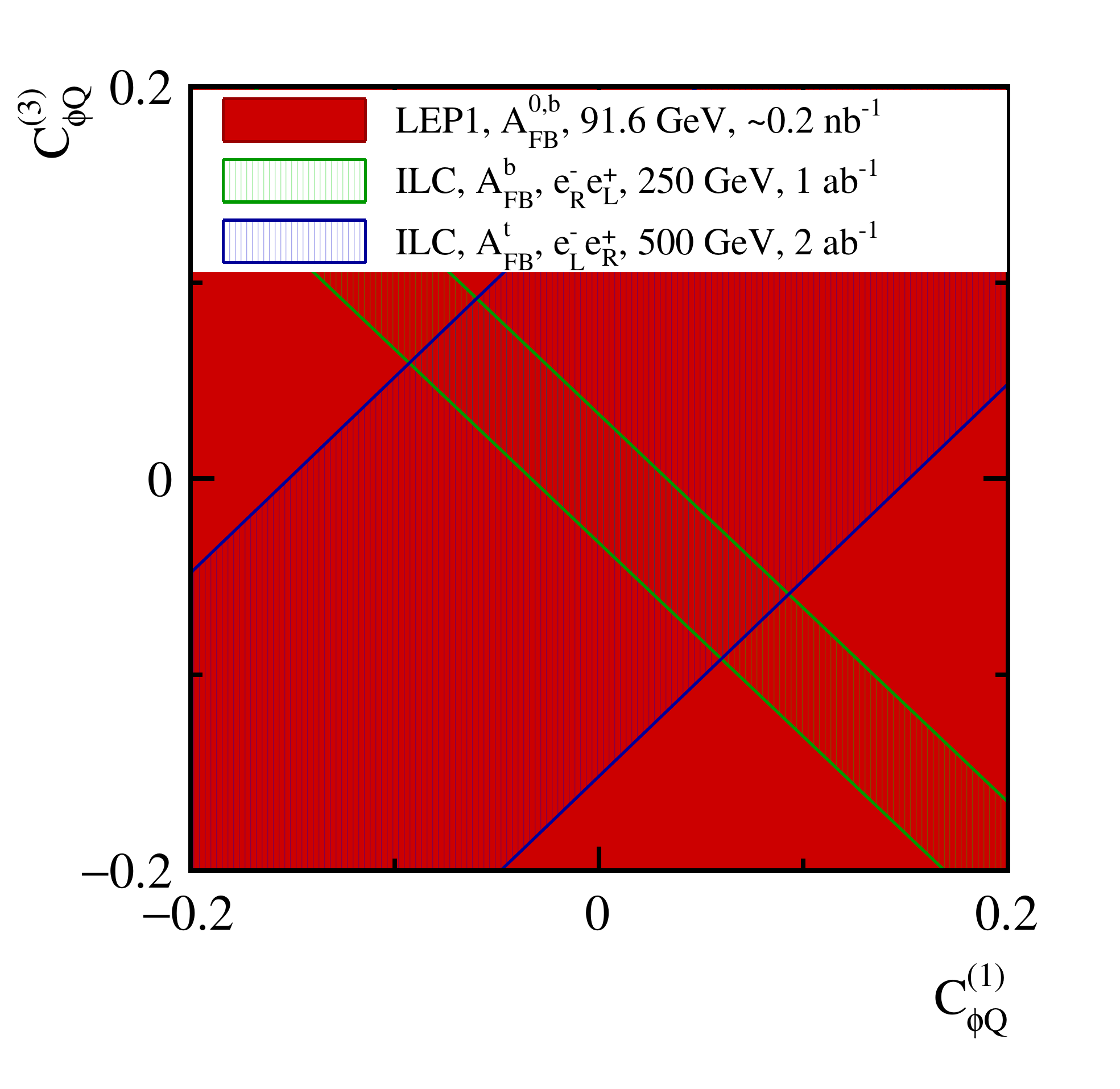}%
\caption{The 68\% C.L. constraints on the $C_{\varphi q}^{1}$ and $C_{\varphi q}^{3}$ operator coefficients which arise from measurements at LEP1 and SLD. Cross section and forward-backward asymmetry measurements are respectively considered in the left and right panel. The LEP1/SLD result corresponds to the combination from Ref.~\cite{Z-Pole}. For each measurement the future lepton collider beam polarization configuration that yields the best limit has been chosen: $P(e^+,e^-)=(+30\%,-80\%)$ for the cross section measurements and the $t\bar{t}$ forward-backward asymmetry, $P(e^+,e^-)=(-30\%,+80\%)$ for the $b\bar{b}$ forward-backward asymmetry.
}%
\label{fig:tt_vs_bb}%
\end{figure}

\subsection{Bottom-quark production}

Measurements in the $\epem\to b\,\bar{b}$ process can provide complementary constraints on several operators. The combination of $t\bar{t}$ and $b\bar{b}$ data can disentangle contributions from $O_{\varphi q}^{1}$ and $O_{\varphi q}^{3}$ operators. The $t\bar{t}$ production process constrains only the difference $C_{\varphi q}^{1} - C_{\varphi q}^{3}$, while $b\bar{b}$ data constrain the sum $C_{\varphi q}^{1} + C_{\varphi q}^{3}$. The combination of both types of measurements provides stringent constraints on both coefficients.
Of course, a combined global fit of $t\bar{t}$ and $b\bar{b}$ data also increases the set of relevant operators. Two additional two-fermion operators of the Warsaw basis must then in principle be considered, $O_{dW}$ and $O_{\varphi d}$, plus three new four-fermion operators of vector Lorentz structure ($O_{ld}$, $O_{lq}^+$, $O_{ed}$) which do interfere with standard-model amplitudes. The real and imaginary components of an additional scalar four-fermion operators $O_{ledq}$ do not.

A combined fit is beyond the scope of the present work, but it is instructive to consider a simple two-parameter fit of the coefficients $C_{\varphi q}^{1}$ and $C_{\varphi q}^{3}$. Constraints of several origins are compared in \autoref{fig:tt_vs_bb}.
The red bands represent existing constraints from the characterization of the $e^+e^-b\,\bar{b}$ process at LEP and SLC. We consider two measurements. The first, in the left panel, is the combination of measurements from the four LEP1 experiments and SLD of the $b\bar{b}$ cross section~\cite{Z-Pole}.\footnote{The LEP experiments and SLD report the ratio $R_b$ of the $b\bar{b}$ and total hadronic cross section. This result can be converted into a cross section under the assumption that no new physics affects the other $Z$-boson decay channels.} The second measurement, in the right panel, is the forward-backward asymmetry of bottom quarks. The latter can be compared to Fig.\,11 of Ref.~\cite{Englert:2017dev}. The combination of LEP1 measurements yields a stronger constraint than the single $A^\text{FB}_{b\bar{b}}$ measurement considered there.

Two additional bands in \autoref{fig:tt_vs_bb} indicate the constraints that can be expected from measurements at the ILC. The green shaded band shows the limit expected from $b\bar{b}$ production at the ILC at $\sqrt{s} = 250\gev$, based on the analysis of Ref.~\cite{Bilokin:2017lco}. The constraint from the cross section is similar in strength to the equivalent result from LEP1. The sensitivity is practically constant as a function of the centre-of-mass energy and the impact of the cross section, which is nearly four orders of magnitude larger at the $Z$-pole than at $250\gev$, roughly cancels the effect of the much greater luminosity at the ILC. The constraint from $A^\text{FB}$, on the other hand, is greatly improved by ILC data. In a global fit, this complementary information is important to simultaneously constrain all operator coefficients. With higher centre-of-mass energies, future lepton colliders would also be much more sensitive to four-fermion operators.

The $e^+ e^- \rightarrow b\,\bar{b}$ data provide very tight constraints on the $O_{\varphi d}\equiv \frac{y_t^2}{2}(\bar{\ges d}\gamma^\mu \ges d)\:\FDF[i]$ operator coefficient. This observation can be related to the results in Ref.~\cite{Bilokin:2017lco} using the relations between the operator coefficients and the left- and right-handed couplings of the $b$-quark to the $Z$-boson they define there: $\delta g_{L} = -(C_{\varphi q}^{1} + C_{\varphi q}^{3}) \frac{m_t^2}{\Lambda^2}$ and $\delta g_{R} = -C_{\varphi d} \frac{m_t^2}{\Lambda^2}$. We see that both studies indeed find that the right-handed coupling is improved considerably, while for the left-handed coupling LEP1 and ILC constraints are of comparable strength.

The blue band indicates the limit that is obtained from $t\,\bar{t}$ production at $\sqrt{s} = 500\gev$. These bands cross the $e^+e^- \rightarrow b\,\bar{b}$ bands at a right angle. A combined fit is thus expected to yield a tight constraints on both operator coefficients.

\begin{figure}[tb]\centering
\includegraphics[width=0.5\linewidth, trim=15 20 30 25, clip]{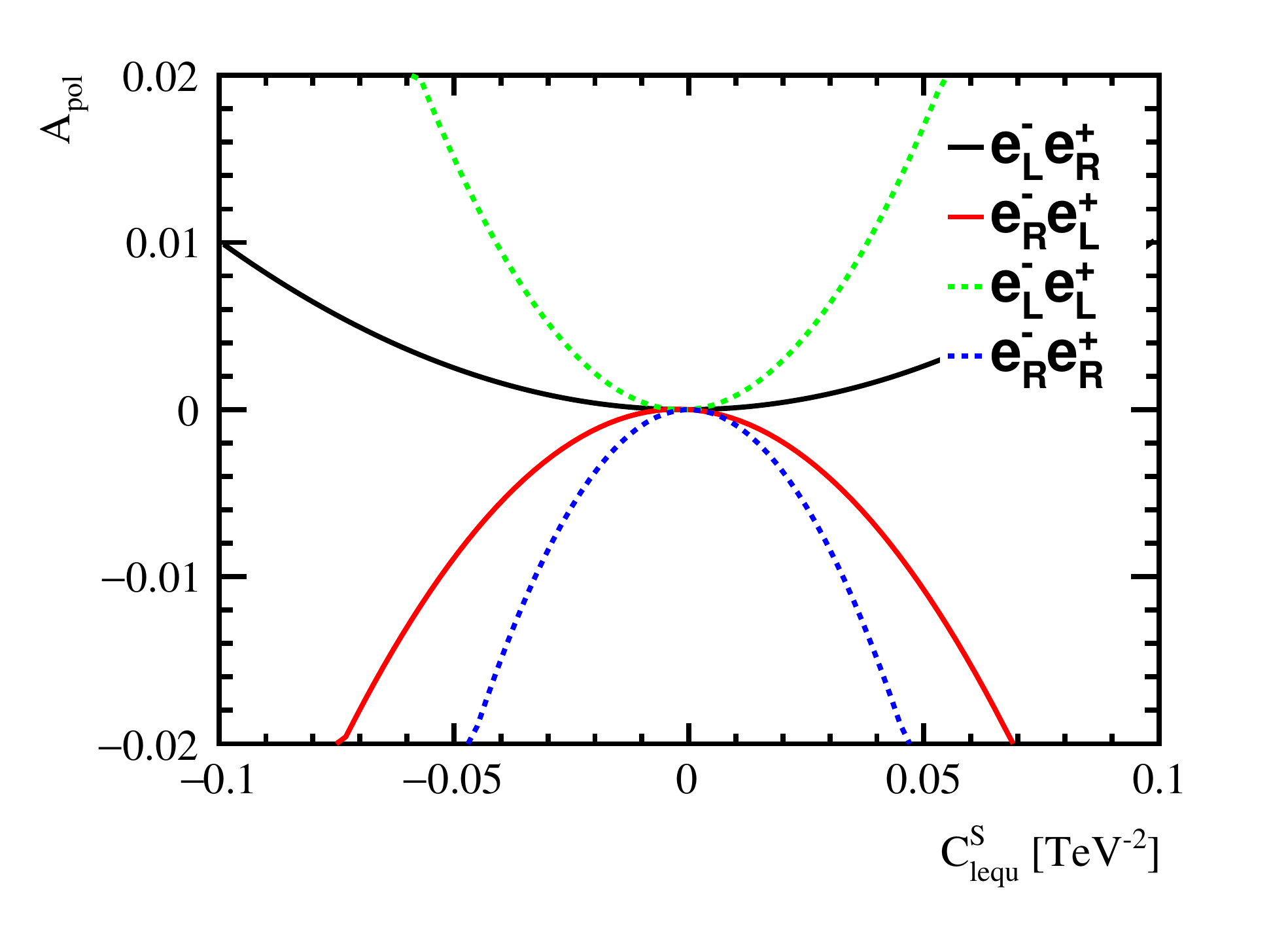}%
\includegraphics[width=0.5\linewidth, trim=15 20 30 25, clip]{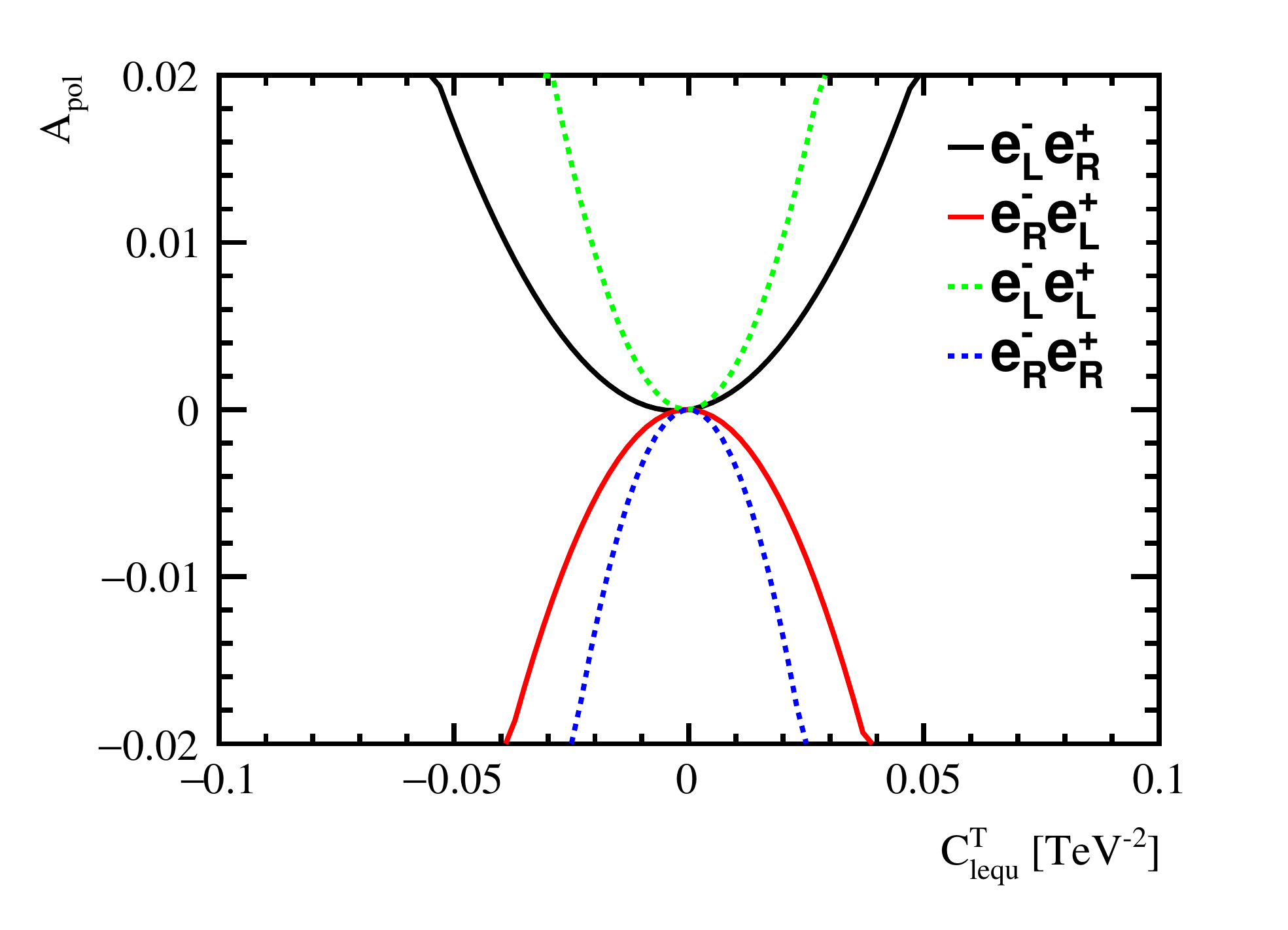}%
\caption{The dependence of $A_{pol} = P_{t} - P_{\bar{t}}$ on the scalar and tensor operator coefficients $C_{lequ}^S$ (left) and $C_{lequ}^T$ (right) for $\sqrt{s} = $ 500\gev. The electron and positron beams are 80\% and 30\% polarized, as envisaged in the ILC design. The four curves represent four different configurations: two standard configurations with opposite electron and polarization ($e^-_Le^+_R$ in black, $e^-_Re^+_L$ in red), and two same-sign configurations ($e^-_Le^+_L$ in green, $e^-_Re^+_R$ in blue.)}
\label{fig:sens_lequs_500}%
\end{figure}
\begin{figure}[tb]\centering
\includegraphics[width=0.5\linewidth, trim=15 20 30 25, clip]{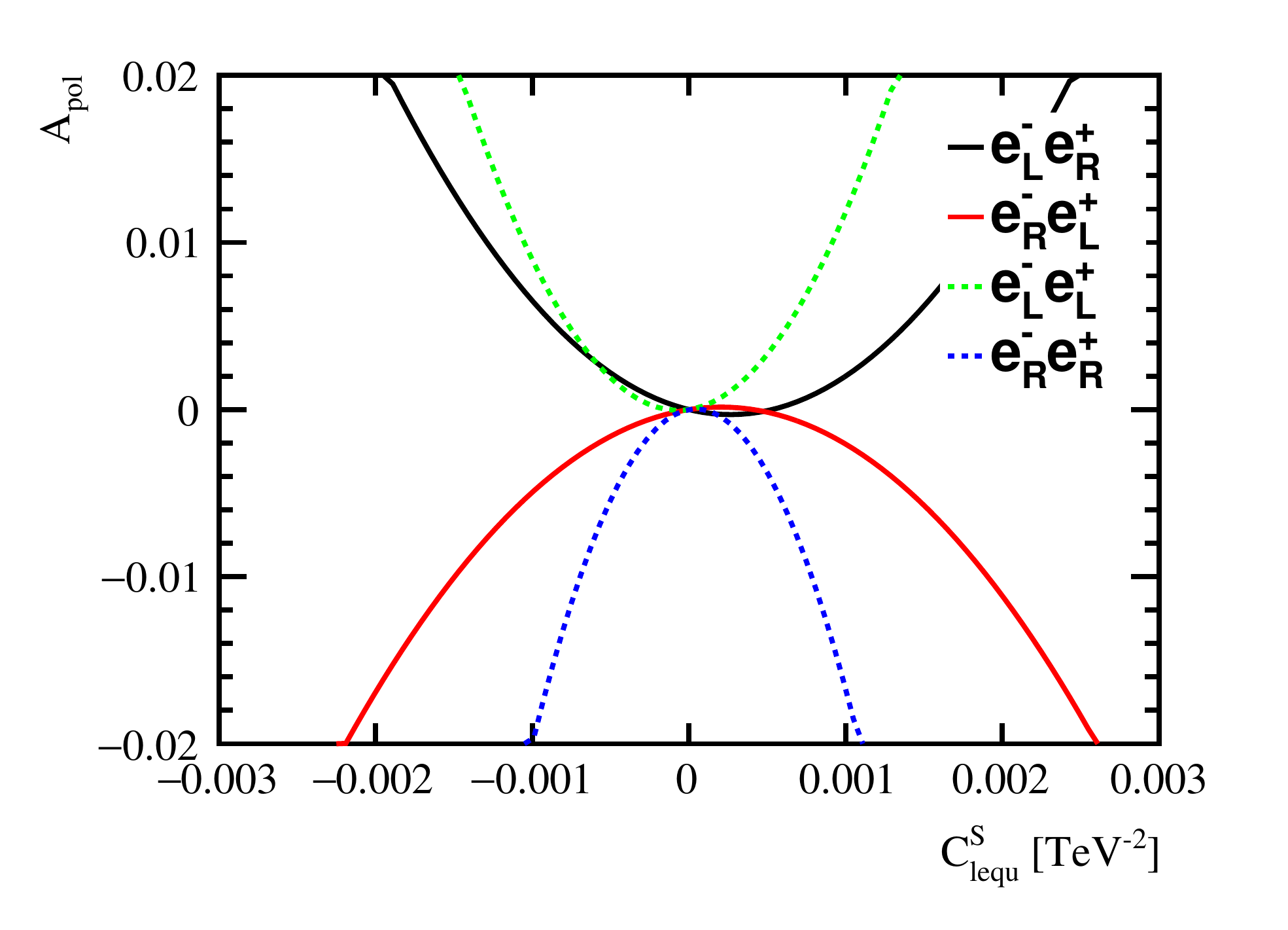}%
\includegraphics[width=0.5\linewidth, trim=15 20 30 25, clip]{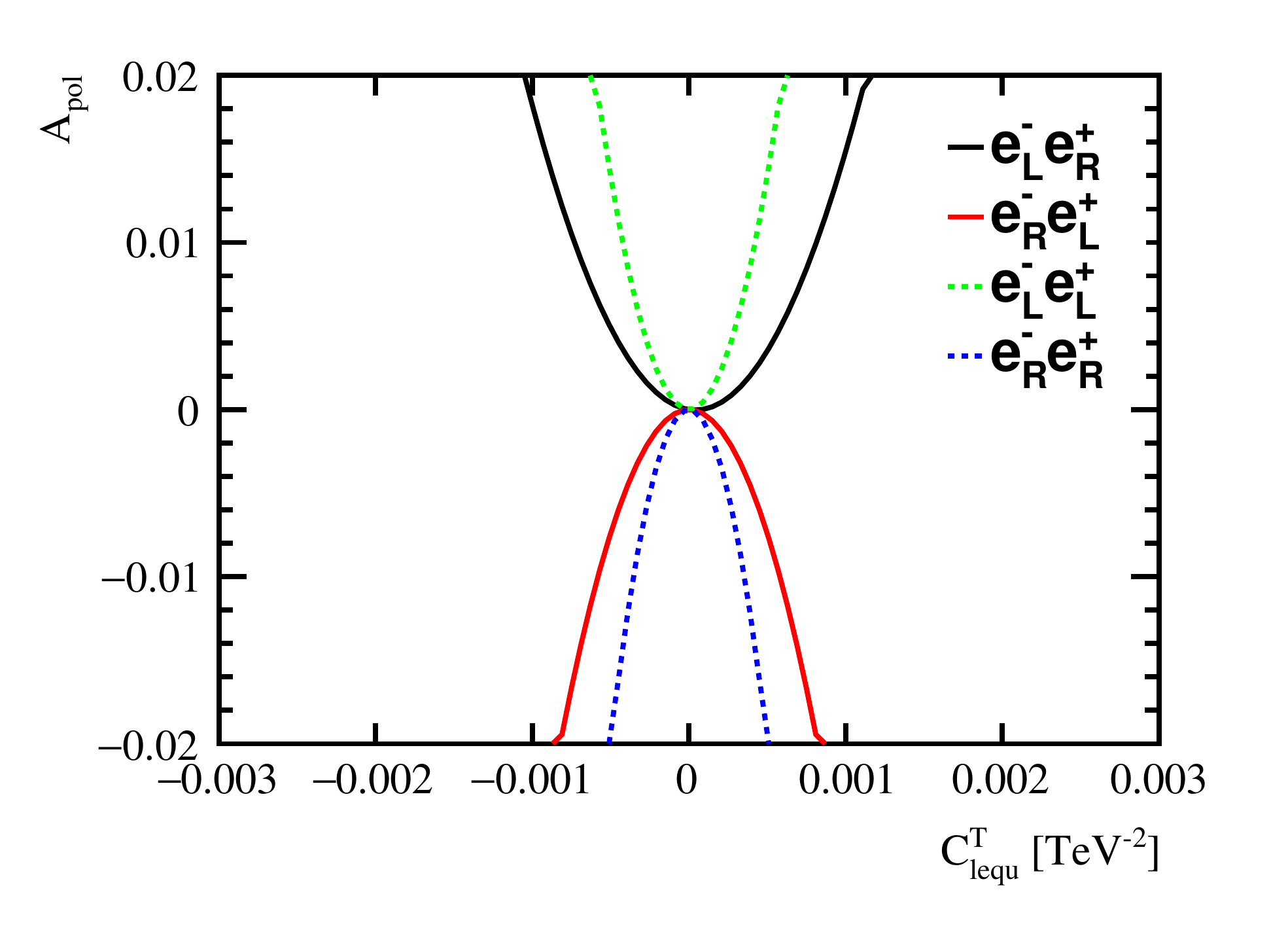}%
\caption{The dependence of $A_{pol} = P_{t} - P_{\bar{t}}$ on the scalar and tensor operator coefficients $C_{lequ}^S$ (left) and $C_{lequ}^T$ (right) for $\sqrt{s} = $ 3\tev. The electron is 80\% polarized, as envisaged in the CLIC design. For the positron beam a 30\% polarization is assumed for comparison with \autoref{fig:sens_lequs_500}, even if positron polarization is not part of the CLIC baseline design. The four curves represent four different configurations: two standard configurations with opposite electron and polarization ($e^-_Le^+_R$ in black, $e^-_Re^+_L$ in red), and two same-sign configurations ($e^-_Le^+_L$ in green, $e^-_Re^+_R$ in blue.)}
\label{fig:sens_lequs_3000}%
\end{figure}
\begin{table}[tb]\centering
\begin{tabular*}{\textwidth}{@{\extracolsep{\fill}}cccc}
\hline
$P(e^+,e^-)$ & $A_{pol}$ uncert. $[\%]$ & $|C_{lequ}^{S}|^2$  & $|C_{lequ}^{T}|^2$ \\ \hline
$( +30\%,-80\%)$ & 1.0 &  $3\times 10^{-3}$ & $1\times 10^{-3}$\\ 
$( -30\%,+80\%)$ & 1.4  & $5\times 10^{-3}$ & $9\times 10^{-4}$\\ 
$( -30\%,-80\%)$ & 2.6 & $1\times 10^{-2}$ & $9\times 10^{-4}$\\ 
$( +30\%,+80\%)$ & 3.4 & $4\times 10^{-3}$ &  $9\times 10^{-4}$\\ \hline
\end{tabular*}
\caption{Individual 68\% C.L. limits on the scalar and tensor four-fermion operator coefficients obtained with an integrated luminosity of $500\,\ifb$ collected at $\sqrt{s} = 500\gev$. Following the ILC scenarios of Ref.~\cite{Barklow:2015tja}, respectively $40\%$ and $10\%$ of the total luminosity is devoted to runs with each of the $P(e^+, e^-) = ( -30\%, +80\%)$, $( +30\%, -80\%)$ and $P(e^+, e^-) = ( -30\%, -80\%)$, $( +30\%, +80\%)$ polarization configurations. We assume $A_{pol}$ is measured in lepton-plus-jets events. A total efficiency to $20\%$ is applied, including top-quark branching fractions and an estimate of the selection efficiency. The scale $\Lambda$ is fixed to $1\,$TeV.}
\label{tab:indv_limits_lequs_500}
\end{table}%
\begin{table}[tb]
\centering
\begin{tabular*}{\textwidth}{@{\extracolsep{\fill}}cccc}
\hline
$P(e^-,e^+)$ & $A_{pol}$ uncert. $[\%]$ & $|C_{lequ}^{S}|^2$  & $|C_{lequ}^{T}|^2$   \\ \hline
$( +30\%, -80\%)$ & 2.2 &  $4\times 10^{-5}$ &  $2\times 10^{-6}$  \\ 
$( -30\%, +80\%)$ &  3.0 & $1\times 10^{-5}$ &  $1\times 10^{-6}$ \\ 
$( -30\%, -80\%)$ & 5.7 & $4\times 10^{-6}$ &$10^{-6}$ \\ 
$( +30\%, +80\%)$ & 7.3 & $5\times 10^{-6}$ & $8\times 10^{-7}$ \\ \hline
\end{tabular*}
\caption{Individual limits on the square of the Wilson coefficients of the scalar and tensor operator coefficients at $\sqrt{s} = 3\tev$. $A_{pol}$ is measured in lepton+jets events in a data sample with an integrated luminosity of $\mathcal{L} = 3\,\iab$. A selection efficiency of 20\% is applied. For comparison to the ILC scenario, a $30\%$ positron polarization is assumed and the sample is divided among the four configurations as in \autoref{tab:indv_limits_lequs_500}}
\label{tab:indv_limits_lequs_3}
\end{table}

\subsection{Observables for scalar and tensor operators}
\label{sec:obs_s_t}

The scalar and tensor four-fermion operators $O_{lequ}^S$ and $O_{lequ}^T$
present a distinctive Lorentz structure. As mentioned earlier,
they do not interfere with standard-model amplitudes in the limit of vanishing lepton
masses. For the range of energy and initial state polarizations we consider,
we find the difference $A_{pol} = P_{t} - P_{\bar{t}}$ of the polarization of
the top and anti-top quark (see \autoref{subsec:dipolereal}) is sensitive to both operators. In a
lepton-plus-jets sample, both polarizations are measured in a
straightforward fashion through \autoref{eq:top-ang-dis}.
The observable $A_{pol}$ vanishes in the SM and is {\em specific} to
the scalar and tensor operators: it has little or no
sensitivity to any of the other operators considered.

The sensitivity to
the scalar and tensor operator moreover increases in runs with electron and positron beam polarizations of the same sign. In \autoref{fig:sens_lequs_500}, the sensitivity of $A_{pol}$
to the $O_{lequ}^S$ and $O_{lequ}^T$ operators at a centre-of-mass energy of
500\gev{} is represented for four different initial-state polarization
configurations. The curves labelled $e^-_Le^+_R$ and $e^-_Re^+_L$ represent
the opposite-sign configurations that are usually considered,
with $P(e^+, e^) = ( +30\%,-80\%)$ and $P(e^+, e^-) = (-30\%, +80\%)$. Two
further curves, labelled as $e^-_Le^+_L$ and $e^-_Re^+_R$ represent same-sign
configurations with $P(e^+, e^-) = ( -30\%,-80\%)$ and
$P(e^+, e^-) = (-30\%,-80\%)$. Clearly, the same-sign configurations
$e^-_Le^+_L$ and $e^-_Re^+_R$ enhance the sensitivity to these operators
significantly. The tensor operator has a larger impact on $A_{pol}$ for
all scenarios considered here.

To get a grasp of the $\sqrt{s}$ dependence of the sensitivity we consider $3\tev$ operation under the same conditions. The CLIC baseline design does not envisage positron polarization. There is however no technical impediment to positron polarization at high energy. We therefore present the sensitivity plots at $\sqrt{s}=3\tev$ under the same conditions as the ILC in \autoref{fig:sens_lequs_3000}. As already observed for the other four-fermion operators, the sensitivity increases strongly with centre-of-mass energy (note the different range on the $x$-axes between \hyperref[fig:sens_lequs_500]{Figs.\,\ref{fig:sens_lequs_500}} and \ref{fig:sens_lequs_3000}).

Operations with same-sign polarizations help constraining the scalar and tensor four-fermion operator coefficients. The ILC operating scenarios envisage a fraction of the integrated luminosity to be collected in the same-sign configurations. At $\sqrt{s}=500\gev$, Ref.~\cite{Barklow:2015tja} equally splits 80\% of the integrated luminosity between the $P(e^+,e^-)=(+30\%,$ $-80\%)$ and $(+30\%,-80\%)$ polarizations. The remaining $20\%$ are shared among $P(e^+,e^-)=(+30\%,+80\%)$ and $(-30\%,-80\%)$ configurations. We provide individual $68\%\,$C.L.\ limits on the scalar and tensor operator coefficients, in \autoref{tab:indv_limits_lequs_500}, assuming a total integrated luminosity of $500\,\ifb$. Statistical uncertainties are based on $20\%$ of the $t\,\bar{t}$ sample, to take into account the branching fraction of the lepton-plus-jets final state and an estimate of the selection efficiency.

The results in \autoref{tab:indv_limits_lequs_500} indicate that the same-sign configurations can indeed offer quite powerful constraints. The higher sensitivity compensates for the smaller integrated luminosity. Individual limits on  $|C_{lequ}^{S}|^2$ and $|C_{lequ}^{T}|^2$ approximately reach the $10^{-3}$ level for all configurations. As for the other four-fermion operators, the sensitivity dramatically improves with the centre-of-mass energy. Individual constraints obtained with $3\,\iab$ of integrated luminosity collected at a centre-of-mass energy of $3\tev$ are displayed in \autoref{tab:indv_limits_lequs_3} and reach the $10^{-6}$ level.

\section{Statistically optimal observables}
\label{sec:oo}

Let us now examine statistically optimal observables exploiting the full \bwbw\ kinematics, still focusing on resonant production, treating the top quarks in the narrow width approximation and in the $m_b/m_t\to 0$ limit. Their analytical construction is based on the decomposition of the differential $\eett\to\bwbw$ cross section in terms of helicity amplitudes, carried out for instance in Ref.~\cite{Schmidt:1995mr}. We extended it to include the dependence on four-fermion operators.
The same technique was already employed in the context of top-quark pair production at lepton colliders in Ref.~\cite{Atwood:1991ka, Grzadkowski:2000nx, Janot:2015yza, Khiem:2015ofa}. Different differential distributions were nevertheless employed: including also the $W$ polarization information~\cite{Atwood:1991ka}, the kinematic information about the $W$ decay products~\cite{Khiem:2015ofa}, or nothing else than the energy and production angle of one final-state charged lepton or bottom quark~\cite{Grzadkowski:2000nx, Janot:2015yza}.

Statistically optimal observables~\cite{Atwood:1991ka, Diehl:1993br} are constructed to maximally exploit the available differential information and extract the tightest constraints on parameters whose dependence is expanded to linear order only.
The total rate information is employed too, in the following.
Perfect optimization naturally requires a model that truly describes the data. For a differential distribution across the phase space $\Phi$ given by
\newcommand{\dps}[1]{\frac{\text{d}#1}{\text{d}\Phi}}%
\newcommand{\vardps}[1]{{\text{d}#1}/{\text{d}\Phi}}%
\newcommand{\ips}{\int\text{d}\Phi}%
\newcommand{\oo}{\bar{O}}%
\newcommand{\eoo}{O}%
\begin{equation*}
\dps{\sigma}
	= \dps{\sigma_\text{SM}}
	+ \sum_i C_i
	\dps{\sigma_i},
\end{equation*}
the observables maximizing the constraints on the $\{C_i\}$ parameter space are shown to be the average values of
$
	\eoo_i = 
	n\:
	\dps{\sigma_i}\Big/\dps{\sigma_\text{SM}}%
$
where $n$ is the number of events observed. They can be computed as
\begin{equation*}
\oo_i = \epsilon\:\mathcal{L}
	\int\text{d}\Phi
	\;\left(
	\dps{\sigma_i}\bigg/\dps{\sigma_\text{SM}}%
	\right)\;
	\dps{\sigma},
\end{equation*}
where $\mathcal{L}$ is the total integrated luminosity and an $\epsilon$ can be introduced to effectively account for finite efficiencies. Defining
\begin{equation*}
	\sigma_i \equiv\ips\dps{\sigma_i}
	,\qquad\text{and}\quad
	d_{ij}\equiv
		\ips\left(
		\;\dps{\sigma_i}\dps{\sigma_j}
		\left/\dps{\sigma_\text{SM}}\right.
		\right),
\end{equation*}
their sensitivity to operator coefficients and the covariance matrix on the
extracted $C_i$ are given by
\begin{equation*}
	S_j^{\eoo_i} \equiv \left. \frac{1}{\oo_i}
		\frac{\partial \oo_i}{\partial C_j}\right|
	_{C_k=0,\forall k}
	= \frac{d_{ij}}{\sigma_i} + \mathcal{O}( C_k)
	,\qquad\text{and}\qquad
	V^{-1}\big|_{ij} =  \epsilon\:\mathcal{L}\;d_{ij} + \mathcal{O}(C_k).
\end{equation*}
In most of our discussion, we will only retain the zeroth order in $C_k$ of these quantities.

We show in \autoref{fig:sensitivities_opt} the $S_i^{\eoo_i}$ sensitivities for
both mostly right-handed and mostly left-handed polarized electron beams. Note
these sensitivities are not necessarily maximized by the statistically optimal
observables which rather minimize (the determinant of) the covariance matrix. Even though such observables can be used to constrain the
$C_{uZ,A}^I$ operator coefficients, the corresponding sensitivities could not be
displayed in \autoref{fig:sensitivities_opt} since the standard-model values of
these observables $\sigma_i$ vanish. In contrast to that of the total and
forward-backward cross sections (see \autoref{fig:sensitivities}), note the
sensitivities of the statistically optimal observables to dipole operators
slowly grow with $\sqrt{s}$ in the range shown. Beyond the energies displayed,
they actually start growing quadratically with the centre-of-mass energy, like the four-fermion operators. This can be understood by examining the $\sqrt{s}$ dependence of $d_{ij}$ and $\sigma_i$. In the large centre-of-mass energy limit, $\vardps{\sigma_\text{SM}}$ scales as $1/s$. Both $\vardps{\sigma_i}$ and $\sigma_i\equiv\ips\;\vardps{\sigma_i}$ tend to constants for four-fermion operators while they respectively scale as $1/\sqrt{s}$ and $1/s$ for dipole operators, given that the former scaling arises from terms having an azimuthal angle dependence which vanishes upon phase-space integration. Then, $d_{ij}$ tends to a constant for dipole operators and grows like $s$ for four-fermion ones, while $\sigma_i$ scales like $1/s$ for the former and tends to a constant for the latter. The ratio of these two quantities therefore behaves as $s$ in both cases.

Beside sensitivities, it is even more instructive to look at the centre-of-mass energy dependence of
the individual one-sigma statistical limits on the $C_i$ operator
coefficients. They are given by $(V^{-1}|_{ii})^{-1/2} = (\epsilon\:\mathcal{L}\:
d_{ii})^{-1/2}$. For definiteness and as in the previous sections, we normalize
them to an integrated luminosity times efficiency of $1\,\iab$. As seen in
\autoref{fig:individual_opt}, a targeted constraint on the four-fermion and
$C_{uZ,A}^I$ operators can be obtained with lower luminosities at higher
centre-of-mass energies. On the contrary, the $C_{\varphi q}^{A,V}$ and
$C_{uZ,A}^R$ operator coefficients are more efficiently constrained at
centre-of-mass energies ranging approximately between $390$ and $550$\,GeV. At
very high energies, the $C_{uZ,A}^{R,I}$ curves saturate to constant
values. Individual limits on two-fermion operators are only slightly better for a mostly left-handed polarized
beam than for a mostly right-handed polarization (respectively by factors of about $1.03$, $1.15$, $1.33$ for
$C_{uA}^{R,I}$, $C_{\varphi q}^A$, $C_{\varphi q}^V$ and $C_{uZ}^{R,I}$,
relatively independently of the centre-of-mass energy).

\begin{figure}
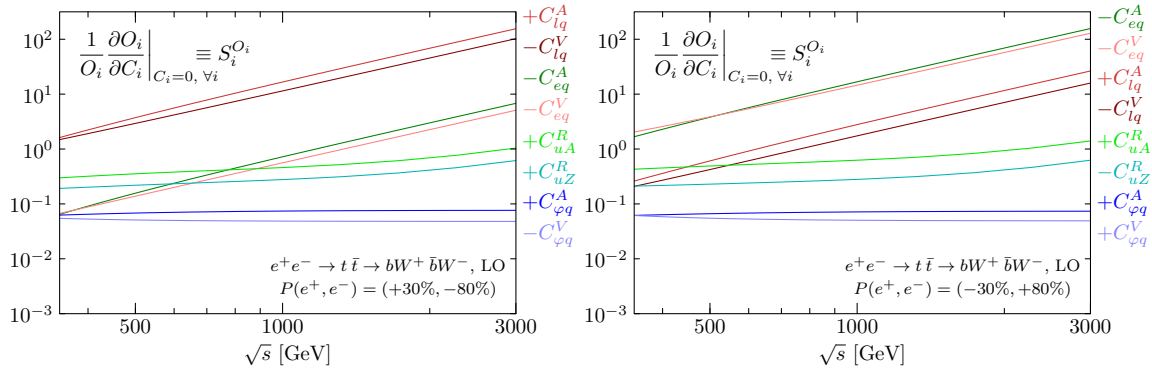
\centering
\includegraphics[width=.5\textwidth]{sensitivities_left_opt.mps}%
\includegraphics[width=.5\textwidth]{sensitivities_right_opt.mps}%
\par\vspace{-3mm}
\caption{Sensitivity of each statistically optimal observable to the
         corresponding operator coefficient. Note that these observables do not
         necessarily have maximal sensitivities but rather, as a set, induce
         minimal statistical uncertainties on the coefficient determination.}
\label{fig:sensitivities_opt}
\end{figure}

\begin{figure}
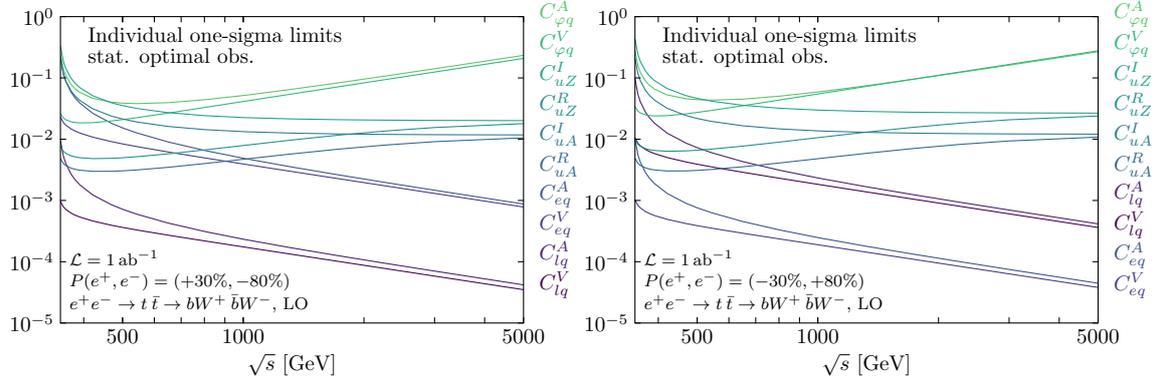
\centering
\includegraphics[width=.5\textwidth]{individual_left_opt.mps}%
\includegraphics[width=.5\textwidth]{individual_right_opt.mps}%
\par\vspace{-3mm}
\caption{Individual statistical one-sigma constraints on the effective operator
         coefficients as functions of the centre-of-mass energy, for either mostly
         left-handed and mostly right-handed electron beam polarizations, and a
         fixed integrated luminosity of $1\,\iab$. Different integrated
         luminosities are trivially obtained through a $(\mathcal{L}\:
         [\iab])^{-1/2}$ rescaling.}
\label{fig:individual_opt}
\end{figure}

\begin{figure}\centering
\includegraphics[scale=1.2]{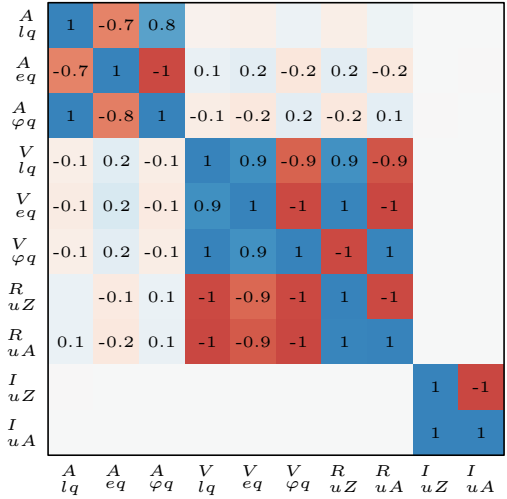}%
\caption{Correlations between the statistically optimal observables measured at a centre-of-mass energy of $380\gev$, for $P(e^+,e^-)=(0,-80\%)$ and $(0,+80\%)$ beam polarizations, respectively below and above the main diagonal. Although they are presented together, note the observable definitions depend on the beam energy and polarization.}
\label{fig:optimal_observable_correlations}
\end{figure}

The covariance matrix between the measurements of these statistically optimal observables is the inverse of the one which applies on the $\{C_i\}$ extractions and is thus given by $\epsilon\mathcal{L}d_{ij}$. For illustration, we display the associated correlation matrices obtained for a centre-of-mass energy of $380\gev$ and $P(e^+,e^-)=(0,\mp80\%)$ polarizations in \autoref{fig:optimal_observable_correlations}. The cases of left- and right-handed electron beams are respectively shown below and above the main diagonal. Note that the definitions of the statistically optimal observables depend on the centre-of-mass energy and polarization. So, although they are presented together, the sets corresponding to left- and right-handed electron beams are not identical.
These correlation matrices clearly indicate that the optimal observables measured in one single run are not independent. Three highly correlated blocs are clearly visible. First, the CP-odd observables relative to the $C_{uZ}^I$ and $C_{uA}^I$ operator coefficients are basically uncorrelated with all others. The bloc of axial vector operators also stands out from the one formed by vector and dipole operators. To derive constraints in all ten directions of the effective-field-theory parameter space, at least two runs are required with different centre-of-mass energies.

\paragraph{Theoretical robustness}
\label{sec:oo_theoretical_robustness}
We briefly examine the robustness of the statistically optimal observables
defined from analytical leading-order amplitudes for the resonant
$\eett\to\bwbw$ process against non-resonant contributions and
next-to-leading-order corrections in QCD. Note however that including non-resonant contribution in their definition could be feasible in the future and could enhance the sensitivity to operators affecting top-quark charged currents (as discussed in \autoref{sec:single_production}). A proper definition of optimal observables beyond leading order is more involved. To assess their theoretical robustness, we evaluate the optimal observables on a
standard-model sample produced for $500\gev$ centre-of-mass energy with
$P(e^+,e^-)=(+30\%,-80\%)$ beam polarization using \mg~\cite{Alwall:2014hca} which implements the complex-mass scheme for the next-to-leading order computation including non-resonant contributions. The corresponding distributions are
shown in \autoref{fig:robust}. The corrections induced on $\oo_i$ by
non-resonant and NLO contributions are dominated by their effects on the total
rate, which are respectively of about $6$ and $24\%$. Residual differences due
to shapes are below the couple of percent level except for the observables
$\oo_{lq}^A$, $\oo_{eq}^A$, $\oo_{\varphi q}^A$ associated with axial-vector operator coefficients for which they can exceed ten
percent. The non-resonant shape effects can naturally be reduced by cuts on both
$bW$ invariant masses. Selecting a $60\gev$ window around the top mass for
instance bring them below the couple of percent. After this cut, NLO QCD shape
corrections on these axial-vector observables can still reach about ten percent.
It is to be noted that the top and anti-top reconstruction from the $W$ and
$b$-jet (anti-$k_t$, $R=0.4$) of appropriate charge misses a sizeable fraction
of the radiation emitted in top-quark decay at NLO in QCD. The reconstructed top and
top--anti-top invariant masses notably develop tails towards low values. More
sophisticated reconstruction techniques could be imagined. No obstacle however
seems to bar the route towards theoretically reliable predictions for the
statistically optimal observables of simpler definitions, like the ones used in
this work. Although predictions can be much more precise, the observable optimization is then only accurate to leading order. This should be sufficient for all effective-field-theory parameters whose sensitivity mostly arise at that order.

\begin{figure}[p]
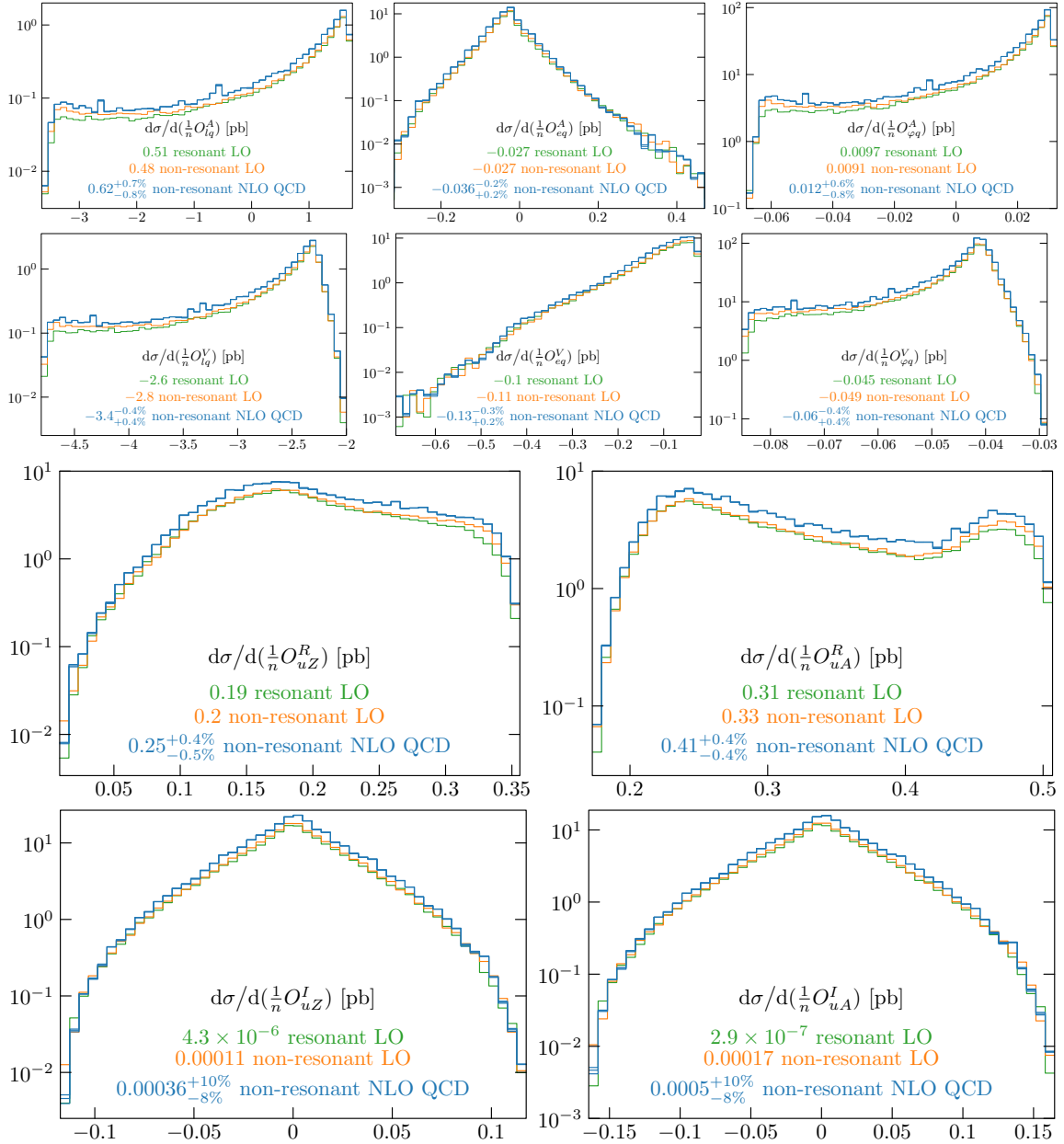
\centering
\adjustbox{max width=\textwidth}{%
\includegraphics{opt_nlo_lqA.mps}%
\includegraphics{opt_nlo_eqA.mps}%
\includegraphics{opt_nlo_pqA.mps}%
}%
\\%
\adjustbox{max width=\textwidth}{%
\includegraphics{opt_nlo_lqV.mps}%
\includegraphics{opt_nlo_eqV.mps}%
\includegraphics{opt_nlo_pqV.mps}%
}%
\\%
\adjustbox{max width=\textwidth}{%
\includegraphics{opt_nlo_ReuZ.mps}%
\includegraphics{opt_nlo_ReuA.mps}%
}%
\\%
\adjustbox{max width=\textwidth}{%
\includegraphics{opt_nlo_ImuZ.mps}%
\includegraphics{opt_nlo_ImuA.mps}%
}%
\caption{Distribution of the various statistically optimal observables measured on standard-model \bwbw\ samples including, successively, resonant top-quark pair production contributions at leading order, non-resonant contributions at leading-order, and additional QCD corrections. The centre-of-mass energy is fixed to $500\gev$ and the beam polarization to $P(e^+,e^-)=(+30\%,-80\%)$. The average values of these distributions, corresponding to $\oo_i/\epsilon\mathcal{L}$ in picobarns, are provided in the legends together with QCD scale variation between $m_t/2$ and $2m_t$ at NLO. That of CP-odd observables are compatible with zero within Monte Carlo uncertainties.}
\label{fig:robust}
\end{figure}

\section{Global reach}
\label{sec:fit}

We present, in this section, the global reach of top-quark pair production  measurements at future lepton colliders. Three benchmark scenarios are considered: circular-collider (CC)-like, ILC-like and CLIC-like. When available, we use the realistic statistical precisions estimated by the corresponding experimental collaborations. To establish a point of comparison, we first present the reach of cross section and forward-backward asymmetry measurements in the ILC-like scenario.
The constraints obtained with statistically optimal observables are then presented. The ten-dimensional parameter space that is accessible in top-quark pair production through interferences with SM amplitudes in the vanishing $m_b$ limit is considered.
The global one-sigma reach is shown, for the three benchmark run scenarios, in \hyperref[fig:fit_opt_benchmark_cc]{Figs.\,\ref{fig:fit_opt_benchmark_cc}}, \ref{fig:fit_opt_benchmark} and \ref{fig:fit_opt_benchmark_clic} which constitute our main results.
The impact of centre-of-mass energy lever arm and beam polarization is then further examined. An extension of the optimal observable capturing the quadratic dependence of scalar and tensor four-fermion operators is performed in \autoref{sec:lequs}.

\subsection{Uncertainty estimates}
\label{sec:run_scenarios}

We summarize here the assumptions and procedures employed to estimate the precision of top-quark pair production measurements at future lepton colliders. The global reach is rather sensitive to the operating scenario, especially to the centre-of-mass range covered by the machine and to the polarization of the electron and positron beams. We adopt the following benchmark scenarios:
\begin{description}
\item[CC-like scenario] As suggested in Ref.\,\cite{Benedikt:2018}, we consider the possibility that a circular lepton collider (CC, for short) would collect $200\ifb$ and $1.5\iab$ at centre-of-mass of $350$ and $365\gev$ respectively, without beam polarization. A specific discussion of this scenario is postponed to \autoref{sec:cc-discussion} where the impact of run parameters is examined.

\item[ILC-like scenario] Basing ourselves on Ref.\,\cite{Barklow:2015tja}, we envisages, in an ILC-like run scenario, the collection of an integrated luminosity of $500\ifb$ at a centre-of-mass energy of $500\gev$ and of $1\iab$ at $\sqrt{s}=1\tev$. The luminosity is shared equally between the mostly left-handed $P(e^+,e^-)=(+30\%,-80\%)$ and mostly right-handed $(-30\%,+80\%)$ beam polarization configurations. Compared to Ref.\,\cite{Barklow:2015tja}, we have modified the scenario in two ways. We ignore the possibility (discussed in \autoref{sec:obs_s_t} and \autoref{sec:lequs}) of colliding electron and positron of like-sign helicities. We also give priority to a 1\tev{} run over a luminosity upgrade, which could enhance the integrated luminosity at $\sqrt{s}=500\gev$ to $4\iab$.

\item[CLIC-like scenario] Following the staging scheme presented in Ref.\,\cite{CLIC:2016zwp}, we consider an integrated luminosity of $500\ifb$ at $\sqrt{s}= 380\gev$, of $1.5\iab$ at $\sqrt{s}=1.4\tev$ and of $3\iab$ at $\sqrt{s}=3\tev$ in a CLIC-like run scenario. These integrated luminosities are equally shared between left-handed $P(e^+,e^-)=(0,-80\%)$ and right-handed $(0,+80\%)$ beam polarization configurations. Positron polarization is not foreseen in the baseline operating scenario.
\end{description}

For estimating measurement uncertainties, we rely on the available full-simulation studies performed by the ILC and CLIC collaborations for the lepton-plus-jets final state obtained after one hadronic and one semi-leptonic top-quark decay~\cite{clictop, Bernreuther:2017cyi, Amjad:2015mma}. The charged lepton is taken to be either an electron or a muon. It allows to efficiently tag top and anti-top quarks, which is key to an accurate reconstruction of observables such as the forward-backward asymmetry. Observables such as the top-quark polarization asymmetries are most efficiently reconstructed in this final state. This restriction renders our prospects conservative. The inclusion of the tau-plus-jets and fully hadronic channels is expected to improve the limits significantly, once tau-tagging and jet-charge techniques are fully deployed\,\cite{Bilokin:2017lco}.

Experimental effects affecting the selection and reconstruction of top quark pairs have been identified in the studies referred to above. The most important ones are:
\begin{list}{$-$}{\itemsep0pt \topsep0pt \parsep0pt \labelwidth5mm \leftmargin\labelwidth}
\item inefficiencies and biases in the selection of lepton-plus-jets events,
\item migrations due to the finite resolution and incorrect pairing of the top-quark decay products,
\item losses due to the presence of significant tails in the luminosity spectrum,
\item uncertainties associated with the subtraction of background processes, notably of single top-quark production at high energies.
\end{list}
We base our analysis for the low centre-of-mass energy runs ($350$ to $500\gev$) on the experimental strategy developed in Refs.\,\cite{Amjad:2015mma, Bernreuther:2017cyi} to address the challenging migrations due to the reconstruction combinatorics. For the high-energy runs, we follow the performance estimates of techniques specifically developed for boosted top-quark reconstruction in Refs.\,\cite{clictop, Boronat:2016tgd}. In this case, the luminosity spectrum and the single-top production background are the main limiting factors.

As inputs to the global fits, we use realistic estimates of the expected statistical uncertainties on all observables. Average acceptances times efficiencies are listed in \autoref{tab:efficiencies}. For CLIC centre-of-mass energies, they are taken from Ref.\,\cite{clictop}. The range of values
indicates the difference between the two polarization configurations. The efficiency is typically higher for right-handed electron beam polarizations. For operation at $\sqrt{s}=380\gev$ and $500\gev$ the selection efficiency before quality cut is quoted.
With a left-handed electron beam, migrations due to poorly reconstructed top-quark candidates are rather pronounced. After
eliminating this effect with a stringent quality cut, the acceptance times efficiency for the configuration with a left-handed electron beam
is less than $40\%$.

\begin{table}
\begin{tabular*}{\textwidth}{@{\extracolsep{\fill}}rccccccc} \hline
$\sqrt{s}\;$ [GeV]                         & 350 & 365 & 380 & 500 & 1000 & 1400 & 3000 \\  \hline
acceptance times efficiency [\%]          &  -   &  - &  64-67\footnotemark  & $\sim50$ &   -  &  37-39 &  33-37 \\ 
equivalent $t\bar{t}$ event fraction   [\%]              & 10  & 10 & 10  & 10 &  6     &    6    &   5   \\ \hline
\end{tabular*}
\caption{Summary of the efficiencies obtained in Refs.\,\cite{clictop, Amjad:2015mma} (first row) and effective rate fractions available for analysis used in this study (second row). When multiplied by the $\eett$ cross section for the nominal centre-of-mass energy and the integrated luminosity, these yield the number of events available for analysis.}
\label{tab:efficiencies}
\end{table}
\footnotetext{The results correspond to the loose requirement on the reconstruction quality in Ref.\,\cite{clictop}. A more stringent cut is needed for some observables for the run with left-handed electron beam polarization, which reduces the efficiency to approximately 40\%.}

We also calculate the equivalent fractions of the theoretical \eett\ rate that is available for analysis, after accounting for efficiency, 
acceptance,
branching ratios and the effect of the luminosity spectrum. Equivalent fractions are averaged for the two polarization configurations
and rounded to the next integer. These numbers, presented in the final row of \autoref{tab:efficiencies}, are used to determine the statistical
uncertainty affecting observable measurements in the following sections.

Systematic uncertainties have been evaluated to some extent in Refs.\,\cite{Amjad:2015mma, Bernreuther:2017cyi}. It is plausible that theoretical and experimental systematics can be controlled to the level of the statistical uncertainties assumed here. We therefore ignore them in the following.

Note that modified optimal observables could be designed to circumvent the reconstruction hurdles mentioned above, symmetrizing their definitions over the two $b$ jets, including single top-quark production contributions in their definitions, or the total invariant mass as additional kinematic variable instead of fixing it to the nominal centre-of-mass energy. We leave such explorations to future work.

\subsection{Cross section and forward-backward asymmetry}

For the sake of comparison with statistically optimal observables, we present in this section the global reach deriving from the measurements of cross sections and forward-backward asymmetries for our ILC-like benchmark scenario. Corresponding results for our CLIC-like scenario are given in \autoref{sec:clic_scenario}.

\begin{figure}
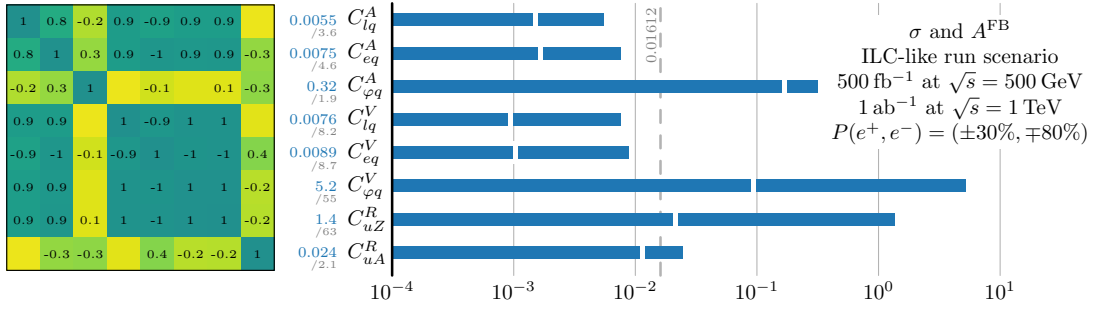
\centering
\scalebox{.8}{%
\includegraphics[scale=1.1]{fit_def_benchmark_corr.mps}\hspace{2mm}%
\raisebox{-4.67mm}{\includegraphics[scale=1]{fit_def_benchmark.mps}}}
\caption{Global one-sigma constraints and correlation matrix (rounded to the first decimal place) arising from the measurement of cross sections and forward-backward asymmetries in the ILC-like run scenario described in \autoref{sec:run_scenarios} and with the effective $t\,\bar{t}$ reconstruction efficiencies given in \autoref{tab:efficiencies}. The white marks stand for the individual constraints obtained when all other operator coefficients are set to zero. The dashed line provides the average constraint strength in terms of GDP (see \autoref{eq:gdp}). Numerical values for the marginalized constraints, and their ratio to individual ones are provided on the left-hand side.  $\Lambda=1\tev$ is assumed.
}
\label{fig:fit_def_benchmark}
\end{figure}

The eight measurements of cross sections and forward-backward asymmetries for two beam polarizations at two centre-of-mass energies are exactly sufficient
to constrain the eight CP-conserving operator coefficients which appear at
the linear level in these observables. Blue bars in \autoref{fig:fit_def_benchmark} cover the one-sigma allowed
ranges for each of those parameters after marginalization over all others.
Numerical values are provided in blue. The associated covariance matrix is also displayed. White marks indicate
individual constraints obtained under the unrealistic assumption that
all but the one operator coefficient considered vanish. Numbers for the
ratios of marginalized over individual constraints are given in grey.
Marginalized constraints are in some cases almost two orders of
magnitude looser than individual ones. The constraints on many operator
coefficients also appear highly correlated with each other.

\begin{figure}
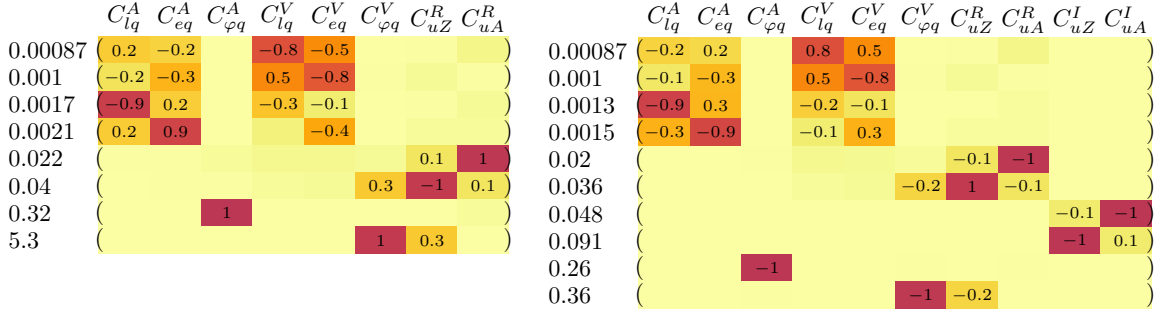
\centering%
\begin{minipage}[t]{.47\textwidth}
	\vspace*{0mm}%
	\includegraphics[scale=.9]{fit_def_benchmark_eig.mps}
\end{minipage}%
\begin{minipage}[t]{.53\textwidth}
	\vspace*{0mm}%
	\includegraphics[scale=.9]{fit_opt_benchmark_eig.mps}%
\end{minipage}
\caption{Eigenvalues and eigenvectors resulting from the diagonalization of the covariance matrix of the global one-sigma constraints deriving from the measurement of cross sections and forward-backward asymmetries (left), or statistically optimal observables (right). In the ILC-like run benchmark scenario described in \autoref{sec:run_scenarios}.
}
\label{fig:fit_def_eig}
\label{fig:fit_opt_eig}
\end{figure}

\begin{figure}\centering
\newcommand*{\ms}{1}
\begin{tabular}{*{4}{@{\;}c}@{}}
\includegraphics[scale=\ms]{mchi_def_0.mps}&%
\includegraphics[scale=\ms]{mchi_def_1.mps}&%
\includegraphics[scale=\ms]{mchi_def_2.mps}&%
\includegraphics[scale=\ms]{mchi_def_3.mps}%
\\%
\includegraphics[scale=\ms]{mchi_def_4.mps}&%
\includegraphics[scale=\ms]{mchi_def_5.mps}&%
\includegraphics[scale=\ms]{mchi_def_6.mps}&%
\includegraphics[scale=\ms]{mchi_def_7.mps}%
\end{tabular}
\caption{Profiled chi-squared for each of the eight effective operator coefficients constrained by the measurement of cross sections and forward-backward asymmetries in the ILC-like benchmark run scenario, when only the linear dependence in the operator coefficient is accounted for (dashed) or when quadratic terms are also included (solid).
}
\label{fig:def_obs_mchi}
%
\vspace{.5cm}
\begin{tabular}{*{5}{@{\;}c}@{}}
\includegraphics[scale=\ms]{mchi_opt_0.mps}&%
\includegraphics[scale=\ms]{mchi_opt_1.mps}&%
\includegraphics[scale=\ms]{mchi_opt_2.mps}&%
\includegraphics[scale=\ms]{mchi_opt_3.mps}&%
\includegraphics[scale=\ms]{mchi_opt_4.mps}%
\\%
\includegraphics[scale=\ms]{mchi_opt_5.mps}&%
\includegraphics[scale=\ms]{mchi_opt_6.mps}&%
\includegraphics[scale=\ms]{mchi_opt_7.mps}&%
\includegraphics[scale=\ms]{mchi_opt_8.mps}&%
\includegraphics[scale=\ms]{mchi_opt_9.mps}%
\end{tabular}
\caption{Profiled chi-squared for each of ten effective operator coefficients
         constrained through the measurement of statistically optimal
         observables in the ILC-like benchmark run scenario. Note an unreasonably large vertical axis scale was used to
         make the difference between the linearised (dashed) and quadratic
         (solid) effective field theories.
}
\label{fig:opt_obs_mchi}
\end{figure}

\subsection{Statistically optimal observables}

Statistically optimal observables defined on the fully differential $\bwbw$ final state are linearly sensitive to two CP-violating effective-field-theory parameters in addition to the eight CP-conserving ones accessible with cross section and forward-backward asymmetry measurements (again at the linear level, in the vanishing $m_b$ limit). Global constraints for our ILC-like scenario are displayed in \autoref{fig:fit_opt_benchmark}.
Although the individual sensitivities are not much improved, the use of statistically optimal observables reduces approximate degeneracies in the multidimensional parameter space and therefore also the correlations between  global constraints in specific directions.
Global limits are then at most a factor of $4$ worse than the individual ones.

A convenient metric to globally quantify the strength of the constraints in the
$n$-dimen\-sional parameter space of effective-operator coefficient is the
so-called \emph{global determinant parameter} defined as the $2n$ root of the Gaussian covariance matrix determinant\,\cite{Durieux:2017rsg}:
\begin{equation}
	\text{GDP}\equiv \sqrt[2n]{\det V}.
\label{eq:gdp}
\end{equation}
It evaluates to the geometric average of the semi axes of the one-sigma
ellipsoid of constraints. Interestingly, ratios of such quantities are
independent of the operator basis used to capture departure from the
standard model. Indeed, they are invariant under rotations and rescalings in the
space of operator coefficients. This measure is quoted for the eight-dimensional parameter space formed by CP-conserving effective-field-theory parameters in both \autoref{fig:fit_def_benchmark} and \autoref{fig:fit_opt_benchmark}. An improvement of by a factor of $1.6$ is obtained with statistically optimal observables. This is equivalent to a factor of $2.5$ increase in integrated luminosity at both $\sqrt{s}=500\gev$ and $1\tev$ (given a $1/\sqrt{\mathcal{L}}$ scaling of the GDP).

Diagonalizing the covariance matrix reveals the linear combinations of operator coefficients that are least and most tightly constrained. The result is shown for the fit based on optimal observables in \autoref{fig:fit_opt_eig} (right) and for the fit relying on the cross section and forward-backward asymmetries in \autoref{fig:fit_def_eig} (left). A linear combination of $C_{\varphi q}^V$ and $C_{uZ}^R$ operator coefficients is least constrained in the two cases. It is also in that direction that the improvement brought by statistically optimal observables is the most significant (more than an order of magnitude). Although the numerical constraints depend on the (arbitrary) operator normalization, the loose bound obtained in that direction when employing cross section and forward-backward asymmetry measurements may lie beyond the range of validity of the linear effective-field-theory approximation.

This is explicitly verified by temporarily including quadratic effective-field-theory dependences. The profiled chi-squared for each of the operator coefficients are displayed in \hyperref[fig:def_obs_mchi]{Figs.\,\ref{fig:def_obs_mchi}} and \ref{fig:opt_obs_mchi}, respectively for cross section plus forward-backward asymmetry and statistically optimal observable measurements. The dashed orange lines are obtained in the linear effective-field-theory approximation. Solid blue lines also include quadratic contributions from dimension-six operators, after expansion of the production, decay, and total width dependences. Note the unreasonably large $y$-axis scale in \autoref{fig:opt_obs_mchi}.
As seen in \autoref{fig:def_obs_mchi}, the quadratic dimension-six operator contributions significantly affect the constraints obtained with cross section and forward-backward asymmetry measurements. Secondary minima develop in the profiled chi-squared for values of the operator coefficients sometimes far away from what would be allowed in the linear effective-field-theory approximation. This situation is very much improved by the use of statistically optimal observables where quadratic contribution become completely negligible. Results obtained are thus cleaner from the effective-field-theory expansion point of view and can readily be translated from one basis of dimension-six operators to the other.


\begin{figure}
\scalebox{.8}{\includegraphics[scale=1.1]{fit_opt_benchmark_corr_cc.mps}\hspace{7.05mm}%
\raisebox{-4.67mm}{\includegraphics[scale=1]{fit_opt_benchmark_cc.mps}}}%
\caption{Global one-sigma constraints and correlation matrix deriving from the measurements of statistically optimal observables in a circular collider (CC-)like benchmark run scenario.}
\label{fig:fit_opt_benchmark_cc}
%
\vspace{1cm}
\scalebox{.8}{\includegraphics[scale=1.1]{fit_opt_benchmark_corr.mps}\hspace{3.3mm}%
\raisebox{-4.67mm}{\includegraphics[scale=1]{fit_opt_benchmark.mps}}}%
\caption{Global one-sigma constraints and correlation matrix deriving from the measurements of statistically optimal observables, in an ILC-like benchmark run scenario.}
\label{fig:fit_opt_benchmark}
%
\vspace{1cm}
\scalebox{.8}{\includegraphics[scale=1.1]{fit_opt_benchmark_corr_clic.mps}\hspace{2mm}%
\raisebox{-4.67mm}{\includegraphics[scale=1]{fit_opt_benchmark_clic.mps}}}%
\caption{Global one-sigma constraints and correlation matrix arising from the measurement of statistically optimal observables in a CLIC-like benchmark run scenario.
\newline
The white marks indicate the constraints that are individually obtained when all other operator coefficients are set to zero. The dashed lines provides the \emph{global determinant parameter} of the constraints on all ten operator coefficients, or on CP-conserving ones only. Numerical values for the marginalized constraints and their ratio to individual ones are provided on the left-hand side. Entries of the covariance matrix are rounded to the first decimal place. $\Lambda=1\tev$ is assumed. 
The overall $t\,\bar{t}$ reconstruction efficiencies quoted in \autoref{tab:efficiencies} are employed at the different centre-of-mass energies.}
\label{fig:fit_opt_benchmark_clic}
\end{figure}

\subsection{Global reach}
\label{sec:global_constraints}

We present, in this section, the global reach offered by statistically optimal observable measurements in the CC-, ILC- and CLIC-like programmes specified in \autoref{sec:run_scenarios} with the overall $t\,\bar{t}$ reconstruction efficiencies quoted in \autoref{tab:efficiencies}.

As seen in \hyperref[fig:fit_opt_benchmark_cc]{Figs.\,\ref{fig:fit_opt_benchmark_cc}}, \ref{fig:fit_opt_benchmark} and \ref{fig:fit_opt_benchmark_clic}, individual constraints on $C_{\varphi q}^V$ and $C_{\varphi q}^A$ operator coefficients are comparable in all three scenarios. The sensitivity to these operators does not grow with energy and arises mostly at low centre-of-mass energies where the top-quark pair production cross section is maximal. They are more efficiently constrained around $\sqrt{s}\simeq 400$ and $550\gev$, as seen in \autoref{fig:individual_opt}.
The limits on these vertex operators and on four-fermion operators of identical Lorentz structures are however correlated.
Beam polarization or angular distributions are unable to disentangle these two types of contributions. Only runs at different energies can. These correlations are therefore reduced in the ILC-like and, even further, in the CLIC-like scenario. Global constraints come close to individual ones in these cases.
The best individual limits on $C_{\varphi q}^V$ and $C_{\varphi q}^A$ are obtained in the ILC-like scenario which features the highest degree of polarisation and runs closest to ideal energies. The CLIC-like scenario however provides slightly stronger global constraints thanks to reduced correlations with four-fermion operators.
The impact of the centre-of-mass energy lever arm will be further examined in the following subsection.
Individual constraints on vertex operators are one to two orders of magnitude stronger than present ones and at least a factor of three better than the most optimistic HL-LHC prospects, as will be discussed in \autoref{sec:hadron_colliders_two_fermion}.
Global constraints are not available for comparison with the ones we derived for future lepton colliders. This is mostly due to limited sensitivities and more involved analyses at hadron colliders. 

Similar observations can be made for the dipole operator coefficients $C_{uA}^R$ and $C_{uZ}^R$ whose sensitivity only mildly grows with energy (see \autoref{fig:sensitivities_opt}) but which are still more efficiently constrained at lower centre-of-mass energies (see \autoref{fig:individual_opt}).
Their CP-violating counterparts are, on the contrary, slightly more easy to constrain at higher energies. They are thus somewhat better bounded in ILC- and CLIC-like scenarios.
In those two cases, $C_{uA}^I$ and $C_{uZ}^I$ are also completely uncorrelated with the eight other CP-conserving coefficients. No difference is then observed between their individual and global limits.
As will be discussed in \autoref{sec:hadron_colliders_two_fermion}, present direct individual constraints on CP-conserving dipole operators are two to three orders of magnitude looser than the prospects we obtain at future lepton colliders.
The most optimistic HL-LHC reach is still about two orders of magnitude lower.
Global constraints comparable to the ones we derive are not available.

Four-fermion operator coefficients benefit greatly from increase in centre-of-mass energy. A clear improvement is therefore seen from CC- to ILC- and CLIC-like scenarios. Four-fermion operators drive the reduction in GDP between ILC- and CLIC-like scenarios, as the constraints obtained on other operator coefficients are similar in these two cases.
GDP ratios between CC, ILC and CLIC constraints are $30:2.2:1$. To match CLIC level of constraints, the corresponding overall increase in luminosity required at the ILC and CC are respectively of $4.8$ and $990$ for all centre-of-mass energies (given our inclusion of statistical uncertainties only, GDPs scale as $1/\sqrt{\mathcal{L}}$).
Although a direct comparison between the two-lepton--two-quark operators of interest here and the (colour octet) four-quark operators presently probed in top-quark pair production at the LHC is not strictly speaking possible,
CC-like scenario would probes four-fermion operator couplings a factor of a few smaller, and a ILC- or CLIC-like scenarios two to four orders of magnitude smaller (see \autoref{sec:hadron_colliders_four_fermion}).
For comparisons in terms of scales probed for unit couplings, the square root of those factors applies.

\subsection{Exploring run scenarios}
\label{sec:run_parameters}

Beside comparing the constraining power of different sets of observables or benchmark scenarios, an
interesting exercise one can perform with GDP ratios (see definition in \autoref{eq:gdp}) is to optimize run
parameters to set the strongest overall constraints. 
Note however that different optimizations would be obtained in specific models of new physics privileging certain directions in the effective-field-theory parameter space.

\begin{figure}
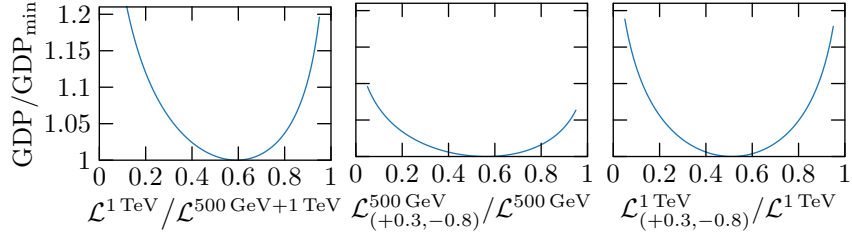
\centering
\adjustbox{max width=\textwidth}{%
\includegraphics{gdp_opt_tev.mps}%
\includegraphics{gdp_opt_x1.mps}%
\includegraphics{gdp_opt_x2.mps}}%
\caption{Variation of the GDP as a function of the share of integrated luminosity spent at the two centre-of-mass energies and with two beam polarization of the ILC-like benchmark run scenario.
}
\label{fig:gdp_opt_var}
\end{figure}

\paragraph{Sharing luminosity}
A first possibility is to vary the share of integrated luminosity spent with each polarization configuration and at the $500$\,GeV and $1$\,TeV centre-of-mass energies of our ILC-like scenario. The optimal repartition of $1.5\,\iab$ in total is the following: 
\begin{equation*}
\begin{array}{r@{\:}l@{\hspace{8mm}}c@{\hspace{8mm}}l}
\sqrt{s}=&500\gev
	& 610\,\ifb
		& 57\% \text{ with }P(e^+,e^-)={(+30\%,-80\%)}
\\
	&1\tev
	& 890\,\ifb
		& 51\% \hspace{2.8cm} {\scriptstyle\prime\prime}
\end{array}.
\end{equation*}
It yields a \text{GDP} of $0.01471$ which scales with the total integrated
luminosity as $1/\sqrt{\mathcal{L}}$. This is only a factor $1.005$
better than in our benchmark scenario. Variation around this minimum with the
fraction of integrated luminosity spent at $\sqrt{s}=1\tev$ and with both
polarization configurations are shown in \autoref{fig:gdp_opt_var}. Adding the
possibility of run at $\sqrt{s}=380\gev$ does not lead to any noticeable
improvement and the strongest constraints are found for a vanishing fraction of
the total integrated luminosity spent at that energy. On the contrary, it
requires about $4.6\,\iab$ shared between $\sqrt{s}=380$ and $500\gev$ runs to
achieve the same overall performance. The optimal run parameters are then:
\begin{equation*}
\begin{array}{r@{\:}l@{\hspace{8mm}}r@{\hspace{8mm}}rl}
\sqrt{s}=&380\gev
	& 1.5\,\iab
		& 57\% &\text{ with }P(e^+,e^-)={(+30\%,-80\%)}
\\
	& 500\gev
	& 3.1\,\iab
		& 51\% &\hspace{2.8cm} {\scriptstyle\prime\prime}
\end{array}.
\end{equation*}
Equivalently, sharing $1.5\,\iab$ between those two lower-energy runs, one would obtain a GDP that is approximately $1.8$ times worse.

\paragraph{Reducing beam polarization}
The impact on the electron and positron beam polarization can also be studied using GDP ratios.
As illustration, we consider again the ILC-like benchmark scenario described earlier with runs at centre-of-mass energies of $500\gev$ and $1\tev$ and decrease the absolute magnitude of each beam polarization separately. \hyperref[fig:gdp_polarization]{Figure\,\ref{fig:gdp_polarization}} shows the ratio of the obtained GDP with the one obtained with $P(e^+,e^-)=(\pm 30\%,\mp 80\%)$. It is seen that the magnitudes of the electron and positron polarizations are equally important. A decrease of $10\%$ in either of them leads to a decrease in GDP of about $5\%$ which could be compensated by an overall $11\%$ increase in luminosity at both centre-of-mass energies.
In particular, vanishing positron polarization leads to a degradation of the statistical constraints by a factor of $0.85$ and could be compensated by an overall $38\%$ increase in integrated luminosities.
The absence of beam polarization degrades the optimal-observable constraints by an average factor of about $2.15$ compared to the benchmark scenario. The global constraints on $C_{uZ}^I$ are the most affected (degraded by a factor of $8$), followed by $C_{eq}^V$, $C_{uA}^I$, $C_{uZ}^R$, $C_{\varphi q}^V$ and $C_{lq}^V$ (degraded, respectively, by factors of $3.5$, $2.7$, $2.6$, $2.5$ and $2.5$).

\begin{figure}\centering
\includegraphics[scale=1]{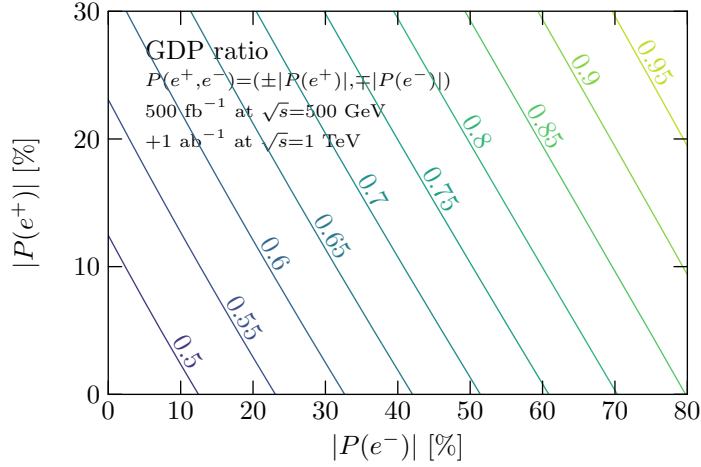}
\caption{Degradation the GDPs obtained with decreased beam polarization magnitudes starting from $P(e^+,e^-)=(\pm 30\%,\mp 80\%)$ configurations. Measurements of statistically optimal observables are employed in an ILC-like scenario.}
\label{fig:gdp_polarization}
\end{figure}

\begin{figure}\centering
\includegraphics[scale=1]{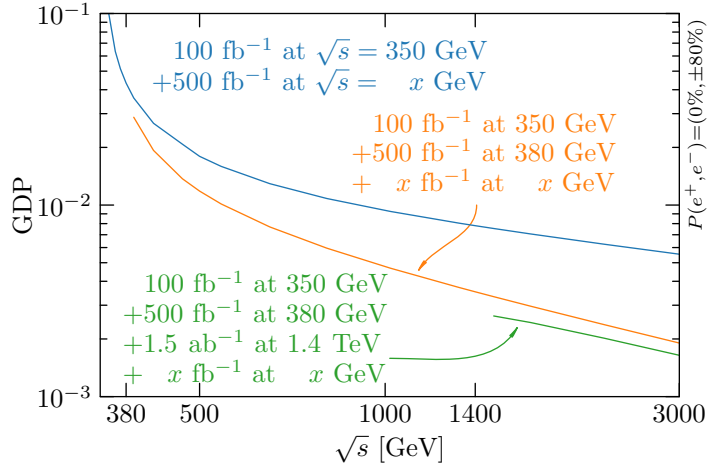}%
\caption{Improvement in the GDP brought by an increasing centre-of-mass energy lever arm. Statistically optimal observables are measured in runs at two different energies, with $P(e^+,e^-)=(0\%,\pm80\%)$ beam polarizations. Perfect efficiency on the $t\,\bar{t}$ reconstruction has been assumed. An overall efficiency of $\epsilon$ would rescale GDPs by a $1/\sqrt{\epsilon}$ factor.
Note the exact top-quark mass ($m_t=172.5\gev$), the leading-order and narrow top-quark width approximations used could affect results obtained near threshold.}
\label{fig:gdp_lever_arm}
\end{figure}

Reference~\cite{Janot:2015yza} stressed that the measurement of observables that are sensitive to the top-quark polarization can compensate for the lack of beam polarization. We indeed find that the observables studied in \autoref{subsec:dipolereal} have good sensitivity to effective operator coefficients and provide complementary information. The analysis of statistically optimal observables nevertheless shows that operation with two beam polarizations provides an important improvement of the results.

\paragraph{Impact of the energy lever arm} Runs at two separate centre-of-mass energies are indispensable to distinguish two- and four-fermion operators. \hyperref[fig:gdp_lever_arm]{Figure~\ref{fig:gdp_lever_arm}} shows that the average constraint strength ---in terms of GDP--- brought by the measurement of statistically optimal observables decreases significantly with the separation between available centre-of-mass energies. Three scenarios are displayed. The first one (blue curve) includes $100\,\ifb$ gathered at $\sqrt{s}=350\gev$ and $500\,\ifb$ at a higher energy point. A dramatic increase is brought by the first few tens of \gev\ of lever arm. A second scenario includes runs at $350$ and $380\gev$ in addition to $x\,\ifb$ collected at a higher $x\gev$ centre-of-mass energy. A third scenario also includes a fixed $1.5\,\iab$ gathered at $1.4\tev$. Two $P(e^+,e^-)=(0\%,\pm80\%)$ beam polarizations and perfect $t\,\bar{t}$ reconstruction efficiencies are assumed.

Compared to our benchmark ILC-like scenario, combining runs at $380$ and $500\gev$ with $500\,\ifb$ equally shared between $P(e^+,e^-)=(\pm30\%,\mp80\%)$ polarizations at both energies would degrade the GDP by a factor of $2.0$. Equal overall statistical performances would require $4.0$ times higher luminosities at both $380$ and $500\gev$. The four-fermion operators are the most affected, with global constraints about five and seven time worse for vector and axial-vector ones. The least affected are $C_{\varphi q}^V$ and $C_{uZ}^R$ constraints with a degradation by factors of about $1.2$.

\paragraph{Circular collider scenario}\label{sec:cc-discussion}
Degrading the polarization magnitude and centre-of-mass energy lever arm to the extreme, one obtains conditions that future circular colliders will be facing for tackling top-quark physics. As mentioned earlier, they could perform runs at the top-quark pair production threshold and, only few $\gev$ higher, at $365\gev$, without beam polarization.
The global constraints deriving from the measurements of statistically optimal observables are displayed in \autoref{fig:fit_opt_benchmark_cc} and were already somewhat discussed in the previous subsection.
They are significantly looser than that achievable with linear colliders, with GDPs respectively $14$ and $31$ times worse than in ILC- and CLIC-like scenarios.
Correlations and the ratios of global to individual constraints are also large. One could therefore expect the linear effective-field-theory approximation to break down. \hyperref[fig:mchi_cc]{Figure~\ref{fig:mchi_cc}} however seems to indicate that quadratic contributions still have a limited impact on one-sigma constraints.

Introducing $P(e^+,e^-)=(\pm30\%,\mp80\%)$ beam polarizations and sharing the same amount of integrated luminosity equally among them improves the GDP by a factor of two.
Global constraints on the $C_{uZ}^I$, $C_{\varphi q}^V$, $C_{eq}^V$, $C_{uA}^I$ and $C_{lq}^V$ coefficients benefit from the strongest improvements with factors of $9.6$, $3.6$, $3.2$, $3.2$ and $2.8$ respectively.
The strengthening of individual constraints is less significant, indicating that beam polarization mostly helps lifting approximate degeneracies instead of improving sensitivities. When exploiting the pair-production threshold region for coupling measurements, it is worth keeping in mind it is usually primarily employed to determine precisely the top-quark mass and width, assuming couplings are sufficiently constrained by runs at higher energies. To our knowledge, no study of the threshold region has so far included top-quark effective-field-theory dependences and analysed their impact on mass and width determinations.

\begin{figure}
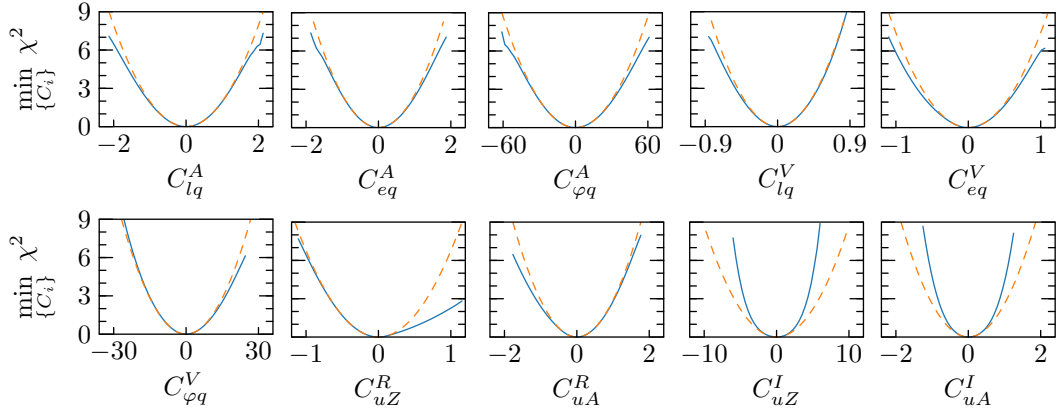

\centering\newcommand*{\ms}{1.}%
\begin{tabular}{*{5}{@{}r}@{}}
\includegraphics[scale=\ms]{mchi_opt_cc_0.mps}&%
\includegraphics[scale=\ms]{mchi_opt_cc_1.mps}&%
\includegraphics[scale=\ms]{mchi_opt_cc_2.mps}&%
\includegraphics[scale=\ms]{mchi_opt_cc_3.mps}&%
\includegraphics[scale=\ms]{mchi_opt_cc_4.mps}%
\\[-1mm]%
\includegraphics[scale=\ms]{mchi_opt_cc_5.mps}&%
\includegraphics[scale=\ms]{mchi_opt_cc_6.mps}&%
\includegraphics[scale=\ms]{mchi_opt_cc_7.mps}&%
\includegraphics[scale=\ms]{mchi_opt_cc_8.mps}&%
\includegraphics[scale=\ms]{mchi_opt_cc_9.mps}%
\end{tabular}
\caption{Profiled chi-squared deriving from the measurements of statistically optimal observables for a circular collider (CC-)like run scenario with the overall $t\,\bar{t}$ reconstruction efficiencies quoted in \autoref{tab:efficiencies}.
Solid blue lines include quadratic effective-field-theory dependences while dashed orange ones are obtained in the linear approximation.}
\label{fig:mchi_cc}
\end{figure}

\section{Comparison with existing limits and HL-LHC prospects}
\label{sec:lhc}

In this section, lepton-colliders prospects are compared to
existing limits on the top-quark effective field theory arising from fit to Tevatron
and LHC data. The indirect limits deriving from low-energy measurements are
briefly discussed. Prospects for the remaining of the LHC programme are also examined.

\subsection{Hadron-collider sensitivity to two-fermion operators}
\label{sec:hadron_colliders_two_fermion}

\begin{table}
\scriptsize
\begin{tabular*}{\textwidth}{@{}lc@{\:}c@{\qquad}*{5}{c@{\:}}@{\qquad}c@{\;\;}c@{\;\;}c} 
\hline\noalign{\vskip.5ex}
& \multicolumn{2}{c}{existing }& \multicolumn{5}{c}{expected at high luminosity } & \multicolumn{3}{c}{expected at $e^+e^-$ }\\
& {\sc TopFitter} & Ref.\,\cite{Birman:2016jhg} & Ref.\,\cite{Birman:2016jhg}  & $t\bar{t}V$\,\cite{Rontsch:2014cca, Rontsch:2015una} & $t\bar{t}V$ 10\% & $t\bar{t}V$ 3\% & $tZj$ \cite{Degrande:2018fog} & CC & ILC & CLIC\\[.5ex]
 \hline\noalign{\vskip.5ex}
$C_{\varphi q}^1$	&$[-12,13]$	&		&		&$[-1.3,1.0]$	&$[-2.0,2.0]$	&$[-0.6,0.6]$	&$[-17,17]$	&$0.14$	&$0.076$	&$0.098$\\
$C_{\varphi q}^3$	&$[-5.3,3.1]$	&		&		&$[-1.0,1.3]$	&$[-2.0,2.0]$	&$[-0.6,0.6]$	&$[-2.8,1.5]$	&$0.14$	&$0.076$	&$0.089$\\
$C_{\varphi u}  $	&$[-20,17]$	&		&		&$[-1.3,3.0]$	&$[-3.4,2.8]$	&$[-0.8,1.0]$	&$[-26,20]$	&$0.29$	&$0.15$		&$0.18$\\
$C_{\varphi ud}$	&		&$[-11,14]$	&$[-8.4,11]$	&		&		&		&$[-8.4,8.4]$	&	&		&	\\
$C_{uB}$		&$[-20,14]$	&		&		&$[-4.8,4.8]$	&$[-12,12]$	&$[-6.6,4.0]$	&$[-12,11]$	&$0.022$&$0.022$	&$0.024$\\
$C_{uW}$		&$[-2.0,2.8]$	&$[-2.7,1.6]$	&$[-1.3,1.3]$	&$[-1.4,1.4]$	&$[-3.6,3.8]$	&$[-2.2,2.2]$	&$[-1.3,1.3]$	&$0.015$&$0.014$	&$0.016$\\
$C_{dW}$		&		&$[-3.4,3.6]$	&$[-2.9,3.1]$	&		&		&		&		&	&		&	\\[.5ex]
\hline
\end{tabular*}
\caption{Individual 95\% C.L.~limits on two-quark operator coefficients deriving from measurements at hadron colliders. The first two columns show the existing limits derived by the \textsc{TopFitter} group~\cite{Buckley:2015lku} and in Ref.~\cite{Birman:2016jhg}.
The next four columns are expected limits with $3\iab$ of integrated luminosity at the LHC, derived from single top and top decay measurements~\cite{Birman:2016jhg}, from differential distributions in $t\bar{t}V$ production~\cite{Rontsch:2014cca, Rontsch:2015una}, and from the total $t\bar{t}V$ cross sections measured with $10\%$ and $3\%$ precision.
The $tZj$ columns show limits expected with $300\ifb$ using a $p_T(t)>250\gev$ selection cut. The last three columns are the individual limits obtained in this work for CC-, ILC- and CLIC-like run scenarios. As discussed in \autoref{sec:global_constraints} individual constraints are similar in those three cases although global ones are less so.}
\label{tab:limits2f}
\end{table}

The Tevatron and LHC experiments are sensitive to top-quark electroweak
couplings through several measurements. Charged-current interactions
of the top quark are sensitive to several of the operators studied in this
paper. The $tbW$ vertex has been characterized thoroughly in studies
of top-quark decay and is also probed in the
electroweak single-top production processes at hadron colliders.

\paragraph{Existing constraints}
ATLAS and CMS published measurements of the longitudinal and left- and 
right-handed helicity fractions $F_0$, $F_L$ and $F_R$ of the
$W$ boson in top-quark decays~\cite{Aad:2012ky,Chatrchyan:2013jna} having
a precision of $2$ to $3\%$. 
%
The rate of single top-quark production in the $t$-channel, first
observed at the Tevatron~\cite{Aaltonen:2009jj,Abazov:2009ii},
is precisely measured at the LHC, with errors on the fiducial cross
section ranging from $6$ to $8\%$~\cite{Aaboud:2017pdi,Khachatryan:2014iya}.
Single top-quark production in the $s$-channel was so far only observed
at the Tevatron~\cite{CDF:2014uma}.

Reference~\cite{Birman:2016jhg} combined the measurements of the single
top production cross sections and the $W$ helicity fractions in top-quark
decay at $\sqrt{s}=8\tev$ in an analysis involving only the three anomalous
couplings affecting the $tbW$ vertex. These constraints can be
converted\footnote{The fit of Ref.~\cite{Birman:2016jhg} yields the
following $95\%\,$C.L.\ limits:
$\Re\{V_R\} \in [-0.33,0.41]$,  
$\Re\{g_L\} \in [-0.19,0.20]$,  
$\Re\{g_R\} \in [-0.15,0.09]$. In our conventions,
$\delta V_L = C_{\varphi q}^{3} \frac{2m_t^2}{\Lambda^2}$,
$V_R = C_{\varphi ud}^* \frac{m_t^2}{\Lambda^2}$,
$g_L = C_{dW}^{*} \frac{4m_tm_W}{\Lambda^2}$,
$g_R = C_{uW} \frac{4m_tm_W}{\Lambda^2}$.}
to the $95\%\,$C.L.\ limits on the effective operator coefficients displayed in the second column of \autoref{tab:limits2f},
for $\Lambda=1\tev$. Dedicated measurements of angular distributions
and asymmetries in single top-quark production can provide slightly
more stringent limits than the inclusive cross section
measurement~\cite{Aaboud:2017yqf,Aaboud:2017aqp,Khachatryan:2016sib}.


More recently, the Tevatron and LHC have gained access
to associated production processes, where a top-quark pair or single
top quark is produced together with a
photon~\cite{Aaltonen:2011sp,Aad:2015uwa,Sirunyan:2017iyh,CMS:2018kma} or $Z$
boson~\cite{Khachatryan:2015sha,Aad:2015eua, Aaboud:2017ylb, Sirunyan:2017nbr}. Measurements
of the production rates yield limits on the $t\bar{t}\gamma$ and
$t\bar{t}Z$ vertices. The existence of these processes is now
firmly established, but the precision of the rate measurements is
still quite limited: the inclusive production cross sections
are currently known to $20-30\%$ precision.

Single top-quark production in association with a $Z$ boson probes both the
charged current interactions (when the $Z$ boson is produced through $W$ fusion)
and the interactions with the $Z$ boson (when the $Z$ boson is radiated off the
top quark). This process has been observed in LHC run~II
and cross-section measurements at $\sqrt{s}=13\tev$ by ATLAS~\cite{Aaboud:2017ylb} and CMS~\cite{Sirunyan:2017nbr} are in good agreement with standard-model
predictions within an experimental uncertainty of approximately $30\%$. At this
level of accuracy, the limits on relevant dimension-six operators are not
competitive with respect to other processes such as $t\bar tV$ and single
top-quark production. Evidence for single top-quark production in association
with a photon has also been found very recently by the CMS
collaboration~\cite{CMS:2018kma}. The associated cross section measurement has
an uncertainty of about $40\%$.

\begin{figure}\centering
\includegraphics[width=\textwidth, trim=15 30 25 20, clip]{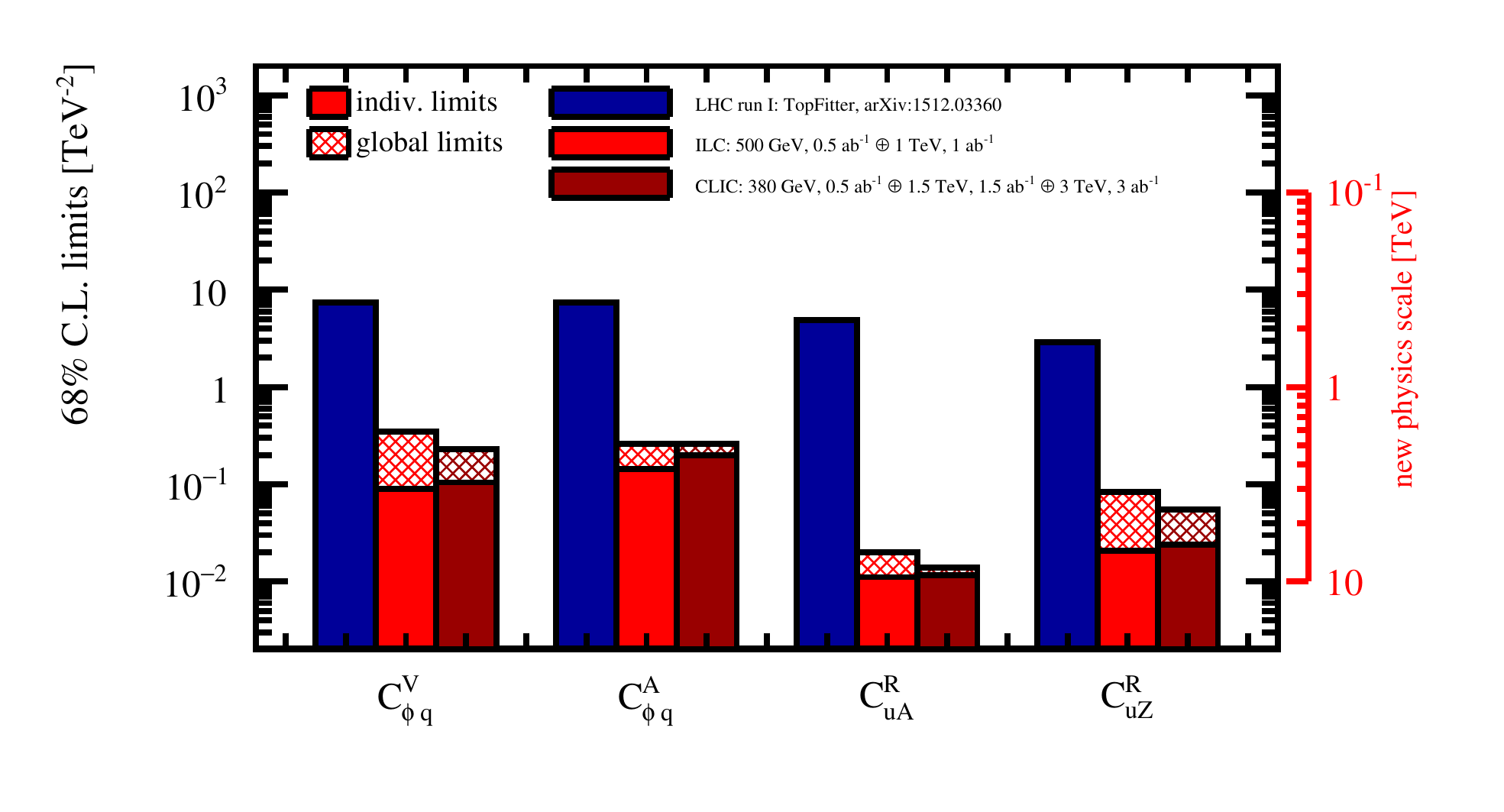}%
\caption{Summary of $68\%\,$C.L.\ limits on the $C_{\varphi q}^V$ and $C_{\varphi q}^A$ coefficients of the two-fermion operators that affect the left- and right-handed couplings of the top quark to the $Z$ boson and the dipole operators $C_{uA}^R$ and $C_{uZ}^R$. Individual limits from Tevatron plus LHC run~I data are taken from Ref.~\cite{Buckley:2015lku} and are converted to the conventions we use by assuming Gaussian probability distributions and by neglecting the unpublished correlations. For the ILC- and CLIC-like scenarios, individual and global constraints are presented. The conversion between our conventions and that of the  LHC TOP WG~\cite{AguilarSaavedra:2018nen} are detailed in \autoref{sec:lhc_top_wg}.}
\label{fig:two_fermion_summary}
\end{figure}

The most complete effective-field-theory analysis of the top-quark sector to date was performed by the \textsc{TopFitter} group. It took into account a large collection
of Tevatron and LHC run~I measurements~\cite{Buckley:2015nca, Buckley:2015lku, Englert:2017dev}, performing a global analysis including nine operator coefficients.
Individual constraints were derived on three others. The conversion of these limits to our normalization is provided in \autoref{app:topfitter}.

We present in \autoref{tab:limits2f} a summary of the individual limits derived from the Tevatron and LHC data by the \textsc{TopFitter} group~\cite{Buckley:2015lku} and in Ref.\,\cite{Birman:2016jhg}. For comparison, prospects obtained in this paper for the CC-, ILC- and CLIC-like run scenarios are given in the last three columns of the same table.
The existing individual constraints obtained by the \textsc{TopFitter} group are visually compared to ILC and CLIC prospects for two-fermion operators in \autoref{fig:two_fermion_summary}. For this figure, the rotation to our basis of operator coefficients is performed by assuming the unpublished correlations between \textsc{TopFitter} limits vanish. Any of the three $\epem$ collider scenarios
considered improves the individual limits by over an order of magnitude.

\paragraph{Prospects}
Several authors have studied the expected evolution of LHC limits during run~II and III as well as after its high-luminosity phase. Reference~\cite{Birman:2016jhg} presented limits expected from helicity fractions and
single top-quark cross section measurements (the latter based partially on Ref.~\cite{Schoenrock:2013jka}) after the complete
LHC programme ($3\iab$ at $\sqrt{s} = 14\tev$) that are only marginally stronger than the existing ones.

Several studies of the prospects for the determination of the top-quark
couplings to the photon and $Z$ boson have been performed. Reference\,\cite{Englert:2017dev}
estimated that constraints arising from inclusive $\sigma(pp\to t\bar{t}Z)$ measurements would improve
by at most a factor of two with the high-luminosity phase of the LHC. 
The precision of the SM prediction, approximately 10\% at NLO, may limit the progress
of the constraints derived from the inclusive cross section.
Reference\,\cite{Schulze:2016qas} considered the construction of two
ratios of cross sections $R_{Z,\gamma} = \sigma_{t\bar{t} Z,\gamma} / \sigma_{t\bar{t}}$
which can be predicted with a precision of $3\%$ and thus avoid this limitation.
An analysis of differential cross sections may also enhance the sensitivity. References\,\cite{Rontsch:2014cca,Rontsch:2015una} derived better limits for $3\iab$ collected at $13\tev$ by considering the $p_T$ distribution of the $Z$ boson and the $\Delta\phi_{ll}$ angular distribution of the leptons arising from the $Z$-boson decay.

We estimate individual $68\%\,$C.L.\ limits
on four operator coefficients under several assumptions for the precision
of the inclusive $pp \rightarrow t\bar{t}V$ cross section measurement at the LHC.
The dependence of the cross section on the coefficients is
taken from Ref.~\cite{Bylund:2016phk} for the first three operators, up to
quadratic order, and estimated in a similar fashion for $C_{uB}$. The first
scenario considers a precision of $30\%$, which corresponds roughly to
the first measurements at the LHC. The limits we obtain are
indeed similar to those of the \textsc{TopFitter} group. The second scenarios
envisages a precision of $10\%$, which matches
the approximate theoretical uncertainty of current next-to-leading order
calculations. The third scenario, with $3\%$ precision, represents the
ultimate precision envisaged in Ref.~\cite{Schulze:2016qas}.
The projections for the latter two scenarios are
displayed in the fifth and sixth columns of \autoref{tab:limits2f}.

With a large integrated luminosity $tZ$ production allows to access
higher-$p_T$ regions, where the sensitivity to dimension-six operators
is enhanced by the breaking of standard-model unitarity cancellations
they induce. In Ref.~\cite{Degrande:2018fog}, this effect is studied
and projections are derived for $300\ifb$ of integrated luminosity.
These results are displayed in the seventh column of \autoref{tab:limits2f}.

The improvement during the remaining LHC program and its high-luminosity stage
is quite substantial for all two-fermion operators, primarily driven by precision
measurements of rare associated production processes. The limits on $C_{\varphi q}^1$,
$C_{\varphi q}^3$ and $C_{\varphi u}$ may improve by nearly an order of magnitude
in the most optimistic scenario.
Improvements are milder for the dipole operators $C_{uB}$ and $C_{uW}$ ---by a
factor of three--- as for these operators
the quadratic term is dominant.

An energy upgrade of the LHC in the existing tunnel~\cite{Bruning:601847} or a new
$100\tev$ $pp$ collider~\cite{Arkani-Hamed:2015vfh, CEPC-SPPCStudyGroup:2015csa}
would dramatically increase the sample size for top-quark associated
production processes and thus present important opportunities for
precision measurements of their rates~\cite{Mangano:2016jyj}. The dependence of the inclusive cross
section on the dimension-six operator coefficients grows only very slowly with
increasing centre-of-mass energy, but differential measurements in the
transverse momentum of the $Z$-boson or photon can strongly enhance the
sensitivity~\cite{Rontsch:2014cca}. A detailed evaluation of the potential
for measurements of the electroweak couplings of the top quark
remains to be performed.

\paragraph{}
We conclude that electron-positron colliders offer excellent potential
to probe new physics through the two-fermion operators considered in
this section, even at a relatively
modest centre-of-mass energy, between $350$ and $500\gev$. Comparison of
the expected individual limits for the CC-, ILC- and CLIC-like scenarios
with the projections for hadron colliders shows that the sensitivity
of $\epem$ machines exceeds that of the Tevatron and LHC run~I by
one to three orders of magnitude. Improvements of the limits during future LHC runs and its high-luminosity phase are limited to factors $1.5$ to $25$,
in the most optimistic scenario. A linear $\epem$ collider therefore
seems the most promising project to reach the ultimate precision in
the determination of two-fermion operator coefficients that affect top-quark electroweak couplings.


\subsection{Hadron collider sensitivity to four-fermion operators}
\label{sec:hadron_colliders_four_fermion}

In this section, we compare the limits on the coefficients of the
$e^+e^+t\,\bar{t}$ operators studied in this paper with limits that can
be set on the coefficients of $q\bar{q}t\bar{t}$ operators at hadron
colliders. As these two sets of operators are of different nature,
the conclusion of such an exercise must be interpreted with caution.\footnote{The four-fermion operators affecting $\eett$ production
can in principle be probed at hadron colliders, in
$pp\to t\bar{t} e^+e^-$ production, but the sensitivity is likely
limited and no such analysis has been performed so far.}

\paragraph{Existing constraints}
Unpolarized top-quark pair production at hadron colliders is only linearly
sensitive to four combinations of $q \bar{q} t \bar{t}$ operator coefficients,
denoted $C^1_u$, $C^2_u$, $C^1_d$, $C^2_d$~\cite{Zhang:2010dr}. The $C^1_q +
C^2_q$ sums and $C^1_q - C^2_q$ differences respectively appear in the total
cross section and production asymmetry ($A^\text{FB}$ at the Tevatron, or $A_C$
at the LHC~\cite{Sirunyan:2017lvd}). The individual and global nine-parameter
$95\%\,$C.L.\
limits obtained from Tevatron and LHC run~I data in Ref.~\cite{Buckley:2015lku}
are reproduced in \autoref{tab:topfitter4f}. They take into account the
measurements of asymmetries, as well as inclusive and differential
cross-sections.\footnote{In the conventions of the LHC TOP WG~\cite{AguilarSaavedra:2018nen},
  the definitions of Ref.~\cite{Buckley:2015lku} correspond to
$C^1_u = \frac12 ( c_{Qq}^{3,8} + c_{Qq}^{1,8} + c_{tu}^8)$,
$C^2_u = c_{Qu}^8 + c_{tq}^8$,
$C^1_d = \frac14 (-c_{Qq}^{3,8} + c_{Qq}^{1,8} + c_{td}^8)$,
$C^2_d = c_{Qd}^8 + c_{tq}^8$,
$C_t   = c_{Qq}^{3,1}$.
Note that $C^1_d$ originally defined in Ref.\,\cite{Zhang:2010dr} is a factor of four larger.}
\begin{table}
\begin{tabular*}{\textwidth}{@{\extracolsep{\fill}}lc@{\,}cc@{\,}c@{\quad}c@{}} \hline
& \multicolumn{4}{c}{95\% C.L. limits from Tevatron + LHC run I} \\
                    & $C^{1}_{u}$   & $C^{2}_{u}$  & $C^{1}_{d}$  & $C^{2}_{d}$  \\ \hline
individual limits   & $[-1.7,0.3]$  & $[-2.9,0.6]$ & $[-1.1,0.2]$ & $[-1.8,1.4]$ \\
marginalised limits & $[-3.9,3.1]$  & $[-7.0,4.5]$ & $[-2.3,2.5]$ & $[-8.4,8.8]$ \\ \hline
\end{tabular*}
\caption{$95\%\,$C.L.\ limits on four-fermion operators obtained in a nine-parameter fit of top physics measurements at the Tevatron and LHC run~I. The results are presented without the factor $v^2/\Lambda^2$ that Ref.~\cite{Buckley:2015lku} applies.}
\label{tab:topfitter4f}
\end{table}
%
The global limits range from $|C|\lesssim 2.5$ to $|C|\lesssim  9$.
For couplings of order one, these correspond to new-physics scales between
$300$ to $600\gev$. The limits on $C_1^u$ and $C_1^d$ are a factor two to
three better than limits on $C_2^u$ and $C_2^d$. Individual limits
are a factor two to five stronger than the results of a global fit.

The global limits on four-fermion operators from the ILC-like scenario
are of order $2 \times 10^{-3}$. The limits in the CLIC-like scenario,
with the powerful constraint from operation at $\sqrt{s}=3\tev$, are of
order $3 \times 10^{-4}$. For couplings of order one, these correspond
to scales between $20$ and $60\tev$.

\paragraph{Prospects}
The LHC limits are expected to improve as run~II and III accumulate a
larger sample. As inclusive cross-section and charge asymmetry measurements
are limited by systematic uncertainties, most of the gain is expected to
come from the boosted regime, where the sensitivity to four-fermion
operators strongly grows. After run~I, the cross
section and charge asymmetry measurements in the boosted
regime yield limits of strength comparable to those deriving from inclusive
measurements~\cite{Englert:2016aei, Rosello:2015sck}. As the integrated
luminosity of the data set at $\sqrt{s}=13\tev$ increases, such measurements
can drive a steady progress of the constraints.
An order of magnitude improvement of the limits is within reach, provided
that the measurements in the boosted regime ($m_{t\bar{t}} > 1$ to $1.5\tev$)
achieve a precision similar to the inclusive measurements in
run~I~\cite{Rosello:2015sck}.


With $300\ifb$, four-top production can provide competitive constraints on the $q\bar{q}t\bar{t}$
operators, in models for which contributions due to powers of dimension-six operator
coefficients as high as four are dominant with respect to that of higher dimensional
operators~\cite{Zhang:2017mls}. With $3\iab$, the estimated precision on the total
four-top production cross section improves by a factor of about two. Limits on
operator coefficients are however only strengthened by the fourth root of this
factor.

The potential of hadron colliders reaching even higher energies remains to be
explored. Reasonably precise measurements of the $t\bar{t}$ production cross section and
charge asymmetry in an extremely boosted regime,
with $m_{t\bar{t}} >10\tev $, seem feasible~\cite{Aguilar-Saavedra:2014iga}.
The sensitivity of such measurements to $q\bar{q} t\bar{t}$ operators
increases by two orders of magnitude.

\paragraph{}
We conclude that the projected bounds on the coefficients of $e^+e^-t\,\bar{t}$ operators in ILC- and CLIC-like run scenarios are approximately
four orders of magnitude more stringent that the current bounds
on the $C^1_u$, $C^2_u$, $C^1_d$, $C^2_d$ coefficients of colour-octet $q\bar{q}t\bar{t}$ operators. 
This represents a gain of about two orders of magnitude in terms
of the scale that is probed, for identical couplings.
While very naive, this comparison shows rather eloquently
how the precision of the measurements in an $\epem$ collider allows to probe high energy
scales, compensating for the relatively low direct energy reach.

\subsection{Indirect limits}

Indirect limits on top-quark operators
can be derived using the renormalization group (RG) running and mixing
of operator coefficients. These arise from loop contributions that are
enhanced by logarithmic $\log\Lambda^2$ terms.
They are however indirect because tree level contributions (into which the
top-quark operators mix) are always present. Whether they can
be safely neglected depends on the model.

RG-induced limits have been derived on a number of operators in Ref.\,\cite{deBlas:2015aea}.
The RG evolution of the entire group of dimension-six operators was used to estimate the
impact of operators involving top quarks on electroweak precision 
measurements at LEP and SLD. The individual $95\%\,$C.L. limits on four-fermion
operators $e^+e^-t\,\bar{t}$ are order $0.1$.
The running of $C_{WB}$ (related to the $S$-parameter~\cite{Peskin:1991sw})
constrains a combination of the top-quark dipole operator coefficients $C_{uW}$ and $C_{uB}$.
The resulting $95\%\,$C.L.\ limits are $[-0.05,0.2]$. The fit to the
electroweak precision data moreover yields strongly correlated constraints on
the left-handed and right-handed electroweak couplings of the top quark
($\delta g_L = (C_{\varphi q}^{1} - C_{\varphi q}^{3}) \frac{v^2}{\Lambda^2}$
and $\delta g_R = C_{\varphi u} \frac{v^2}{\Lambda^2}$). The individual limits
on $C_{\varphi q}^{1}$, $C_{\varphi q}^{3}$ and $C_{\varphi u}$ are of the
order of $0.05$. A two-parameter fit yields marginalized limits
that are $5$ to $10$ times looser.

Electroweak precision observables, namely $\delta g^b_L$ and the $T$ parameter, are considered
in Ref.\,\cite{Brod:2014hsa} together with the $K^+\to\pi^+\nu\bar\nu$ and $B_s\to\mu^+\mu^-$ rare meson decays.
The authors assume $C_{\varphi q}^1+ C_{\varphi q}^3=0$ to satisfy the $Z\to b\bar b$ constraints at LEP. The loop-level
contributions from $C_{\varphi q}^3$ and $C_{\varphi u}$ are proportional to $\log\mu_W/\Lambda$, where $\mu_W$ is the
electroweak scale. The coefficients $C_{\varphi q}^3\frac{v^2}{\Lambda^2}\log\frac{\mu_W}{\Lambda}$ and
$C_{\varphi u}\frac{v^2}{\Lambda^2}\log\frac{\mu_W}{\Lambda}$ can be constrained to the order of a few percent, which
translates to about $0.5$ on $C_{\varphi q}^3$ and $C_{\varphi u}$ assuming $\Lambda=1\tev$.

The above limits are based on strong assumptions, namely that only one operator
receives a non-zero initial condition at the scale $\Lambda$. In particular,
tree-level contributions from other operators into which the
top-quark operators mix are set to zero. To obtain more reliable limits, one
should include the finite (i.e.~non-logarithmic) term from the loop
contribution. This allows limits to be extracted by marginalising over the tree
level contributions. For example, a global fit to all electroweak precision observables was performed in Ref.\,\cite{Zhang:2012cd},
including loop contributions from dimension-six top-quark operators, and marginalising over the $O_{\varphi WB}$ and $O_{\varphi D}$ operators, which are often identified as the $S$ and $T$
parameters. The resulting limits are scale-independent, in contrast to the RG-induced
ones. The individual $68\%\,$C.L.\ constraints are:
\begin{equation*}
	C_{\varphi q}^1-C_{\varphi q}^3=2.0\pm2.8,\quad
	C_{\varphi u}=1.8\pm1.9,\quad
	C_{uW}=-0.6\pm1.9,\quad
	C_{uB}=14\pm15 .
\end{equation*}

Finally, at the LHC, top-quark operators enter Higgs boson production and decay processes
through top- and bottom-quark loops. Higgs signal strength measurements thus constrain top-quark
operators indirectly. The corresponding limits were derived
in Ref.~\cite{Vryonidou:2018eyv} using projected Higgs measurements at the LHC, with $3\iab$, including
differential distributions. The resulting $95\%\,$C.L.\ individual limits on the
relevant operators in this work are:
\begin{gather*}
	C_{\varphi q}^1+C_{\varphi q}^3\in[-1.3,1.3],\quad
	C_{\varphi q}^1-C_{\varphi q}^3\in[-3.3,3.3],\quad
	C_{\varphi u}\in[-2.5,2.5],\\
	C_{uW}\in[-0.23,0.23],\quad
	C_{uB}\in[-0.20,0.20].
\end{gather*}
These are based on the assumption that other Higgs operators do not modify the
Higgs signal strengths. However the EFT scale is set to $m_H$. Therefore, the log-enhanced
terms are not used and the results can be considered as an estimate of the indirect sensitivity
at the high-luminosity LHC. Also note that marginalised bounds are much weaker.%

Almost all individual indirect limits presented in this section are much stronger than the existing bounds
deriving from LHC and Tevatron measurements and shown in \autoref{tab:limits2f}.
They however rely on model-dependent assumptions.
Nevertheless, they are still weaker than direct limits expected from linear colliders, 
by about one order of magnitude. Therefore, our conclusion that a linear collider is the
most promising project for determining top-quark electroweak couplings holds even
in comparison with these indirect bounds.

\section{Conclusions}
\label{sec:conclusions}

We evaluated the potential of future lepton colliders to reveal new physics effects in precision measurements of top-quark pair production. A broad set of dimension-six operators was considered, affecting notably the electroweak interactions of the top quark. Four-fermion $e^+e^-t\,\bar{t}$ operators and the CP-violating imaginary parts of electroweak dipole operators were also included. Predictions at next-to-leading order in QCD for these operators were made available in the \mg\ framework.

We studied the sensitivity of a large number of observables, as well as the impact of the centre-of-mass energy and beam polarization. Combining measurements of the top-quark polarization and of CP-odd observables with that of the cross section and forward-backward asymmetry increases the sensitivity to the real and imaginary parts of the dipole operators, respectively. We also examined the power of additional constraints, such as the measurement of the top-quark width in a threshold scan or of bottom-quark pair production.
We observed that even with an extended set of observables, control over the beam polarization remains an important handle to simultaneously constrain the contributions of vector and axial-vector operators. Operation at high centre-of-mass energy provides tight bounds on four-fermion operators whose contributions grows quadratically with the energy, with respect to standard-model ones. The inclusion of data acquired at two centre-of-mass energies is crucial in a global fit of two-fermion and four-fermion operators.

To effectively and simultaneously cover all considered directions of the effective-field-theory parameter space, we constructed a set of statistically optimal observables that maximally exploits the information contained in the fully differential $\bwbw$ distribution. The global reach achieved with their measurements considerably exceeds the one obtained through the determination of cross sections and forward-backward asymmetries only. A combination of statistically optimal observable measurements at two different centre-of-mass energies is sufficient to simultaneously constrain the ten operator coefficients considered. Larger separations between the two centre-of-mass energies resolve approximate degeneracies and bring global limits closer to individual ones. Beam polarization helps increasing individual sensitivities and reducing global correlations.

Our study takes into account experimental effects in the selection of lepton-plus-jets events, the reconstruction of the top-quark candidates, the presence  of background processes and of significant tails in the luminosity spectrum. We use conservative estimates for the effective efficiency needed to reproduce the statistical uncertainties obtained in full-detector simulation studies. It is expected that systematic uncertainties can be controlled to a similar level.

Improved statistically optimal observables could help addressing identified reconstruction challenges in the future. Definitions symmetrized over the two final-state $b$ jets could mitigate the impact of migrations due to the mis-pairing of top-quark decay products that are important at low energies, especially with a left-handed electron beam polarizations. In an inclusive $\epem\to\bwbw$ analysis, accounting for the effective-field-theory dependence of non-resonant contributions would effectively turn the single top-quark production background which becomes large at high energies into a signal component. Considering the centre-of-mass energy as an additional kinematic variable instead of fixing it to its nominal value in the optimal observable definitions could also help exploiting the significant lower tail of the beam-energy spectrum which develops in $\!\tev$ operation.

The projected reach of circular-collider-, ILC- and CLIC-like operation scenarios was presented in \hyperref[fig:fit_opt_benchmark_cc]{Figs.\,\ref{fig:fit_opt_benchmark_cc}}, \ref{fig:fit_opt_benchmark} and \ref{fig:fit_opt_benchmark_clic}, respectively. The individual limits on the coefficients of the operators modifying top-quark electroweak couplings are one to three orders of magnitude better than present constraints. Improvements by factors of three to two hundred are also expected compared to the most optimistic prospect for the individual reach of the HL-LHC. Clean global analyses are moreover readily achievable at future lepton colliders, yielding robust direct constraints of limited model dependence.

\section*{Acknowledgements}
The studies reported here have greatly benefited from interaction with colleagues involved in the ILC, CLIC, FCCee and CEPC projects. In particular, we
acknowledge the effort of the ILC and CLIC collaborations to develop the Monte Carlo simulation infrastructure and to perform realistic studies of
detector effects. We would like to thanks the authors of Ref.\,\cite{Buckley:2015lku} for providing us the likelihood leading to their Fig.\,7, and Michael Russell for updates on \textsc{TopFitter} results.
MP and MV are supported by the Spanish national program for particle physics project FPA2015-65652-C4-3-R (MINECO/FEDER).
MP is supported by the ``Severo Ochoa'' Grant SEV-2014-0398-05, reference BES-2015-072974.
CZ is supported by IHEP under Contract No. Y7515540U1 and by the United States Department of Energy under Grant Contracts de-sc0012704.

\appendix

\section{Effective-field-theory expressions for anomalous vertices}
\label{sec:anomalous_vertices}

Anomalous vertices have been widely used to parametrize interactions of the top
quark beyond the standard model. Following Ref.~\cite{Schmidt:1995mr}, the
$Z,\gamma\to t\bar{t}$ and $t\to bW^+$ vertices can be written as
\begin{gather*}
	ie
	\big[
	\gamma^\mu (F_{1V}^{\gamma,Z}+\gamma_5 F_{1A}^{\gamma,Z})
	+\frac{i\sigma^{\mu\nu}q_\nu}{2m_t}
	(F_{2V}^{\gamma,Z} + \gamma_5 F_{2A}^{\gamma,Z})
	\big]
\\
	\frac{ig}{\sqrt{2}}
	\big[
	\gamma^\mu (F_{1L}^W P_L+ F_{1R}^W P_R)
	+\frac{i\sigma^{\mu\nu}q_\nu}{2m_t}
	(F_{2L}^W P_L + F_{2R}^W P_R)
	\big]
\end{gather*}
When it exists, the correspondence with the fully gauge invariant effective
field theory described in \autoref{sec:eft} is given the following:
\begin{align*}
F^\gamma_{1V} &= \frac{2}{3}
,\\
F^\gamma_{1A} &= 0
,\\
F^\gamma_{2V} &
	= 4\frac{m_t^2}{\Lambda^2}
	\left[ C_{uA}^R = \Re\{ C_{uW}^{(33)} + C_{uB}^{(33)} \} \right]
,\\
F^\gamma_{2A} &
	= 4\frac{m_t^2}{\Lambda^2}
	i
	\left[ C_{uA}^I = \Im\{ C_{uW}^{(33)} + C_{uB}^{(33)} \} \right]
,\\[4mm]
F^Z_{1V} &
	= \frac{\frac{1}{4}-\frac{2}{3} s_W^2}{s_Wc_W}
	-\frac{m_t^2}{\Lambda^2} 
	\frac{1}{2s_Wc_W}
	\left[
	C_{\varphi q}^V =
	C_{\varphi u}^{(33)}+(C_{\varphi q}^{1(33)}-C_{\varphi q}^{3(33)})
	\right]
,\\
F^Z_{1A} &
	= \frac{-\frac{1}{4}}{s_Wc_W}
	-\frac{m_t^2}{\Lambda^2} \frac{1}{2s_Wc_W}
	\left[ C_{\varphi q}^A =
	C_{\varphi u}^{(33)}-(C_{\varphi q}^{1(33)}-C_{\varphi q}^{3(33)})
	\right]
,\\
F^Z_{2V} &
	= 4\frac{m_t^2}{\Lambda^2}
	\left[ C_{uZ}^R = \Re\{ c^2_W C_{uW}^{(33)} - s^2_W C_{uB}^{(33)} \}/{s_Wc_W}
	\right]
,\\
F^Z_{2A} &
	= 4\frac{m_t^2}{\Lambda^2}
	i
	\left[ C_{uZ}^I = \Im\{ c^2_W C_{uW}^{(33)} - s^2_W C_{uB}^{(33)} \}/{s_Wc_W}
	\right]
,\\[4mm]
F^W_{1R} &
	= \frac{m_t^2}{\Lambda^2}
	C_{\varphi ud}^{(33)*}
,\\
F^W_{1L} &
	= 1
	+\frac{m_t^2}{\Lambda^2}
	2\left[
	C_{\varphi q}^{3(33)} = \frac{1}{2}(C_{\varphi q}^+ -\frac{1}{2}[C_{\varphi q}^V-C_{\varphi q}^A])
	\right]
,\\
F^W_{2L} &
	= 8\frac{m_t^2}{\Lambda^2}
	C_{dW}^{(33)*}
,\\
F^W_{2R} &
	= 8\frac{m_t^2}{\Lambda^2}
	\left[
	C_{u W}^{(33)} = s_W^2 C_{uA} + s_W c_W C_{uZ}
	\right]
.
\end{align*}
It is however worth stressing a few shortcomings of the anomalous coupling
approach. An important one is that it completely misses four-fermion operators
which are often generated at tree-level when introducing new heavy states beyond
the standard model. The $F^\gamma_{1V,1A}$ anomalous couplings also break
electromagnetic gauge invariance. Their inclusion would thus render the theory
sick and preclude the computation of quantum corrections. Finally, the anomalous couplings of the top quark to a photon or $Z$ boson are
in general allowed to be complex although they then break unitarity.


\section{Additional results for the CLIC-like scenario}
\label{sec:clic_scenario}

\begin{figure}[p]\raggedleft
{\includegraphics[width=.5\textwidth]{sensitivities_left_0_s.mps}}%
{\includegraphics[width=.5\textwidth]{sensitivities_right_0_a.mps}}%
\vspace*{-2mm}
\caption{Sensitivity of the total (left) and forward-backward (right)
\eett\ cross sections to various operators, as a function of the centre-of-mass
energy, and for a mostly left-handed (left) and right-handed (right) electron beam
polarization. The dashed black line indicates the slope of a sensitivity scaling
as the centre-of-mass energy squared.}
\label{fig:sensitivities_def_clic}
\vspace*{5mm}

\includegraphics[width=.5\textwidth]{individual_left_s_clic.mps}%
\includegraphics[width=.5\textwidth]{individual_right_a_clic.mps}
\vspace*{-8mm}
\caption{Statistical one-sigma constraints on the effective operator
         coefficients as functions of the centre-of-mass energy, with either
         mostly left-handed (left) and mostly right-handed (right) electron beam polarizations,
         from either cross section (left) or forward-backward asymmetry (right) measurements
         for a fixed integrated luminosity times efficiency of $1\,\iab$.
         Different integrated luminosities are trivially obtained through a
         $1/\sqrt{\mathcal{L}\: [\iab]}$ rescaling.}
\label{fig:individual_def_clic}
\vspace*{5mm}
\scalebox{.9}{\includegraphics[scale=1.1]{fit_def_benchmark_corr_clic.mps}\hspace{2mm}%
\raisebox{-4.67mm}{\includegraphics[scale=1]{fit_def_benchmark_clic.mps}}}
\par\vspace{-2mm}
\caption{One-sigma constraints and correlation matrix (rounded to the first decimal place) arising from the measurement of cross sections and forward-backward asymmetries in a CLIC-like run scenario. The $t\,\bar t$ reconstruction efficiencies of \autoref{tab:efficiencies} are assumed and $\Lambda$ is fixed to $1\tev$.}
\label{fig:fit_def_benchmark_clic}
\end{figure}

\begin{figure}\raggedleft
{\includegraphics[width=.5\textwidth]{sensitivities_left_opt_clic.mps}}%
{\includegraphics[width=.5\textwidth]{sensitivities_right_opt_clic.mps}}%
\vspace*{-2mm}
\caption{Sensitivity of each statistically optimal observable to the
         corresponding operator coefficient, as a function of the centre-of-mass
         energy, and for $P(e^+,e^-)=(0\%,\pm80\%)$ beam polarizations.
}
\label{fig:sensitivities_opt_clic}
\vspace*{5mm}
\par
\includegraphics[width=.5\textwidth]{individual_left_opt_clic.mps}%
\includegraphics[width=.5\textwidth]{individual_right_opt_clic.mps}
\vspace*{-8mm}
\caption{Individual one-sigma constraints on the effective operator
         coefficients as functions of the centre-of-mass energy, with
         $P(e^+,e^-)=(0\%,\pm80\%)$ beam polarizations, from the measurements of
         statistically optimal observables, for a fixed integrated luminosity
         times efficiency of $1\,\iab$. Different integrated luminosities are
         trivially obtained through a $1/\sqrt{\mathcal{L}\: [\iab]}$
         rescaling.}
\label{fig:individual_opt_clic}
\end{figure}

We provide here additional results corresponding to the CLIC-like benchmark run
scenario introduced in \autoref{sec:run_scenarios}.
The sensitivities of the total and forward-backward-integrated cross sections,
as functions of the centre-of-mass energy and for a vanishing positron
polarization are shown in \autoref{fig:sensitivities_def_clic}. Individual
constraints deriving from the measurements of the cross section or
forward-backward asymmetry are shown in \autoref{fig:individual_def_clic}. They
are normalized for an integrated luminosity times efficiency of $1\,\iab$.
With the overall $t\,\bar{t}$ reconstruction efficiencies of \autoref{tab:efficiencies}, the global constraints resulting from the measurements of
cross sections and forward-backward asymmetries are displayed in
\autoref{fig:fit_def_benchmark_clic}. For $C_{\varphi q}^V$ and $C_{uZ}^R$, they
are almost an order of magnitude looser than the individual constraints.
Although we do not study the impact of the quadratic effective-field-theory
dependences, it can be expected to be non
negligible, as in the ILC-like scenario.

The sensitivities of each statistically optimal observable to the corresponding
operator coefficient, as functions of the centre-of-mass energy, are displayed
in \autoref{fig:sensitivities_opt_clic}. The individual constraints deriving
from optimal observable measurements, normalized for
a $1\iab$ integrated luminosity times efficiency, are shown in
\autoref{fig:individual_opt_clic}. The global constraints were already presented in
\autoref{fig:fit_opt_benchmark_clic}, on \autopageref{fig:fit_opt_benchmark_clic}. The global determinant parameter on the
eight-dimensional space of CP-conserving parameters is improved by a factor of
about $1.4$ compared to the one obtained from the measurements of cross sections
and forward-backward asymmetries. Most importantly, the correlation between
operator coefficients is reduced and their global limits approach their
individual ones. Sensitivity to the quadratic dependence of the dimension-six
operator coefficients is thus expected to be limited, so that
clearner effective-field-theory results are obtained.


\section{Selected results at NLO in QCD}
\label{app:numerics_nlo}

\mg~\cite{Alwall:2014hca} is employed with the following input parameters:
\begin{gather*}
\begin{aligned}
	m_b	&= 0 \text{ GeV}	\\
	m_t	&= 172.5 \text{ GeV}
\end{aligned}
\qquad
\begin{aligned}
	m_Z	&= 91.1876 \text{ GeV}	\\
	\Gamma_Z&= 2.4952 \text{ GeV}
\end{aligned}
\qquad
\begin{aligned}
	m_W	&= 79.82436 \text{ GeV}	\\
	\Gamma_W&= 2.085 \text{ GeV}
\end{aligned}
\\
	s_W^2		= 0.2337		\qquad
	\alpha		= 1/127.9		\qquad
	\alpha_S(m_t)	= 0.1080
\end{gather*}
which satisfy tree-level electroweak relations. Note however that, in the complex mass scheme necessary for the non-resonant $\epem\to\bwbw$ simulation at next-to-leading order in QCD of \autoref{sec:oo_theoretical_robustness}, a value of $m_W=80.385$\,GeV, hardcoded in lines \texttt{1645} and \texttt{1692} of \texttt{madgraph/core/base\_objects.py}, is employed (corresponding to $G_F=1.205895 \times 10^{-5}\,\text{GeV}^{-2}$).

The \texttt{TEFT\_EW} UFO model~\cite{Bylund:2016phk} is extended to include four-fermion operators and imaginary coefficients for the electroweak dipole operators. The necessary UV and R2 counterterms required for simulation at NLO in QCD are implemented.
The effective-field-theory dependence of the top-quark width is also taken into account.

\newcommand{\val}[5]{%
	\def\central{#1}
	\def\myzero{0}
	\ifx\central\myzero%
		\text{---}%
	\else%
		\underset{#2}{#1}%
		\begin{array}{@{\:}l@{}}
			\scriptstyle #4\%\\[-1.3mm]
			\scriptstyle \pm #5\%\\[-1.3mm]
			\scriptstyle #3\%
		\end{array}%
	\fi%
}%
\newcommand{\var}[5]{%
	\def\central{#1}
	\def\myzero{        0}
	\def\myozero{0}
	\ifx\central\myzero%
		\text{---}%
	\else%
	\ifx\central\myozero%
		\text{---}%
	\else
		\underset{#2}{#1}%
		\begin{array}{@{\:}l@{}}
			\scriptstyle #5\%\\[-1.3mm]
			\scriptstyle \pm #3\%\\[-1.3mm]
			\scriptstyle #4\%
		\end{array}%
	\fi\fi%
}%
\newcommand{\no}[1]{\text{---}}
\newcommand{\lintab}[2][1]{%
	\par\noindent%
	\scalebox{#1}{%
	\ensuremath{%
	\begin{array}{
		*{2}{c}
			@{\quad}|@{\quad}
		*{1}{c}
			@{\quad}|@{\quad}
		*{11}{c}
		}
	\text{pol}	& \sqrt{s} \text{ [GeV]}
	& \text{SM}
		& C_{lq}^A
		& C_{eq}^A
		& C_{\varphi q}^A
		& C_{lq}^V
		& C_{eq}^V
		& C_{\varphi q}^V
		& C_{uZ}^R
		& C_{uA}^R
		& C_{uZ}^I
		& C_{uA}^I
		& C_{uG}^R
	\\[2mm]
	\input{#2}
	\end{array}%
	}}%
}

The total and forward-backward-integrated cross sections for which we now provide the linear effective field theory dependence at next-to-leading order in QCD are parametrized as follows:
\begin{equation*}
	\sigma = 
		\sigma_\text{SM} 
		+ \left(\frac{1\text{ TeV}}{\Lambda}\right)^2 \sum_i C_i\; \sigma_i
		+ \left(\frac{1\text{ TeV}}{\Lambda}\right)^4 \sum_{i\le j} C_i C_j\; \sigma_{ij}.
\end{equation*}
The linear $\sigma_i$ dependences are provided in \hyperref[tab:nlo_xsec]{Tables\,\ref{tab:nlo_xsec}} and \ref{tab:nlo_afb}.
Central values, $k$-factors, scale and Monte-Carlo uncertainties will be formatted as
\begin{equation*}
\val	{\text{central value}}
	{\text{k-factor}}
	{-\text{scale down}}
	{+\text{scale up}}
	{\text{Monte Carlo}}
	.
\end{equation*}
The scale uncertainty is computed from the running of $\alpha_S(\mu)$ between
$\mu=m_t/2$ and $2m_t$.

\begin{sidewaystable}
	\par\noindent%
	\scalebox{.75}{%
	\ensuremath{%
	\begin{array}{
		*{2}{c}
			@{\quad}|@{\quad}
		*{1}{c}
			@{\quad}|@{\quad}
		*{11}{c}
		}
	\text{pol}	& \sqrt{s} \text{ [GeV]}
	& \text{SM}
		& C_{lq}^A
		& C_{eq}^A
		& C_{\varphi q}^A
		& C_{lq}^V
		& C_{eq}^V
		& C_{\varphi q}^V
		& C_{uZ}^R
		& C_{uA}^R
		& C_{uZ}^I
		& C_{uA}^I
		& C_{uG}^R
	\\[2mm]
	\input{nlo_s_lin.tex.bis}
	\end{array}%
	}}%

\caption{Linear effective field theory dependence of the total \eett\ cross section [fb]. The $+-$, and $-+$ labels specify the helicities of the electron and positron, respectively. Any mixted polarization can be optained through $(1-P_{e^-})(1+P_{e^+}) [-+] + (1+P_{e^-})(1-P_{e^+}) [+-]$. In particular, for unpolarized beams, denoted as $00$, the sum of $+-$ and $-+$ contributions is obtained. Note the large Monte-Carlo uncertainties affecting most of the smallest values.}
\label{tab:nlo_xsec}
\end{sidewaystable}

\begin{sidewaystable}
	\par\noindent%
	\scalebox{.75}{%
	\ensuremath{%
	\begin{array}{
		*{2}{c}
			@{\quad}|@{\quad}
		*{1}{c}
			@{\quad}|@{\quad}
		*{11}{c}
		}
	\text{pol}	& \sqrt{s} \text{ [GeV]}
	& \text{SM}
		& C_{lq}^A
		& C_{eq}^A
		& C_{\varphi q}^A
		& C_{lq}^V
		& C_{eq}^V
		& C_{\varphi q}^V
		& C_{uZ}^R
		& C_{uA}^R
		& C_{uZ}^I
		& C_{uA}^I
		& C_{uG}^R
	\\[2mm]
	\input{nlo_a_lin.tex.bis}
	\end{array}%
	}}%

\caption{Similarly as \autoref{tab:nlo_xsec}, linear effective field theory dependence of the forward-backward \eett\ cross section [fb].}
\label{tab:nlo_afb}
\end{sidewaystable}


\section{Optimal observables for scalar and tensor four-fermion operators}
\label{sec:lequs}

Some observables sensitive $O_{lequ}^{S,T}$ were already discussed in
\autoref{sec:obs_s_t}. The set of statistically optimal observables
can also be extended to achieve sensitivity to scalar and tensor four-fermion
operators. Due to their chirality, these operators do not
interfere with standard-model amplitudes (in the limit of vanishing lepton
masses) and yield no linear dependence in the \bwbw\ differential
distribution. The standard prescription for the construction of statistically
optimal observables can therefore not be applied. One can however still
define observables that are optimally sensitive to their quadratic dependences.
Because these two operators do not interfere with their own Hermitian
conjugates, only dependences on $|C_{lequ}^S|^2$, $|C_{lequ}^T|^2$,
$\Re\{C_{lequ}^SC_{lequ}^{T*}\}$, and $\Im\{C_{lequ}^SC_{lequ}^{T*}\}$ are
generated. Four optimal observables corresponding to each of these terms
in the $\eett\to\bwbw$ differential distribution can thus be defined.
Three new physical degrees of freedom are introduced, namely two norms and
one relative phase. The mapping between these quantities and the optimal
observables associated to $|C_{lequ}^S|^2$, $|C_{lequ}^T|^2$,
$\Re\{C_{lequ}^SC_{lequ}^{T*}\}$, $\Im\{C_{lequ}^SC_{lequ}^{T*}\}$ is both
non-linear and subject to constraints: norms are positive and the
phase has a period of $2\pi$. The additional four statistically optimal
observables are still defined at leading order and in the narrow width
approximation, on the \bwbw\ final state. A numerical method based
on amplitudes computed by \mg~\cite{Alwall:2014hca} in its standalone
\texttt{c++} mode is used to define them. Unlike the analytical approach used so
far, it does not include the subleading effective-field-theory dependence
arising in top-quark decays. The good agreement obtained cross-validates the two
methods.

A ILC-like benchmark scenario is assumed as before, with runs at centre-of-mass
energies of $500\,$GeV and $1\,$TeV. The global Gaussian constraints on the set
of $14$ operator coefficient combinations which include $|C_{lequ}^{S,T}|^2$ and
$\Re,\Im\{C_{lequ}^SC_{lequ}^{T*}\}$ in addition to the $10$ coefficients
previously considered are displayed as blue bands in \autoref{fig:ext_opt}. For
comparison, dark blue arrows show the bounds obtained without including scalar
and tensor operators. At this stage no relation is assumed between the different
quadratic combinations of scalar and tensor four-fermion operators. Expressing
them as functions of two norms and one relative phase as well as imposing
constraints on the domain of those quantities (norms are positive and phases
have a period of $2\pi$) leads to the limits shown as arrows of a lighter shade
of blue. The largest absolute value is retained as constraints become
non-Gaussian and asymmetric around zero after this non-linear transformation.
After marginalization over other degrees of freedom the relative phase between
$C^S_{lequ}$ and $C^T_{lequ}$ is left unconstrained and is therefore not
displayed in \autoref{fig:ext_opt}. By default, all four
$P(e^+,e^-)=(\pm30\%,\pm80\%)$ beam polarizations are used. The $+-$ and $-+$
($++$ and $--$) polarizations receive each $40\%$ ($10\%$) of the total
luminosity at each centre-of-mass energy. Splitting the integrated luminosity equally
between the $+-$ and $-+$ ones, as before, leads to the constraints shown with
arrows of the lightest shade of blue.

It is seen that introducing scalar and tensor operators mildly loosen the
constraints on other four-fermion operators. Runs with $++$ and $--$ beam
polarizations help mitigating this effect and strengthen the constraints
on $|C_{lequ}^{S,T}|^2$, as expected. The resulting reduction of integrated
luminosity spent on $+-$ and $-+$ beam polarization configurations causes
a limited degradation of constraints in some other directions of the
effective-field-theory parameter space. Overall, the use of like-sign
beam polarizations yields an improvement by a factor of about $1.1$,
in terms of GDP defined on the Gaussian constraints in the space of
$14$ combinations of operator coefficients artificially treated as independent degrees
of freedom without constraints on their domain.

\begin{figure}
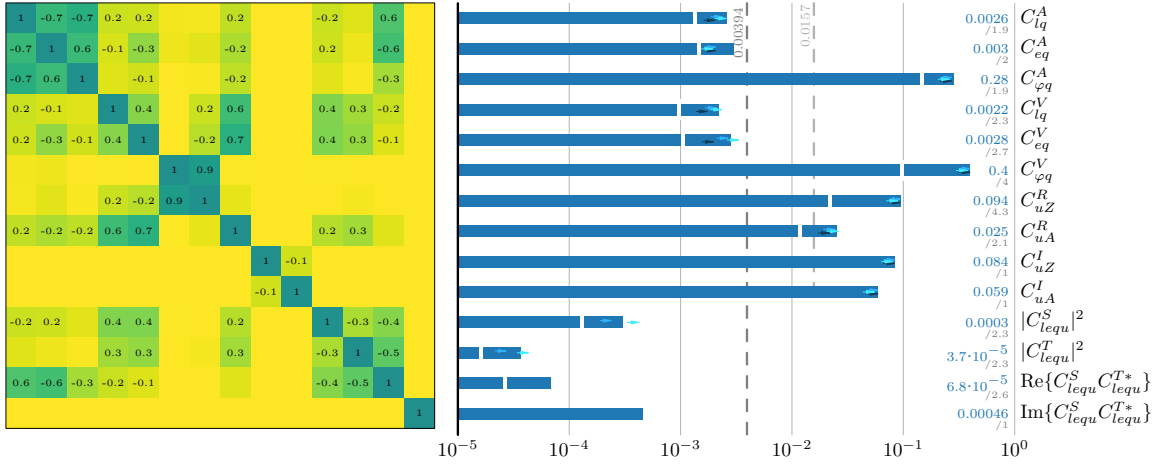

\adjustbox{max width = \textwidth}{%
\includegraphics{fit_opt_benchmark_corr_14.mps}%
\raisebox{-5mm}{\includegraphics{fit_opt_benchmark_14.mps}}}
\caption{Global constraints deriving from the measurements of a set of statistically optimal observables extended to include dependences on the scalar and tensor four-fermion operators for a modified ILC-like run scenario. Runs collecting $500\,\ifb$ at $500\gev$ and $1\,\iab$ at $1\,$TeV, as well as the overall $t\,\bar{t}$ reconstruction efficiencies of \autoref{tab:efficiencies} are assumed. The four $P(e^+,e^-)=(\pm30\%,\pm80\%)$ beam polarizations are exploited, the $+-$ and $-+$ ($++$ and $--$) receiving each $40\%$ ($10\%$) of the integrated luminosity.}
\label{fig:ext_opt}
\end{figure}


\section{Conversion to LHC TOP WG EFT conventions}
\label{sec:lhc_top_wg}
The effective-field-theory conventions employed in this paper are somewhat different than the standards recently established by the LHC TOP WG in Ref.~\cite{AguilarSaavedra:2018nen}. We provide here the conversion to those standards:
\begin{small}%
\begin{equation}
\begin{pmatrix}
c_{\varphi Q}^-\\
c_{\varphi t}  \\
c_{uW}^{[I]}   \\
c_{tZ}^{[I]}   \\
c_{Ql}^{-(1)}  \\
c_{Qe}^{(1)}   \\
c_{tl}^{(1)}   \\
c_{te}^{(1)}   \\
\end{pmatrix}
=
\left(\begin{array}{*{10}{c}}
0	& 0	&-y_t^2/2	& 0	& 0	& y_t^2/2	& 0	& 0	\\
0	& 0	& y_t^2/2	& 0	& 0	& y_t^2/2	& 0	& 0	\\
0	& 0	& 0	& 0	& 0	& 0	& c_W y_t e	& s_W y_t e	\\
0	& 0	& 0	& 0	& 0	& 0	& y_t e	& 0	\\
 - 1/2	& 0	& 0	& 1/2	& 0	& 0	& 0	& 0	\\
0	&  - 1/2	& 0	& 0	& 1/2	& 0	& 0	& 0	\\
1/2	& 0	& 0	& 1/2	& 0	& 0	& 0	& 0	\\
0	& 1/2	& 0	& 0	& 1/2	& 0	& 0	& 0
\end{array}\right)
\begin{pmatrix}
C_{lq}^A	\\
C_{eq}^A	\\
C_{\varphi q}^A	\\
C_{lq}^V	\\
C_{eq}^V	\\
C_{\varphi q}^V	\\
C_{uZ}^{R[I]}	\\
C_{uA}^{R[I]}	\\
\end{pmatrix}
\end{equation}\end{small}%
Our main results, the constraints deriving from the measurements of statistically optimal observables are translated to those conventions for the CC-, ILC- and CLIC-like scenarios in \autoref{fig:lhc_top_wg}. Because these conventions are less natural to the description of the top-quark pair production process at lepton colliders, larger correlations are present. The qualitatively different sensitivities to vector and axial vector operators would not have been manifest either with these degrees of freedom.

\begin{figure}[p]
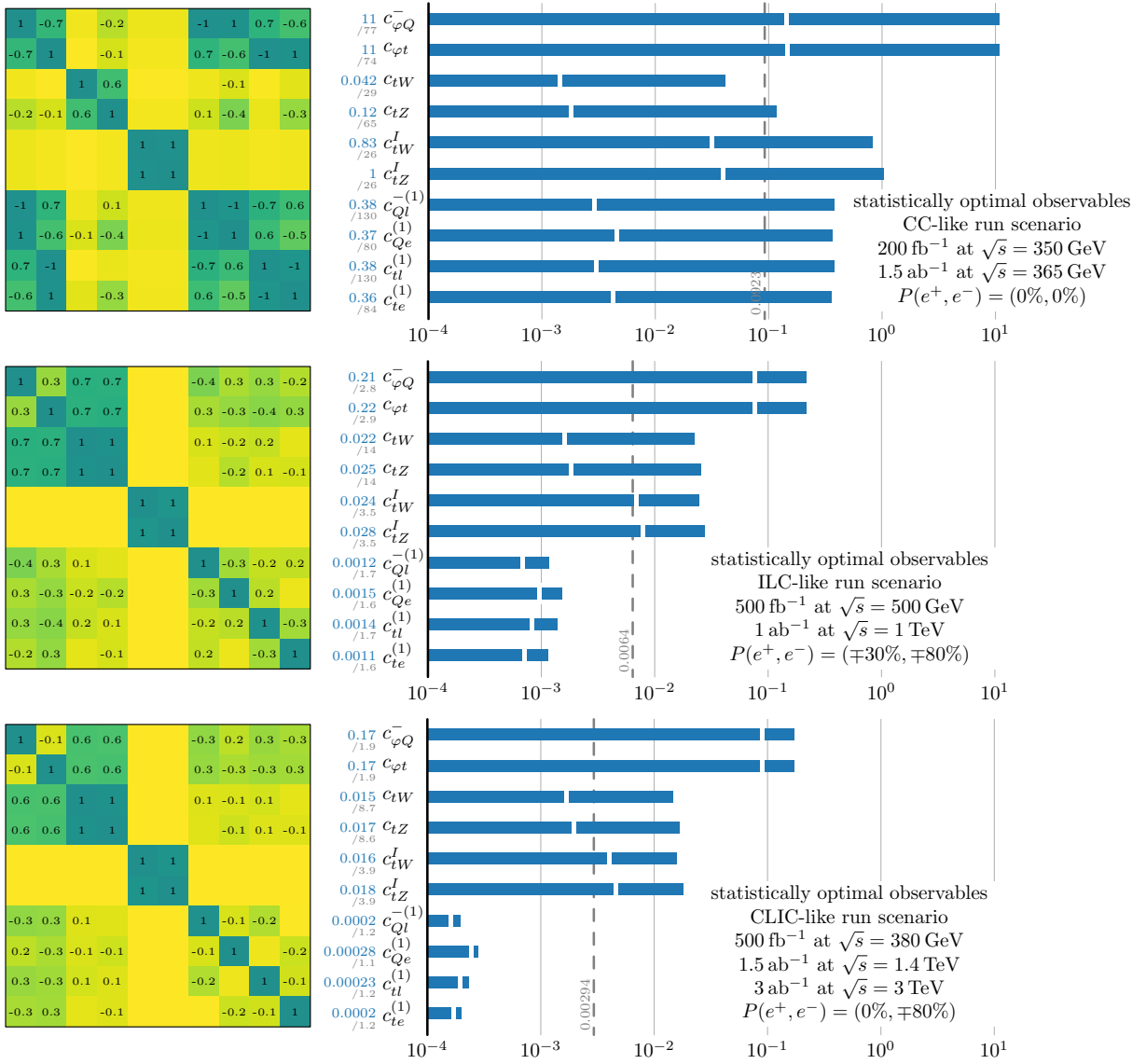

\scalebox{.8}{\raisebox{5.25mm}{%
\includegraphics[scale=1.075]{fit_opt_benchmark_corr_cc_r.mps}}\hspace{5mm}%
\includegraphics[scale=1]{fit_opt_benchmark_cc_r.mps}}%
\\[2mm]
\scalebox{.8}{\raisebox{5.25mm}{%
\includegraphics[scale=1.075]{fit_opt_benchmark_corr_r.mps}}\hspace{3.5mm}%
\includegraphics[scale=1]{fit_opt_benchmark_r.mps}}%
\\[2mm]
\scalebox{.8}{\raisebox{5.25mm}{%
\includegraphics[scale=1.075]{fit_opt_benchmark_corr_clic_r.mps}}\hspace{2mm}%
\includegraphics[scale=1]{fit_opt_benchmark_clic_r.mps}}%
\caption{Global one-sigma constraints and correlation matrices deriving from the measurements of statistically optimal observables for circular-collider- (top), ILC- (centre) and CLIC-like (bottom) run scenarios, in the effective-field-theory conventions established by the LHC TOP WG~\cite{AguilarSaavedra:2018nen}. The overall $t\,\bar{t}$ reconstruction efficiencies of \autoref{tab:efficiencies} are employed and $\Lambda$ is fixed to $1\tev$. White marks stand for the individual constraints obtained when all other operator coefficients vanish. Dashed vertical lines provide the average constraint strengths in terms of GDP.}
\label{fig:lhc_top_wg}
\end{figure}


\section{Conversion of {TopFitter} limits}
\label{app:topfitter}

This appendix details the conversion to our normalizations of the limits obtained by the \textsc{TopFitter} group~\cite{Buckley:2015nca, Buckley:2015lku, Englert:2017dev} on the operator coefficients relevant for top-quark production at lepton colliders. The conversion factors are the following:
\begin{equation*}
\begin{aligned}
  \bar{c}_{\varphi q} &= -\frac{1}{2}y_t^2 \frac{v^2}{\Lambda^2} C_{\varphi q}^-, \\
  \bar{c}_{\varphi u} &= \frac{1}{2} y_t^2 \frac{v^2}{\Lambda^2} C_{\varphi u}, \\
\end{aligned}
\qquad
\begin{aligned}
  \bar{C}_{\varphi q}^1 &= \frac{1}{2} y_t^2 \frac{v^2}{\Lambda^2} C_{\varphi q}^1, \\
  \bar{C}_{\varphi q}^3 &= \frac{1}{2} y_t^2 \frac{v^2}{\Lambda^2} C_{\varphi q}^3, \\
\end{aligned}
\qquad
\begin{aligned}
  \bar{C}^{33}_{uB} &= y_t g'\, \frac{v^2}{\Lambda^2} C_{uB}, \\
  \bar{C}^{33}_{uW} &= y_t g\, \frac{v^2}{\Lambda^2} C_{uW},
\end{aligned}
\end{equation*}
where $v\simeq 246\gev$, $y_t\simeq 0.99$, $g\simeq0.65$ and $g'\simeq 0.35$.
The individual and global limits obtained in these references are displayed in \autoref{tab:topfitter}. The first five columns correspond to the analysis of Tevatron and LHC run~I measurements performed in Ref~\cite{Buckley:2015lku, Englert:2017dev}. Prospects for LHC run~III or its high-luminosity phase are given in the sixth column. They are extracted from Fig.\,2 of Ref.\,\cite{Englert:2017dev}. In its last four columns, \autoref{tab:topfitter} also presents the individual and global prospects made in Ref.\,\cite{Englert:2017dev} for cross section and forward-backward asymmetry measurements in $\eett$ production. The measurements are assumed to have total relative uncertainties of $2\%$ for each of the three $P(e^+,e^-)=(\pm30\%,\mp80\%)$, $(0\%,0\%)$ beam polarizations at either $\sqrt{s}=500\gev$ or $3\tev$.

\begin{table}
\adjustbox{max width = \textwidth}{%
\tiny
\begin{tabular*}{1.1\textwidth}{@{\extracolsep{\fill}}lc@{\,}cc@{\,}c@{\hspace{5ex}}*{3}{@{\;}c@{\,}c}@{}}
\hline
& \multicolumn{4}{c}{Ref.\,\cite{Buckley:2015lku}}
& \multicolumn{6}{c}{Ref.\,\cite{Englert:2017dev}}
\\
& \multicolumn{2}{c}{Fig.\,12: all, $95\%\,$C.L.}
& \multicolumn{2}{c}{\hspace*{-3ex}Fig.\,7: $t\bar{t}V$, ind.}
& \multicolumn{2}{c}{\hspace*{-2ex}Fig.\,2: $t\bar{t}Z$, ind., $95\%\,$C.L.\hspace*{-2ex}}
& \multicolumn{2}{c}{Fig.\,6, $\epem$, $500\gev$}
& \multicolumn{2}{c}{Fig.\,7, $\epem$, $3\tev$}
\\
& glo.	& ind.\ 	& $95\%\,$C.L.	&$68\%\,$C.L.
& present	& run\,III or HL
& ind.\	$95\%\,$C.L.	& glo.\ $95\%\,$C.L.
& ind.\	$95\%\,$C.L.	& glo.\ $95\%\,$C.L.
\\[.5ex]\hline\noalign{\vskip.5ex}
$C_{\varphi q}^-$	&			&			&			&			& $[ -17,  18]$	& $[-6.9,5.7]$
& $[-0.51,0.51]$	& $[-9.0,8.9]$		& $[-0.60,0.59]$	& $[-6.4,5.8]$
\\
$C_{\varphi q}^1$	&			& $[-12, 13]$		& $[ -11,  11]$		& $[-7.5, 7.7]$		&		&
&
\\
$C_{\varphi q}^3$	& $[-8.5, 4.1]$		& $[-5.3, 3.1]$		& $[-6.2, 6.2]$		& $[-4.2, 4.2]$		&		&
&
\\
$C_{\varphi u}$		&			& $[ -20,  17]$		& $[ -24,  17]$		& $[ -18,  10]$		& $[ -20,  20]$	& $[-12,9.4]$
& $[-0.45,0.45]$	& $[-6.8,7.2]$		& $[-0.52,0.53]$	& $[-1.7,1.9]$
\\
$C_{uB}$		&			& $[ -20,  14]$		& $[ -20,  13]$		& $[ -15, 8.2]$		& $[ -22,  22]$	& $[-13,13]$
& $[-0.035,0.035]$	& $[-0.48,0.47]$	& $[-0.037,0.030]$	& $[-0.23,0.085]$
\\
$C_{uW}$		& $[-6.2, 5.2]$		& $[-2, 2.8]$		& $[-2.5, 1.0]$		& $[-2.1,0.57]$		& $[-6.5, 6.5]$	& $[-3.9,3.9]$
& $[-0.028,0.028]$	& $[-0.46,0.45]$	& $[-0.028,0.022]$	& $[-0.22,0.11]$
\\[.5ex]
\hline
\end{tabular*}}
\caption{Individual and global constraints derived by the \textsc{TopFitter} group, in the normalization employed in this paper. The results in the first six columns correspond to an analysis of Tevatron and LHC run~I measurements. The last four columns the individual and global prospects made in Ref.\,\cite{Englert:2017dev} for the constraints from measurements of the $\eett$ cross section and forward-backward asymmetry measurements.
We set $\Lambda=1\tev$. The numerical values for Fig.\,7 of Ref.\,\cite{Buckley:2015lku} have been obtained from a quartic fit to the likelihood provided by the authors. Others are directly obtained from the figures.
}
\label{tab:topfitter}
\end{table}

\bibliographystyle{apsrev4-1_title}
\bibliography{EW_EFT}

\end{fmffile}
\end{document}